\documentclass[useAMS,usenatbib,usegraphics,fleqn]{mn2e}
\usepackage{epsf}
\usepackage{epsfig}
\usepackage{amssymb}
\usepackage{amsmath}

\newcommand{\refsim}{\textsc{ref}}
\newcommand{\agn}{\textsc{agn}}
\newcommand{\nocool}{\textsc{nocool}}

\newcommand{\planck}{\textit{Planck}}
\newcommand{\wmap}{\textit{WMAP}}

\newcommand{\calsim}{\textsc{bahamas}}
\defcitealias{LeBrun2014}{L14}

\title[Scatter and evolution of hot gas properties]{The scatter and evolution of the global hot gas properties of simulated galaxy cluster populations}

\author[A. M. C. Le Brun et al.]{Amandine~M.~C.~Le Brun,$^{1,2}$\thanks{E-mail: amandine.le-brun@cea.fr}
Ian~G.~McCarthy,$^{2}$
Joop Schaye$^3$
and Trevor~J.~Ponman$^4$
\\
$^{1}$Laboratoire AIM, IRFU/Service d'Astrophysique -- CEA/DRF -- CNRS -- Universit\'e Paris Diderot, B\^at. 709, CEA-Saclay,\\ 91191 Gif-sur-Yvette Cedex, France\\
$^{2}$Astrophysics Research Institute, Liverpool John Moores University, 146 Brownlow Hill, Liverpool L3 5RF\\
$^{3}$Leiden Observatory, Leiden University, P. O. Box 9513, 2300 RA Leiden, the Netherlands\\
$^{4}$Astrophysics and Space Research Group, School of Physics and Astronomy, University of Birmingham, Edgbaston, Birmingham\\ B15 2TT, UK
}

\date{Accepted 2016 December 22. Received 2016 December 16; in original form 2016 June 14.}

\pubyear{2016}

\begin{document}
\label{firstpage}
\pagerange{\pageref{firstpage}--\pageref{lastpage}}
\maketitle

\begin{abstract}
We use the cosmo-OWLS suite of cosmological hydrodynamical simulations to investigate the scatter and evolution of the global hot gas properties of large simulated populations of galaxy groups and clusters.  Our aim is to compare the predictions of different physical models and to explore the extent to which commonly-adopted assumptions in observational analyses (e.g.\ self-similar evolution) are violated.  We examine the relations between (true) halo mass and the X-ray temperature, X-ray luminosity, gas mass, Sunyaev--Zel'dovich (SZ) flux, the X-ray analogue of the SZ flux ($Y_X$) and the hydrostatic mass. For the most realistic models, which include AGN feedback, the slopes of the various mass--observable relations deviate substantially from the self-similar ones, particularly at late times and for low-mass clusters. The amplitude of the mass--temperature relation shows negative evolution with respect to the self-similar prediction (i.e.\ slower than the prediction) for all models, driven by an increase in non-thermal pressure support at higher redshifts.  The AGN models predict strong positive evolution of the gas mass fractions at low halo masses. The SZ flux and $Y_X$ show positive evolution with respect to self-similarity at low mass but negative evolution at high mass.  The scatter about the relations is well approximated by log-normal distributions, with widths that depend mildly on halo mass. The scatter decreases significantly with increasing redshift. The exception is the hydrostatic mass--halo mass relation, for which the scatter {\it increases} with redshift. Finally, we discuss the relative merits of various hot gas-based mass proxies.
\end{abstract}



\section{Introduction}

Galaxy clusters are potentially powerful systems for measuring fundamental cosmological parameters, such as the overall matter density of the Universe, the amplitude of the matter power spectrum, as well as the evolution of dark energy (for recent reviews, see \citealt*{Voit2005,Allen2011,Kravtsov2012}).  

The classical test is to compare the theoretically predicted and observed evolutions of the abundance of galaxy clusters. Since the abundance of dark matter haloes is a strong function of mass, to exploit clusters for cosmological purposes, one usually requires an accurate method for estimating the masses of individual clusters using (generally speaking) quite limited observational information (but see \citealt{Caldwell2016} and \citealt{Ntampaka2016} for mass-independent tests).  Furthermore, the scatter and covariance of the adopted mass--observable relations must be properly included in the cosmological modelling, and a detailed knowledge of the selection function of the survey is also necessary.  

The use of theoretical models/simulations has become commonplace to calibrate mass--observable relations and to assess their scatter and biases \citep*[e.g.][]{Kravtsov2006}.  Aside from assisting in the calibration of absolute mass measurements, theoretical models are required to predict the abundance of clusters as a function of their mass for a given set of cosmological parameters (e.g.\ \citealt{Jenkins2001,Tinker2008}).  Here, we note that a number of recent studies have shown that the predicted abundance is sensitive to the details of feedback processes associated with galaxy formation (e.g.\ \citealt{Cui2014,Cusworth2014,Martizzi2014,Velliscig2014}), as energetic feedback can eject baryons from collapsed structures (e.g.\ \citealt{McCarthy2011}) and lower their total mass, thereby reducing the number of haloes above a given mass threshold.

Ongoing and upcoming galaxy cluster surveys, such as the Dark Energy Survey \citep{DES2005}, \textit{eRosita} \citep{Merloni2012}, \textit{Euclid} \citep{Laureijs2011}, SPT-3G \citep{Benson2014}, and Advanced ACTpol \citep{Henderson2016}, will deliver samples of tens of thousands of galaxy clusters.  With such large samples becoming available, the limiting uncertainties in the cosmological analyses will be due to systematic (as opposed to statistical) errors, the largest of which are likely to be associated with absolute mass calibration on the observational side and our incomplete knowledge of the effects of galaxy formation physics on the total masses and observable properties (e.g.\ X-ray luminosity, Sunyaev--Zel'dovich (SZ) flux, etc.) of clusters on the theoretical side.  

Cosmological hydrodynamical simulations can help to address both of these issues, by calibrating  the mass--observable relations (or, alternatively, by providing a testbed for direct observational mass reconstruction algorithms, such as synthetic weak lensing maps, which observers can use to calibrate the mass--observable relations empirically) and by providing a self-consistent framework for capturing the effects of galaxy formation physics on the predicted halo mass distribution.  However, an important caveat is that predictions of current simulations are often sensitive to the details of the `sub-grid' modelling of important feedback processes \citep{LeBrun2014,Sembolini2016,McCarthy2016}, as we will demonstrate here as well.  Therefore, continual confrontation of the simulations with the observations (via production of realistic synthetic observations of the simulations) is also needed to test the realism of the former.  

Encouragingly, the realism of simulations of galaxy clusters, in terms of their ability to reproduce the observed properties of local clusters, has been improving rapidly in recent years and is due in large part to the inclusion of energetic AGN feedback \citep[e.g.][]{Sijacki2007,Bhattacharya2008,Puchwein2008,Fabjan2010,McCarthy2010,Planelles2013,LeBrun2014,Planelles2014,McCarthy2016}.  However, much less is known about the realism of such simulations at higher redshifts, where there are fewer high-quality observations with which to compare the simulations, and there are significantly greater uncertainties in the role of important selection effects for the observed systems.  The sparseness of high-quality observations of high-redshift systems means that cosmological tests using distant clusters will have to rely much more heavily on simulations, both to help calibrate the mass--observable relations and self-consistently predict the abundance of haloes in the presence of baryons.  It is therefore crucial to examine what current simulations predict in terms of the evolution of the hot gas properties of clusters.

In the present study, we use the cosmo-OWLS \citep{LeBrun2014,McCarthy2014} suite of large-volume cosmological hydrodynamical simulations to conduct a study of the scatter and evolution of the global hot gas properties of large populations of galaxy groups and clusters as a function of the important non-gravitational physics of galaxy formation.  We will examine to what extent the predicted scaling relations evolve self-similarly (both in terms of amplitude and slope), can be characterised by simple power-laws with log-normal scatter, and whether or not the scatter itself depends on mass and redshift.  We will examine a large number of commonly-used scaling relations, including the dependencies on the true total halo mass of gas mass, (soft) X-ray luminosity, SZ flux, $Y_X$ (the X-ray analog of the SZ flux), the mass-weighted and spectroscopic temperatures and the X-ray hydrostatic mass.  We also present an analysis of the evolution and scatter of the X-ray luminosity--temperature relation (in Appendix~\ref{app:LT}).

This work extends upon an already relatively large body of previous studies of the impact of the non-gravitational physics of galaxy formation on the evolution of scaling relations such as those of \citet{Short2010,Stanek2010,Battaglia2010,Battaglia2012,Fabjan2011,Planelles2013} and \citet{Pike2014} which will be examined in the context of the results of the present study in Section~\ref{sec:comp}.

The present paper is organised as follows. We briefly introduce the cosmo-OWLS simulation suite used here in Section~\ref{sec:sims} and summarise self-similar theory in Section~\ref{sec:sstheory}.  We then describe how we fit the mass--observable scaling relations and the scatter about them in Section~\ref{sec:fitting}.  In Sections~\ref{sec:slopeevo},~\ref{sec:normevo} and~\ref{sec:scatter}, we examine the evolution of the mass slopes, amplitude, and scatter, respectively, in the various mass--observable relations.  We then examine the scatter and evolution of the hydrostatic bias in Section~\ref{sec:HSE}.  Finally, we conduct a short comparison to previous studies in Section~\ref{sec:comp} and summarise and discuss our findings in Section~\ref{sec:disc}. 

Throughout the paper, masses, luminosities, temperatures, integrated Sunyaev--Zel'dovich signal and X-ray analogues of the integrated Sunyaev--Zel'dovich signal are quoted in physical $\textrm{M}_{\odot}$, $\textrm{erg s}^{-1}$, keV, $\textrm{Mpc}^2$ and $\textrm{M}_{\odot}~\textrm{keV}$, respectively; $\ln$ denotes natural logarithm, while $\log_{10}$ corresponds to decimal logarithm.


\section{Simulations}
\label{sec:sims}

\begin{table*}
\centering
\caption{cosmo-OWLS runs presented here and their included sub-grid physics. Each model has been run in both the \wmap7 and \planck~cosmologies.}
\begin{tabular}{|l|l|l|l|l|l|l|}
\hline
Simulation & UV/X-ray background & Cooling & Star formation & SN feedback & AGN feedback & $\Delta T_{\rm heat}$ \\
\hline
\nocool & Yes & No & No & No & No & ...\\
\refsim & Yes & Yes & Yes & Yes & No & ...\\
\agn~8.0 & Yes & Yes & Yes & Yes & Yes & $10^{8.0}$ K\\
\agn~8.5 & Yes & Yes & Yes & Yes & Yes & $10^{8.5}$ K\\
\hline
\end{tabular}
\label{table:owls}
\end{table*}

We take advantage of the cosmo-OWLS suite of cosmological simulations described in detail in \citet{LeBrun2014} (hereafter \citetalias{LeBrun2014}; see also \citealt{McCarthy2014,vanDaalen2014,Velliscig2014}; \citealt*{LeBrun2015}). 

The cosmo-OWLS simulations constitute an extension to the OverWhelmingly Large Simulation project \citep[OWLS;][]{Schaye2010}.  cosmo-OWLS was conceived with cluster cosmology in mind and is composed of large volume ($400~h^{-1}$ Mpc on a side) periodic box simulations with $2\times1024^3$ particles using updated initial conditions derived from the \planck~data\footnote{We also ran simulations using initial conditions derived from the 7-year \textit{Wilkinson Microwave Anisotropy Probe} (\wmap) data. We will only present results from the \planck~cosmology ones here, but will comment on any notable differences with the equivalent runs in the \wmap7 cosmology.} (\citeauthor{Planck_cosmology}) \{$\Omega_{m}$, $\Omega_{b}$, $\Omega_{\Lambda}$, $\sigma_{8}$, $n_{s}$, $h$\} = \{0.3175, 0.0490, 0.6825, 0.834, 0.9624, 0.6711\}. This results in dark matter and (initial) baryon particle masses of $\approx4.44\times10^{9}~h^{-1}~\textrm{M}_{\odot}$ and $\approx8.12\times10^{8}~h^{-1}~\textrm{M}_{\odot}$, respectively. The gravitational softening of the runs presented here is fixed to $4~h^{-1}$ kpc (in physical coordinates below $z=3$ and in comoving coordinates at higher redshifts). We use $N_{ngb}=48$ neighbours for the smoothed particle hydrodynamics (SPH) interpolation and the minimum SPH smoothing length is fixed to one tenth of the gravitational softening. 

The simulations were conducted with a version of the Lagrangian TreePM-SPH code \textsc{gadget3} \citep[last described in][]{Springel2005a}, which was significantly modified to incorporate new `sub-grid' physics as part of the OWLS project. All the runs used here were started from identical initial conditions and only the included non-gravitational physics and some of its key parameters were methodically altered. We use four of the five physical models presented in \citetalias{LeBrun2014} here: 
\begin{description}
\item \textbf{{\small NOCOOL} :} A non-radiative model. It includes net heating from the \citet{Haardt2001} X-ray and ultra-violet photoionising background, whose effects on the intracluster medium (ICM) are however negligible.
\item \textbf{{\small REF} :} This model also incorporates prescriptions for metal-dependent radiative cooling \citep*{Wiersma2009a}, star formation \citep{Schaye2008}, stellar evolution, mass loss and chemical enrichment \citep{Wiersma2009b} from Type II and Ia supernovae and Asymptotic Giant Branch stars, and kinetic stellar feedback \citep{DallaVecchia2008}.
\item \textbf{{\small AGN} 8.0} and \textbf{{\small AGN} 8.5}: In addition to the physics included in the {\small REF} model, these models include a prescription for supermassive black hole growth (through both Eddington-limited Bondi--Hoyle--Lyttleton accretion and mergers with other black holes) and AGN feedback (\citealt{Booth2009}, which is a modified version of the model originally developed by \citealt{Springel2005b}). The black holes accumulate the feedback energy until they can heat neighbouring gas particles by a pre-determined amount $\Delta T_{\rm heat}$. As in \citet{Booth2009}, 1.5 per cent of the rest-mass energy of the gas which is accreted on to the supermassive black holes is employed for the feedback. This yields a satisfactory match to the normalisation of the black hole scaling relations (\citealt{Booth2009}; see also \citetalias{LeBrun2014}) which is insensitive to the exact value of $\Delta T_{\rm heat}$. The two AGN models used here only differ by their value of $\Delta T_{\rm heat}$, which is the most important parameter of the AGN feedback model in terms of the gas-phase properties of the resulting simulated population of groups and clusters (\citealt{McCarthy2011}; \citetalias{LeBrun2014}). It is fixed to $\Delta T_{\rm heat}=10^{8.0}$ K for \agn~8.0 and  $\Delta T_{\rm heat}=10^{8.5}$ K for \agn~8.5. Note that since the same quantity of gas is being heated in these models, more time is required for the black holes to accrete a sufficient amount of gas for heating the adjacent gas to a higher temperature. Hence, increased heating temperatures result into more episodic and more violent feedback episodes.
\end{description}

Table~\ref{table:owls} provides a list of the runs used here and the sub-grid physics that they include. 

Haloes are identified with a standard friends-of-friends (FoF) percolation algorithm on the dark matter particles with a linking length of 0.2 in units of the mean interparticle separation. The baryonic content of the haloes is identified by locating the nearest DM particle to each baryonic (i.e.\  gas or star) particle and connecting it with the FoF group of the DM particle. Artificial haloes are removed by carrying out an unbinding calculation with the \textsc{subfind} algorithm \citep{Springel2001,Dolag2009}: any FoF halo that does not have at least one self-bound substructure (called subhalo) is removed from the FoF groups list. 

Spherical overdensity masses $M_{\Delta}$ (where $M_{\Delta}$ is the total mass within a radius $r_{\Delta}$ that encloses a mean internal overdensity of $\Delta$ times the critical density of the Universe) with $\Delta=200$, 500 and 2500 have been computed for all the FoF haloes. The spheres are centred on the position of the minimum of the gravitational potential of the main subhalo (the most massive subhalo of the FoF halo). Then, all galaxy groups and clusters with $M_{500}\ge10^{13}~\textrm{M}_{\odot}$ are extracted from each snapshot for analysis. There are roughly $25,000$ such systems at $z=0$ in the \nocool~run with the \planck~cosmology, for example.  

The Sunyaev--Zel'dovich signal is characterised by the value of its spherically integrated Compton parameter $d_{A}^{2}Y_{500}=(\sigma_{T}/m_{e}c^{2})\int PdV$ where $d_{A}$ is the angular diameter distance, $\sigma_{T}$ the Thomson cross-section, $c$ the speed of light, $m_{e}$ the electron rest mass, $P=n_{e}k_{B}T_{e}$ the electron pressure and the integration is done over the sphere of radius $r_{500}$. The X-ray equivalent of the Sunyaev--Zel'dovich signal is characterised by $Y_{X,500}=M_{gas,500}T_{spec,cor}$ where $M_{gas,500}$ is the gas mass enclosed within $r_{500}$ and $T_{spec,cor}$ is the core-excised X-ray spectroscopic temperature (computed within the $[0.15-1]r_{500}$ annulus).

Note that contrary to what was done in \citetalias{LeBrun2014}, in the present study, we use true halo masses (as opposed to halo masses computed under the assumption of hydrostatic equilibrium) and that the X-ray luminosities, spectral temperatures, gas masses and integrated Sunyaev--Zel'dovich signals were computed within the true $r_{500}$ aperture and in the case of core-excised quantities within the annulus $[0.15-1]r_{500}$ (as opposed to within $r_{500,hse}$ and $[0.15-1]r_{500,hse}$ where $r_{500,hse}$ is the value of $r_{500}$ obtained when the halo masses are computed assuming hydrostatic equilibrium). The spectral temperatures and X-ray luminosities were computed using the synthetic X-ray methodology presented in \citetalias{LeBrun2014} though. The rationale behind these choices is that we aim to elucidate the relations between the hot gas observables and true halo mass, since those are useful for: (i) calibrating the mass--observable relations whose use is of paramount importance when doing (precision) cosmology with galaxy clusters, and (ii) making large synthetic surveys by applying template methods to large dark matter-only simulations (e.g.\ \citealt{Bode2007,Sehgal2007,Sehgal2010}).

The various cosmo-OWLS models have been compared to a wide range of observational data in \citetalias{LeBrun2014} and \citet{McCarthy2014} (see also \citealt{Hojjati2015,LeBrun2015}). In \citetalias{LeBrun2014}, we concentrated on the comparison to low-redshift properties such as X-ray luminosities and temperatures, gas mass fractions, entropy and density profiles, integrated Sunyaev--Zel'dovich signal, $I$-band mass-to-light ratio, dominance of the brightest cluster galaxy and central black hole masses and concluded that the fiducial AGN model (\agn~8.0) produces a realistic population of galaxy groups and clusters, broadly reproducing both the median trend and, for the first time, the scatter in physical properties over approximately two decades in mass ($10^{13}~\textrm{M}_{\odot} \la M_{500} \la 10^{15}~\textrm{M}_{\odot}$) and 1.5 decades in radius ($0.05 \la r/r_{500} \la 1.5$). In \citet{McCarthy2014}, we investigated the sensitivity of the thermal Sunyaev--Zel'dovich power spectrum to important non-gravitational physics and found that while the signal on small and intermediate scales is highly sensitive to the included galaxy formation physics, it is only mildly affected on large scales. 

We note that no explicit attempt was made in cosmo-OWLS to calibrate the simulations to reproduce the hot gas properties of groups and clusters.  As a consequence, some of the models perform better than others in terms of their comparison with observational data.  \citet{McCarthy2016} have recently calibrated the same simulation code to better reproduce the stellar masses of galaxies, while also reproducing the observed trend between hot gas mass fraction and halo mass as inferred from X-ray observations.  The calibrated model is referred to as \calsim. In terms of the local gas-phase properties, the \calsim~model is similar to the cosmo-OWLS \agn~8.0 model.  We will briefly comment below on any significant differences in the predicted evolution of the mass--observable relations of the \calsim~and \agn~8.0 models, but defer a detailed study of the \calsim~model to future work \citep{Barnes2017}.


\section{Self-similar scalings}
\label{sec:sstheory}

The dominant force in the formation and evolution of galaxy clusters is gravity.  Since gravity is scale free, galaxy clusters are, to `zeroth order', expected to obey self-similarity, such that the properties only depend upon the cluster mass, and more massive galaxy clusters are just scaled up versions of less massive ones with a scaling factor that depends only upon the mass ratios \citep[e.g.][]{White1978,Kaiser1986,Voit2005}.  While not expected to be strictly valid in a universe with a significant baryonic component\footnote{Note that even if one completely neglects baryons and their associated non-gravitational physics (as in the case of a dark matter only simulation), haloes still do not strictly obey self-similarity.  That is because in cold dark matter models the smallest objects collapse first, while the most massive (galaxy clusters) are still collapsing today.  The internal structure of haloes is sensitive to the time of collapse via the evolution of the background density (e.g.\ \citealt{Wechsler2002}).  Thus, while gravity may be scale free, the finite age of the Universe imprints a scale in structure formation.} (since baryons can radiate, which leads to star and black hole formation and then feedback), the self-similar model is still quite useful as a baseline for the interpretation of simulations and observations alike.

If one defines the total cluster mass (denoted as $M_\Delta$) as that contained within a region that encloses a mean overdensity $\Delta\rho_{crit}(z)$\footnote{as is often the convention. One could also use a multiple of the mean matter density $\rho_{m}(z)$ which leads to predictions for self-similar evolution in terms of powers of $1+z$ instead of $E(z)$. As discussed later in the text, we have tried fitting for evolution using powers of $1+z$ and found that it decreases the qualify of the fits.}, then, under the assumption of self-similarity, one can predict both the redshift evolution (which is in this case only due to the increasing mean density of the Universe) and the slope of a given mass--observable relation. The redshift dependence comes from the evolution of the critical density for closure: 
\begin{equation}
\rho_{crit}(z)\equiv\frac{3H(z)^2}{8\pi G}=E(z)^{2}\frac{3H_0^2}{8\pi G}=E(z)^2\rho_{crit}(z=0)
\end{equation}
\noindent where
\begin{equation}
E(z)\equiv \frac{H(z)}{H_0}=\sqrt{\Omega_m (1+z)^3+\Omega_\Lambda}
\end{equation}
\noindent gives the evolution of the Hubble parameter, $H(z)$, in a flat $\Lambda$CDM Universe. For instance, since  
\begin{equation}
M_\Delta\propto \Delta \rho_{crit}(z) r_\Delta^3
\end{equation} 
\noindent by definition, the cluster size scales as 
\begin{equation}
r_{\Delta}\propto M^{1/3}_{\Delta}E(z)^{-2/3}.
\label{eq:r-M} 
\end{equation}

Gas falling into a cluster potential well is heated via shocks and will eventually settle and achieve approximate virial equilibrium within that potential. The gas is then expected to have a temperature that is close to the virial temperature: 
\begin{equation}
k_BT_\Delta\propto-\frac{1}{2}\Phi=\frac{GM_\Delta\mu m_p}{2r_\Delta}
\label{eq:Tvir-M}
\end{equation}
where $k_B$ is the Boltzmann constant and $\mu$ is the mean molecular weight. Thus, the self-similar temperature--total mass relation can be obtained by combining equations~\eqref{eq:r-M} and~\eqref{eq:Tvir-M} and is as follows:
\begin{equation}
T_\Delta\propto M_\Delta^{2/3}E(z)^{2/3}.
\label{eq:ssT-M}
\end{equation}

The bolometric X-ray emission of massive clusters is dominated by thermal bremsstrahlung, implying that it scales as $L_X^{bol}\propto \rho^2\Lambda(T)r_\Delta^3\propto \rho^2T^{1/2}r_\Delta^3$ as the cooling function $\Lambda(T)\propto T^{1/2}$, which combined with equations~\eqref{eq:r-M} and~\eqref{eq:ssT-M} gives the self-similar bolometric X-ray luminosity--total mass and bolometric X-ray luminosity--temperature relations (we examine the luminosity--temperature relation in Appendix~\ref{app:LT}):
\begin{equation}
L_{X,\Delta}^{bol}\propto M_\Delta^{4/3}E(z)^{7/3}
\label{eq:ssLxbol-M}
\end{equation}
and
\begin{equation}
L_{X,\Delta}^{bol}\propto T_\Delta^{2}E(z).
\label{eq:ssLxbol-T}
\end{equation}
Hereafter, the bolometric X-ray luminosity $L_X^{bol}$ will be simply denoted as $L_X$.

Finally, $Y_{SZ,\Delta}\propto Y_{X,\Delta}\equiv M_{gas,\Delta}T_{\Delta}\propto M_\Delta T_\Delta$ assuming a constant gas fraction. Thus, the self-similar integrated Sunyaev--Zel'dovich signal--total mass and $Y_X$--total mass relations follow from equation~\eqref{eq:ssT-M}:
\begin{equation}
Y_{X/SZ,\Delta}\propto M_\Delta^{5/3}E(z)^{2/3}.
\label{eq:ssY-M}
\end{equation}

With the launch of the first X-ray telescopes, such as the Einstein Observatory, EXOSAT and ROSAT, in the 1980s-1990s, it was quickly realised that the self-similar model was incompatible with the observations of the X-ray luminosity--temperature relation \citep[e.g.][]{Mushotzky1984,Edge1991,Markevitch1998,Arnaud1999,Lumb2004,Osmond2004,Pratt2009,Hilton2012}, which was found to be significantly steeper than the self-similar expectation ($L_X\propto T^\alpha$ with $\alpha\simeq 2.5-3$ for clusters and likely even steeper for groups).  This led to the conclusion that some non-gravitational processes, most likely connected to galaxy formation, must be breaking the self-similarity by introducing some physical scales \citep[see for instance][for the first proposed solutions to this puzzle]{Evrard1991,Kaiser1991}.  

However, it should be noted that the X-ray luminosity is likely to be particularly sensitive to non-gravitational physics, since the luminosity is dominated by dense, centrally-concentrated gas with short cooling times and is therefore likely to be significantly affected by radiative cooling and feedback. Other quantities, such as the mean temperature or integrated SZ flux, are not expected to deviate as strongly from the self-similar prediction, owing to their stronger contribution from gas at large radii which is less affected by non-gravitational processes. Furthermore, even if (some of) the local relations do not strictly obey self-similarity with mass, the {\it evolution} can still be close to the self-similar prediction (and self-similarity with mass may be a better approximation to the truth at higher redshift than in the local Universe).  At present, there are precious few constraints on the evolution of the hot gas properties of clusters that are robust to uncertainties in selection effects. For these reasons, self-similar evolution is still commonly adopted in cosmological analyses (e.g.\  \citealt{Allen2008}; \citealt{Vikhlinin2009b}; \citeauthor{Planck2013a}; \citealt{Mantz2014,Mantz2015}; \citeauthor{Planck2015a}).


\section{Fitting of relations}
\label{sec:fitting}

\begin{table*}
\centering
\caption{Results of fitting the evolving power-law (equation~\ref{eq:power}) to both the median relation and the log-normal scatter about it for the \agn~8.0 simulation. The scatter is the log-normal scatter in the natural logarithm of the $Y$ variable. The results for the other simulations are given in Appendix~\ref{app:fits}. Note that $\alpha_{\rm SS}$ and $\beta_{\rm SS}$ represent the self-similar predictions for the evolution and mass slope, respectively.}
\begin{tabular}{lp{1.217cm}ccccccc}
\hline
Scaling relation & Median or scatter & $A$ & $\alpha$ & $\beta$ & $\chi^2$ & d.o.f. & $\alpha_{\rm SS}$ & $\beta_{\rm SS}$ \\ 
\hline
$T_{spec,cor}-M_{500}$        	    & Median & \phantom{$--$}0.280$\pm$0.004      & \phantom{$-$}0.356$\pm$0.024 & \phantom{$-$}0.577$\pm$0.006  & 0.044 & 30 & 2/3 & 2/3 \\ 
$L_{bol}-M_{500}$   		& Median & \phantom{$-$-}43.440$\pm$0.015    & \phantom{$-$}2.920$\pm$0.083 & \phantom{$-$}1.812$\pm$0.019  & 0.496 & 30 & 7/3 & 4/3 \\ 
$M_{gas,500}-M_{500}$        & Median & \phantom{$-$-}12.851$\pm$0.012   & \phantom{$-$}0.576$\pm$0.066  & \phantom{$-$}1.317$\pm$0.015 & 0.322 & 30 & 0 & 1\\ 
$Y_{X,500}-M_{500}$	     & Median & \phantom{$-$-}13.137$\pm$0.014   & \phantom{$-$}0.909$\pm$0.077  & \phantom{$-$}1.888 $\pm$0.018 & 0.436 & 30 & 2/3 & 5/3 \\ 
$d_{A}^{2}Y_{500}-M_{500}$ & Median & \phantom{$-$}$-$5.754$\pm$0.014 & \phantom{$-$}0.981$\pm$0.077  & \phantom{$-$}1.948$\pm$0.018 & 0.431 & 30 & 2/3 & 5/3 \\ 
$M_{500,hse,spec}-M_{500}$        & Median & \phantom{$-$-}13.902$\pm$0.009   & $-$0.027$\pm$0.047  & \phantom{$-$}0.952$\pm$0.011 & 0.161 & 30 & 0 & 1 \\ 
\hline
$T_{spec,cor}-M_{500}$        	    & Scatter & \phantom{$-$}$-$0.967$\pm$0.026   & $-$0.492$\pm$0.142  		      & $-$0.174$\pm$0.033		  	 & 1.473 & 30 & ... & ...\\ 
$L_{bol}-M_{500}$   		& Scatter & \phantom{$-$}$-$0.430$\pm$0.028  & $-$0.464$\pm$0.155 		       & $-$0.133$\pm$0.036 		 & 1.740 & 30 & ... & ...\\ 
$M_{gas,500}-M_{500}$        & Scatter & \phantom{$-$}$-$0.976$\pm$0.013  & $-$0.422$\pm$0.069		       & $-$0.412$\pm$0.016			 & 0.349 & 30 & ... & ...\\ 
$Y_{X,500}-M_{500}$	     & Scatter & \phantom{$-$}$-$0.750$\pm$0.027  & $-$0.441$\pm$0.148 		       & $-$0.250$\pm$0.034		 & 1.599 & 30 & ... & ...\\
$d_{A}^{2}Y_{500}-M_{500}$ & Scatter & \phantom{$-$}$-$0.862$\pm$0.025 & $-$0.163$\pm$0.136 &$-$0.269$\pm$0.032 			 & 1.344 & 30 & ... & ...\\ 
$M_{500,hse,spec}-M_{500}$        & Scatter & \phantom{$-$}$-$0.646$\pm$0.018  & \phantom{$-$}0.352$\pm$0.096		       & $-$0.118$\pm$0.022			 & 0.678 & 30 & ... & ...\\ 
\hline
\end{tabular}
\label{table:powerlaw}
\end{table*}

\begin{table*}
\centering
\caption{Results of fitting the evolving broken power-law fitting (equation~\ref{eq:brokenextra}) to both the median relation and the log-normal scatter about it for the \agn~8.0 simulation. The scatter is the log-normal scatter in the natural logarithm of the $Y$ variable. The results for the other simulations are given in Appendix~\ref{app:fits}.}
\begin{tabular}{lp{1.217cm}ccccccc}
\hline
Scaling relation & Median or scatter & $A''$ & $\alpha''$ & $\beta''$ & $\gamma''$ & $\delta''$ & $\chi^2$ & d.o.f. \\ 
\hline
$T_{spec,cor}-M_{500}$         & Median & \phantom{$--$}0.314$\pm$0.005      & \phantom{$-$}0.257$\pm$0.020 	& \phantom{$-$}0.703$\pm$0.018   & \phantom{$-$}0.514$\pm$0.009  & $-$0.063$\pm$0.011		       & 0.013 & 28 \\ 
$L_{bol}-M_{500}$    	      & Median & \phantom{$-$-}43.469$\pm$0.018 & \phantom{$-$}2.590$\pm$0.078 & \phantom{$-$}2.163$\pm$0.069  & \phantom{$-$}1.846$\pm$0.036 & $-$0.259$\pm$0.042 & 0.189 & 28 \\ 
$M_{gas,500}-M_{500}$         & Median & \phantom{$-$-}12.940$\pm$0.010  & \phantom{$-$}0.227$\pm$0.042 & \phantom{$-$}1.738$\pm$0.037  & \phantom{$-$}1.181$\pm$0.020 & $-$0.240$\pm$0.023 & 0.056 & 28 \\ 
$Y_{X,500}-M_{500}$	      & Median & \phantom{$-$-}13.246$\pm$0.010    & \phantom{$-$}0.504$\pm$0.044 & \phantom{$-$}2.383$\pm$0.039 & \phantom{$-$}1.712$\pm$0.020 & $-$0.275$\pm$0.024		       & 0.060 & 28 \\ 
$d_{A}^{2}Y_{500}-M_{500}$  & Median & \phantom{$-$}$-$5.637$\pm$0.011 & \phantom{$-$}0.689$\pm$0.045 	& \phantom{$-$}2.336$\pm$0.040  & \phantom{$-$}1.710$\pm$0.021 & $-$0.175$\pm$0.024 & 0.062 & 28 \\ 
$M_{500,hse,spec}-M_{500}$ & Median & \phantom{$-$-}13.895$\pm$0.017   & $-$0.011$\pm$0.072 		       	& \phantom{$-$}0.930$\pm$0.063  & \phantom{$-$}0.967$\pm$0.033 & \phantom{$-$}0.009$\pm$0.039 & 0.160 & 28 \\ 
\hline
$T_{spec,cor}-M_{500}$         & Scatter & \phantom{$-$}$-$1.171$\pm$0.024   & $-$0.027$\pm$0.103	 	        & $-$0.807$\pm$0.091 		  	  & \phantom{$-$}0.256$\pm$0.048 & \phantom{$-$}0.269$\pm$0.055 & 0.327 & 28 \\ 
$L_{bol}-M_{500}$    	      & Scatter  & \phantom{$-$}$-$0.623$\pm$0.014  & $-$0.313$\pm$0.060		        & $-$0.448$\pm$0.053 		  & \phantom{$-$}0.381$\pm$0.028 & \phantom{$-$}0.013$\pm$0.032 & 0.112 & 28 \\ 
$M_{gas,500}-M_{500}$         & Scatter  & \phantom{$-$}$-$1.025$\pm$0.021  & $-$0.390$\pm$0.088		        & $-$0.485$\pm$0.077 		  & $-$0.280$\pm$0.041 		    & $-$0.002$\pm$0.047 		       & 0.240 & 28 \\ 
$Y_{X,500}-M_{500}$	      & Scatter  & \phantom{$-$}$-$0.936$\pm$0.032 & $-$0.084$\pm$0.135  		       	& $-$0.762$\pm$0.119		  	  & \phantom{$-$}0.171$\pm$0.062 & \phantom{$-$}0.189$\pm$0.072 & 0.562 & 28 \\
$d_{A}^{2}Y_{500}-M_{500}$  & Scatter & \phantom{$-$}$-$1.037$\pm$0.031 & \phantom{$-$}0.273$\pm$0.130 	& $-$0.849$\pm$0.115  		  & \phantom{$-$}0.085$\pm$0.060 & \phantom{$-$}0.263$\pm$0.070 & 0.525 & 28 \\ 
$M_{500,hse,spec}-M_{500}$ & Scatter & \phantom{$-$}$-$0.756$\pm$0.024 & \phantom{$-$}0.566$\pm$0.102 	& $-$0.424$\pm$0.090 		  & \phantom{$-$}0.127$\pm$0.047 		    & \phantom{$-$}0.115$\pm$0.055 		       & 0.324 & 28 \\
\hline
\end{tabular}
\label{table:brokenpowerlaw}
\end{table*}

Our characterisation of the scaling relations is a two-stage process, in which we first derive the median mass--observable relations (i.e.\  $T-M$, $L_X-M$, $M_{gas}-M$, $Y_X-M$, $Y_{SZ}-M$ and $M_{hse}-M$) as a function of redshift and then measure the scatter about these relations. Specifically, we first compute the median values of the observable $Y$ (where hereafter $Y$ denotes one of $T$, $L_{X}$, $M_{gas}$, $Y_{X}$, $Y_{SZ}$ and $M_{500,hse}$) in ten equal-width logarithmic mass bins over the range $13.0\le\log_{10}[M_{500}(\textrm{M}_\odot)]\le15.5$ at nine different redshifts ($z=0$, 0.125, 0.25, 0.375, 0.5, 0.75, 1.0, 1.25 and 1.5).  To characterise the scatter about the median relations, we simply divide each cluster's observable value by that expected from (a cubic spline interpolation of) the median mass--observable relation, $Y_{spline}$.  We then fit the residuals (i.e.\ measure the scatter) in four bins of $M_{500}$ (chosen to be $\log_{10} [M_{500}(\textrm{M}_\odot)]=13.0-13.5$, $13.5-14.0$, $14.0-14.5$ and $14.5-15.5$) with a log-normal distribution of the form: 
\begin{equation}
P(X)=\frac{binsize}{X\sqrt{2\pi\sigma^2}}\exp\left(-\frac{(\ln X-\mu)^2}{2\sigma^2}\right),
\label{eq:lognormal}
\end{equation}
where $X=Y/Y_{spline}$.

When fitting the log-normal distribution, we fix $\mu=0$ (i.e.\ the central value of the histogram is imposed to be the median observable in the corresponding mass bin i.e.\ $Y=Y_{spline}$) and use the \textsc{mpfit} least-square minimization package in \textsc{idl} \citep{Markwardt2009} to perform the fitting. Note that the $binsize$ prefactor is needed to convert the log-normal PDF into a histogram whose bin width is equal to $binsize$. The $binsize$ is set to the width (in linear units) of the distribution enclosing the central 90 per cent of the simulated systems divided by a factor of ten (i.e.\ the width of the distribution is resolved with ten bins).

For comparison, we have also computed the root mean square (RMS) dispersion about the median scaling relations as a function of mass and redshift as follows:
\begin{equation}
\sigma_{j,rms}(z)=\sqrt{\frac{\sum_{i=1}^{N_j} \left(\ln Y_i(z)-\ln Y_{spline,i}(z)\right)^2}{N_j(z)}}
\label{eq:rms}
\end{equation}
where $N_j(z)$ is the number of system in mass bin $j$ at redshift $z$ and $Y_{spline,i}(z)$ is the value of $Y(z)$ obtained for $M_{500,i}(z)$ using the best-fitting spline. Since the trends obtained using the RMS dispersion are nearly identical to the ones obtained with the log-normal scatter (the RMS scatter tends to be only slightly larger than the log-normal estimate), in the remainder of the paper, we only present the results obtained for the log-normal scatter described above.

From the above procedure, we obtain a median mass--observable scaling relation and the log-normal scatter about it (and its dependence on mass) as a function of redshift for each of the simulation models.  We then fit an evolving power-law of the form:
\begin{equation}
Y=10^AE(z)^\alpha\left(\frac{M_{500}}{10^{14}~\textrm{M}_\odot}\right)^\beta,
\label{eq:power}
\end{equation}
to the median relations and the log-normal scatter about them.  

Note that self-similar theory predicts that the mass--observable relations should be single power-laws in both mass and $E(z)$.  However, as we will demonstrate below, a single power-law in mass does not describe the mass--observable relations particularly well.  For this reason, we also consider an evolving broken power-law of the form:
\begin{equation}
Y=10^{A'}E(z)^{\alpha'}\left(\frac{M_{500}}{10^{14}~\textrm{M}_\odot}\right)^{\epsilon'}
\label{eq:broken}
\end{equation}
where 
\begin{equation}
\epsilon'=
\begin{cases}
\beta' & \text{if } M_{500}\le10^{14}~\textrm{M}_\odot \\
\gamma' & \text{if } M_{500}>10^{14}~\textrm{M}_\odot, 
\end{cases}
\end{equation}

Finally, we also try an evolving broken power-law with a redshift dependent low-mass power-law index of the form:
\begin{equation}
Y=10^{A''}E(z)^{\alpha''}\left(\frac{M_{500}}{10^{14}~\textrm{M}_\odot}\right)^{\epsilon''}
\label{eq:brokenextra}
\end{equation}
where 
\begin{equation}
\epsilon''=
\begin{cases}
\beta''+\delta''E(z) & \text{if } M_{500}\le10^{14}~\textrm{M}_\odot \\
\gamma'' & \text{if } M_{500}>10^{14}~\textrm{M}_\odot 
\end{cases}
\end{equation}

The motivation for a redshift-dependent low-mass power-law index is that groups are more sensitive to non-gravitational physics than more massive clusters. Consequently, their evolution deviates more strongly from the self-similar prediction, which we demonstrate in detail below. The functional form of the redshift dependence of the low-mass power law index was empirically inferred by looking at the redshift dependence of $\beta'$ (and $\gamma'$; see Fig.~\ref{fig:slopeevobroken} and corresponding text). 

Note that in all three cases (equations~\ref{eq:power},~\ref{eq:broken} and~\ref{eq:brokenextra}), we held the mass pivot point fixed to $10^{14}~\textrm{M}_\odot$. There are two main reasons for this: virtually all of the scaling relations appear to break at $M_{500}\sim10^{14}~\textrm{M}_\odot$ for the radiative models, and having a fixed pivot point makes the comparison between different physical models and fitting formulae straightforward. 

Note that here $\chi^2$ is defined as 
\begin{equation}
\chi^2\equiv\sum_{i=1}^{N_{bin}}(Y_{i}-Y_{bf,i})^2
\label{eq:chi2}
\end{equation}
where $Y_{bf,i}$ is given by one of the following equations:~\eqref{eq:power}, or \eqref{eq:broken} or \eqref{eq:brokenextra}; as no errors can straightforwardly be assigned to the variables (this is especially true when we fit the median scaling relations since, in this case $Y_{i}$ is held fixed to $Y_{i,spline}$ as is explained at the beginning of this Section)\footnote{Therefore, the values of $\chi^2$ should \emph{only} be used to compare the respective quality of the fits for the evolving power law and broken power-law models at fixed scaling relation and physical model, as their differences and/or ratios are otherwise meaningless.}. For the same reason, the quoted errors for the best-fitting parameters should be taken `with a pinch of salt' as they have been computed by rescaling the diagonal values of the covariance matrix computed by \textsc{mpfit} so that the $\chi^2$ per degree of freedom is equal to 1 (according to the suggestion in the documentation of \textsc{mpfit}). It should nevertheless be noted that they are, most of the time, much smaller than the differences between the different physical models. Additionally, error bars are not relevant when comparing different physical models which were started from the same initial conditions as is the case here.
 
The results of fitting the evolving power-law and broken power-law with an evolving low-mass power-law index for the \agn~8.0 simulation (our most realistic model, see \citetalias{LeBrun2014}) are presented in Tables~\ref{table:powerlaw} and~\ref{table:brokenpowerlaw}, respectively\footnote{The results for the other simulations are given in Appendix~\ref{app:fits}.}.  Hereafter, $\beta$, $\epsilon'$ and $\epsilon''$ are called the mass slope.

It is worth mentioning that we also experimented with characterising the evolution of median relations and the scatter about them using powers of $1+z$ instead of $E(z)$, as is sometimes done in the literature for simulations and observations alike \citep[e.g.][]{Ettori2004,Maughan2006,Kay2007,Sehgal2011,Lin2012}. However, we found that this generally leads to poorer fits and we will thus not discuss the results obtained using powers of $1+z$ any further.

\begin{figure}
\begin{center}
\includegraphics[width=1.0\hsize]{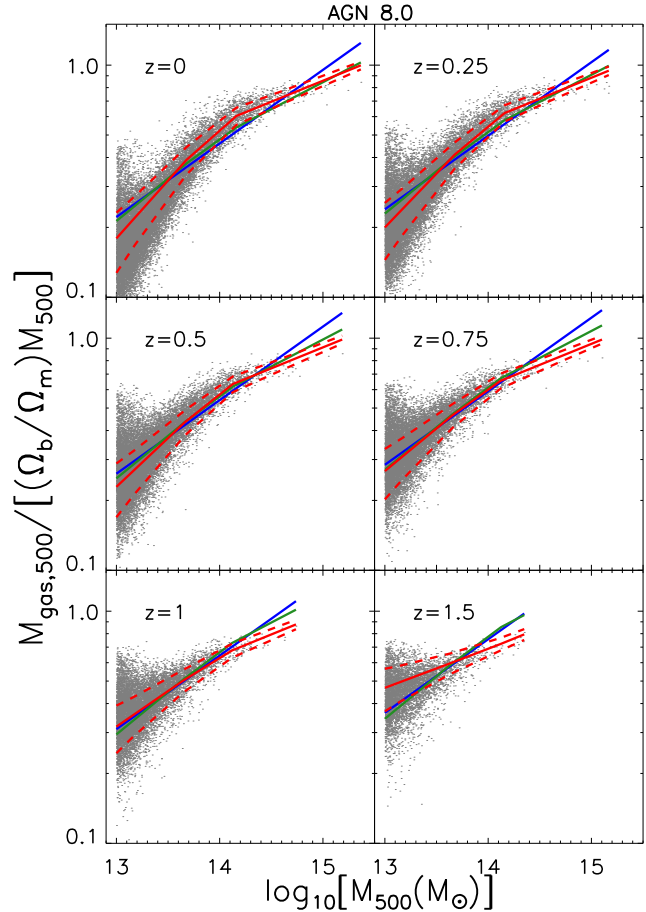}
\caption{Gas fraction (in units of the universal baryon fraction $\Omega_b/\Omega_m$)--$M_{500}$ relation at six different redshifts ($z=0$, 0.25, 0.5, 0.75, 1.0 and 1.5 from \emph{top left} to \emph{bottom right}). In each panel, the grey dots correspond to the individual simulated groups and clusters with $\log_{10}[M_{500}(\textrm{M}_\odot)]\ge13.0$, the solid blue, green and red lines respectively correspond to the best-fitting evolving power-law (equation~\ref{eq:power}), broken power-law (equation~\ref{eq:broken}) and broken power-law (equation~\ref{eq:brokenextra}) with a redshift dependent low-mass power-law index to the median gas mass. The dashed red lines correspond to the best-fitting evolving broken power-law with a redshift dependent low-mass power-law index to the log-normal scatter in bins of mass and redshift.  The median relations and the scatter about them are reasonably well modelled by evolving broken power-laws with redshift dependent low-mass power-law indices of the form given by equation~\eqref{eq:brokenextra}, whereas the other functional forms fail to reproduce the median relations, especially at the low-mass end.}
\label{fig:reconstructed}
\end{center}
\end{figure}

In Fig.~\ref{fig:reconstructed}, we show examples of our fits at various redshifts for one of the variables studied here: the gas mass, which is a representative variable. In each panel, the grey dots correspond to the individual simulated groups and clusters with $\log_{10}[M_{500}(\textrm{M}_\odot)]\ge13.0$ taken from the \agn~8.0 model, the solid blue, green and red lines respectively correspond to the best-fitting evolving power-law (equation~\ref{eq:power}), broken power-law (equation~\ref{eq:broken}) and broken power-law with a redshift dependent low-mass power-law index (equation~\ref{eq:brokenextra}) to the median gas mass in bins of mass and redshift and the dashed red lines correspond to the best-fitting evolving broken power-law with a redshift dependent low-mass power-law index to the log-normal scatter in bins of mass and redshift. For $z\le 1.5$, the median relations and the scatter about them are reasonably well modelled by evolving broken power-laws with redshift dependent low-mass power-law indices of the form given by equation~\eqref{eq:brokenextra}, whereas power-laws and broken power-laws of the form given by equations~\eqref{eq:power} and~\eqref{eq:broken} fail to reproduce the median relations, especially at the low-mass end.

\begin{figure}
\begin{center}
\includegraphics[width=1.0\hsize]{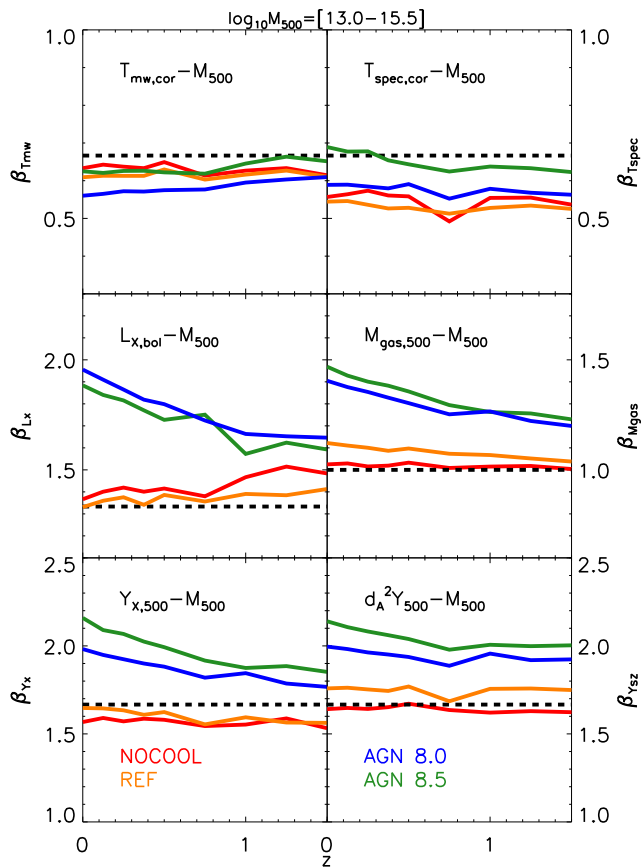}
\caption{Evolution of the mass slope of the scaling relations between total mass and core-excised temperature (for both mass-weighted and X-ray spectroscopic temperature), bolometric X-ray luminosity, gas mass, $Y_X$ and the integrated Sunyaev--Zel'dovich signal (from \emph{top left} to \emph{bottom right}). In each panel, we plot the evolution of the best-fitting power-law indices obtained by fitting the power-law given by equation~\eqref{eq:power} at each individual redshift. The solid curves (red, orange, blue and green) correspond to the different simulations and the horizontal dashed lines to the self-similar expectation, respectively.  The AGN models show significant deviations from self-similarity for all the scaling relations, except for the mass--temperature relation for which only a mild deviation is predicted (independent of the included sub-grid physics).  The deviations from self-similarity increase with decreasing redshift for the AGN models.  The models which lack efficient feedback (i.e.\ \nocool~and \refsim) show approximately self-similar behaviour for most scaling relations.}
\label{fig:slopeevo}
\end{center}
\end{figure}


\section{Evolution of the mass slope}
\label{sec:slopeevo}

\begin{figure*}
\begin{center}
\includegraphics[width=0.49\hsize]{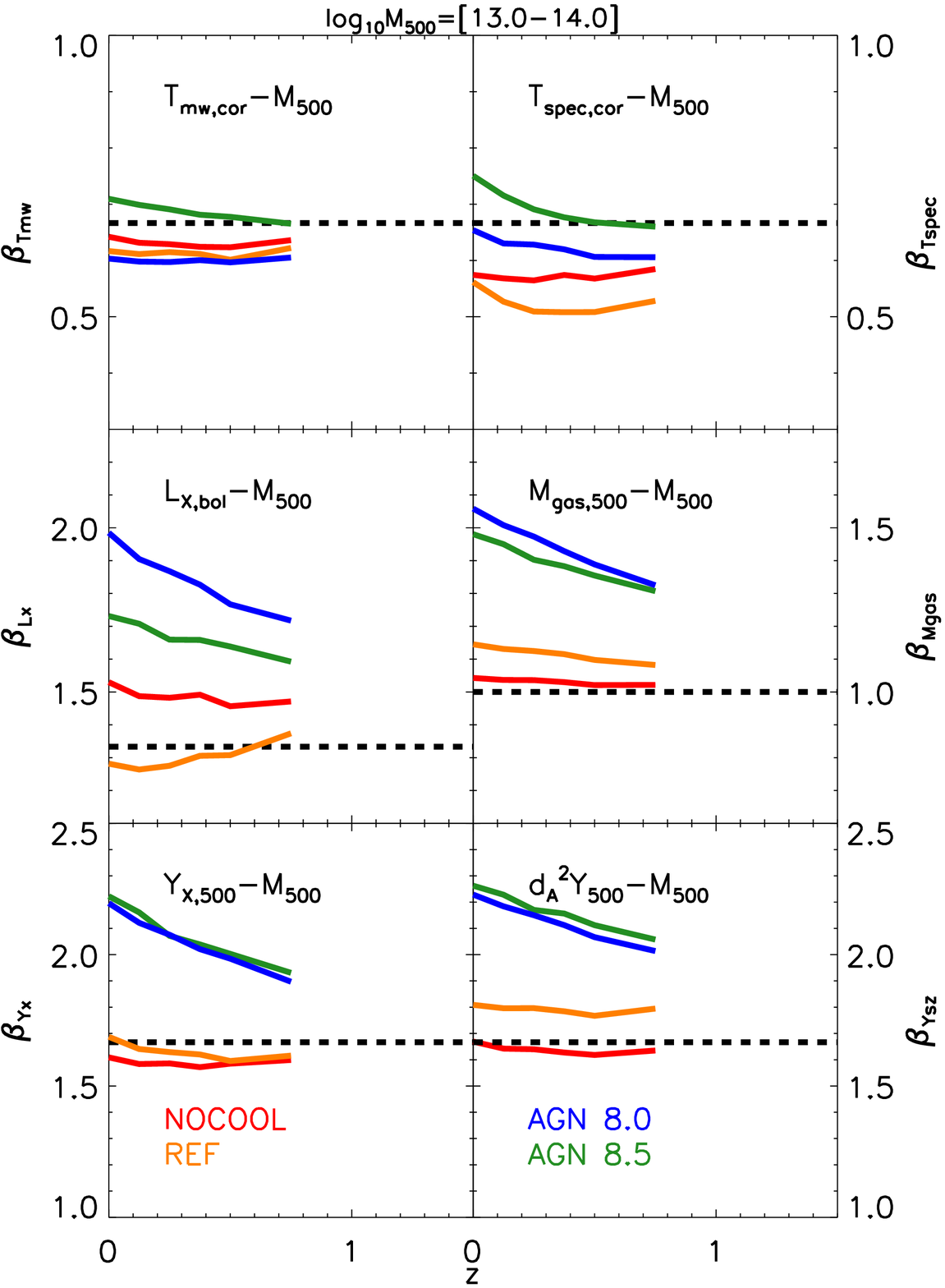}
\includegraphics[width=0.49\hsize]{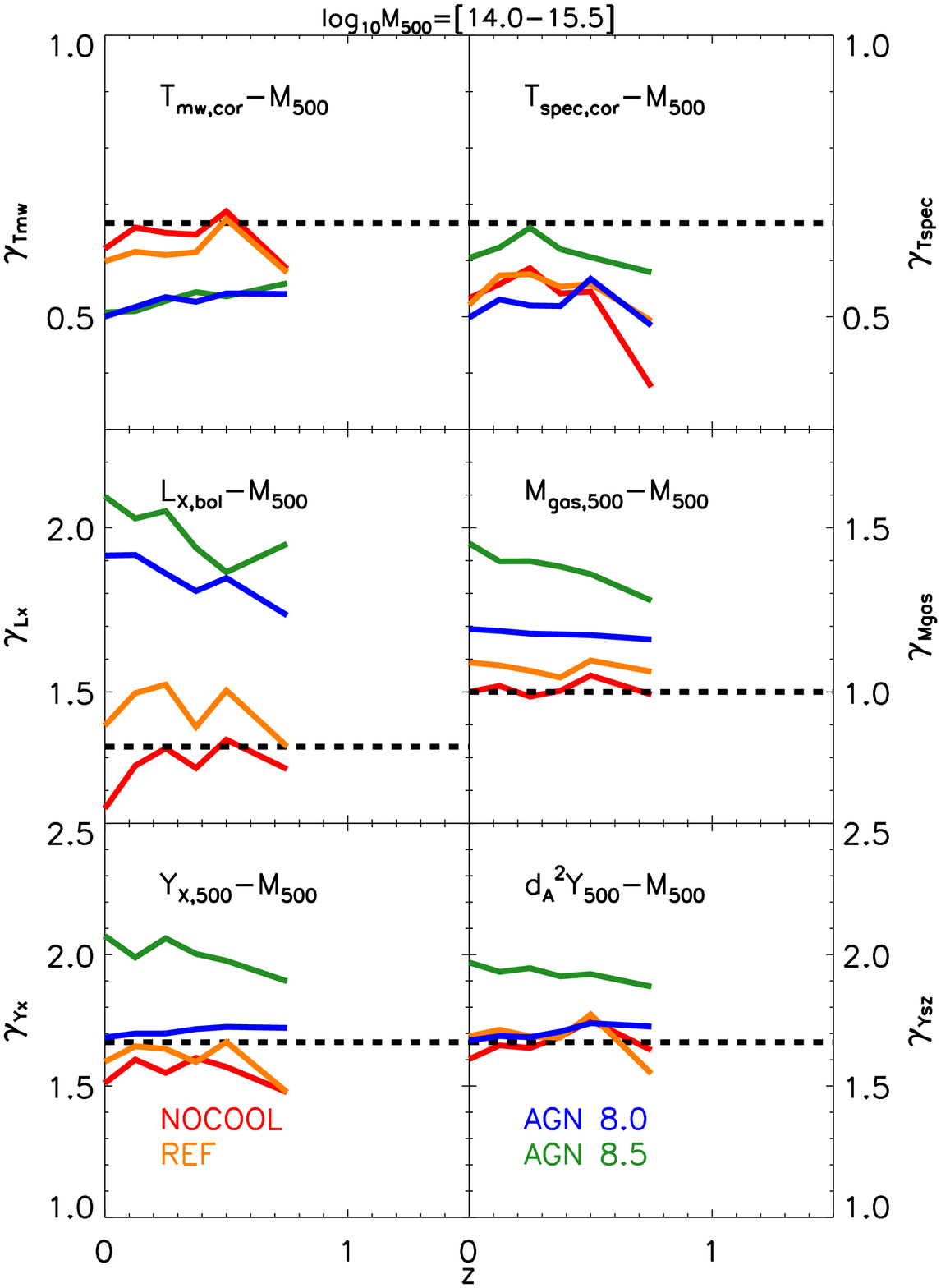}
\caption{Evolution of the mass slopes of the scaling relations between total mass and core-excised temperature (for both mass-weighted and X-ray spectroscopic temperature), bolometric X-ray luminosity, gas mass, $Y_X$, and the integrated Sunyaev--Zel'dovich signal (from \emph{top left} to \emph{bottom right}).  In each subpanel, we plot the evolution of the low-mass (\emph{left} panel) and high-mass (\emph{right} panel) best-fitting power-law indices obtained by fitting the broken power-law given by equation~\eqref{eq:broken} at each redshift independently.  The solid curves (red, orange, blue and green) correspond to the different models and the horizontal dashed lines to the self-similar expectation, respectively.  With the exception of the X-ray temperature and the soft X-ray luminosity, for both the low-mass and the high-mass slopes, the non-radiative (\nocool) model and the \refsim~model (which neglects AGN feedback) are approximately consistent with the self-similar expectation, whereas the models that include AGN feedback deviate significantly from self-similarity. The main differences between the two power-law indices at fixed scaling relation and physical model is that the low-mass one displays a stronger redshift dependence and tends to deviate more strongly from self-similarity.}
\label{fig:slopeevobroken}
\end{center}
\end{figure*}

We start by examining the evolution of the logarithmic slope (the mass slope) of the total mass--observable relations.  Note that, by definition, no evolution of the mass slope is expected in the context of the self-similar model.  The slope of a particular relation is fixed and can be predicted assuming only Newtonian gravity and that the gas is in virial equilibrium (see Section~\ref{sec:sstheory}).  Any evolution or deviation at any redshift from the predicted mass slope signals that either some non-gravitational physics is at play, or that the gas is not virialised (or both). 

In Figs.~\ref{fig:slopeevo} and~\ref{fig:slopeevobroken}, we show the evolution of the mass slopes from $z=0$ to $z=1.5$ for the scaling relations between total mass and core-excised\footnote{The results for non-core excised temperatures are presented in Appendix~\ref{app:Tnocor}.} temperature (for both mass-weighted and X-ray spectroscopic temperature), bolometric X-ray luminosity, gas mass, $Y_X$, and the integrated Sunyaev--Zel'dovich signal for each of the four physical models (different coloured curves).  Fig.~\ref{fig:slopeevo} shows the evolution of the mass slope obtained when we fit a single power-law (equation~\ref{eq:power}) at each individual redshift (and so omitting the $E(z)$ factor for the moment), while  Fig.~\ref{fig:slopeevobroken} shows the evolutions of the low-mass (\emph{left} panel) and high-mass (\emph{right} panel) mass slopes resulting from the fitting of the broken power-law in equation~\eqref{eq:broken} to the median scaling relations at each individual redshift. In each panel, the solid curves (red, orange, blue and green) correspond to the different simulations and the horizontal dashed lines to the self-similar expectation.

Starting first with the mass--X-ray temperature relation, the fitted mass slope is slightly shallower than the self-similar expectation of 2/3 for all the models.  This result is mostly independent of redshift and mass, but does depend somewhat upon the included sub-grid physics.  The sensitivity to sub-grid physics is stronger when using the observable spectroscopic temperature, as opposed to the mass-weighted temperature.  

We note that changing the sub-grid physics can affect the mean temperature in several ways.  First, the mean temperature profile can be altered, because the mean entropy of the gas can be raised or lowered by including feedback and radiative cooling.  The degree of scatter about the mean temperature profile (i.e.\ `multiphase' structure) will also be affected.  Energetic feedback processes can drive outflows and introduce turbulence, so that the temperature of the gas is no longer just determined by the entropy configuration of the gas and the potential well depth.   In addition, the degree of clumpiness of the gas, which is affected by feedback, will influence the observable mean spectroscopic temperature, since denser gas contributes more to the X-ray emissivity.  (The same is true for the gas-phase metallicity.)  However, in spite of all of these possible effects on the mean temperature, fundamentally, the mean temperature of the gas cannot deviate very strongly from the virial temperature, otherwise it will not be close to equilibrium with the total potential well, in which case the gas will re-adjust itself in a few sound-crossing or dynamical times (in practice, the minimum of the two) in order to achieve equilibrium.

The picture is not as clear-cut in the case of the bolometric X-ray luminosity--total mass relation, for which the mass slope displays strong simultaneous redshift and non-gravitational physics dependencies. When the self-similar prediction of 4/3 (see equation~\ref{eq:ssLxbol-M}) is considered, even the non-radiative simulation (\nocool) and the simulation which neglects AGN feedback altogether (\refsim) have bolometric X-ray luminosity--mass scaling relations that are slightly steeper than self-similar, and this nearly independently of redshift for both models but only at high-mass for \refsim. The deviations from self-similarity are probably due to the fact that the gas does not trace the dark matter (it has e.g.\ a different mass--concentration relation compared to the dark matter) which affects both the density and the temperature, and hence the X-ray luminosity. The temperature and density are also potentially affected by non-thermal pressure support. It is worth mentioning that this scaling relation is steeper in the calibrated \calsim~model than for both of the \agn~models discussed here and that this is only the case at the low-mass end.

The total mass--gas mass relation is steeper than the self-similar slope of 1 (which corresponds to gas tracing total mass, with a constant gas fraction) for all of the radiative models, whereas it is consistent with self-similarity for the non-radiative (\nocool) model.  These results are approximately independent of mass and redshift for the \nocool~and \refsim~models.  For the models that include AGN feedback, the mass slope steepens with decreasing redshift and mass, moving away from the self-similar expectation at low redshifts.  The low-mass regime shows a particularly large deviation with respect to the self-similar result (see Fig.~\ref{fig:slopeevobroken}), with the slope approaching 3/2 at late times.  The fact that the mass slope deviates most strongly from the self-similar result at low masses makes sense, since the AGN feedback is more efficient at ejecting gas from the high-redshift progenitors of groups than those of massive clusters \citep{McCarthy2011}.   That the mass slope moves further away from the self-similar result with decreasing redshift is also understandable, since haloes of fixed mass (as we consider here) have shallower potential wells at low redshift, as we discuss further below. It is noteworthy that the gas mass--total mass relation is the only other scaling (besides the soft X-ray luminosity--total mass one) whose slope has been affected by the calibrations undertaken as part of the \calsim~project: it is slightly steeper at the low-mass end than both \agn~8.0 and 8.5. 

\begin{figure*}
\begin{center}
\includegraphics[width=0.49\hsize]{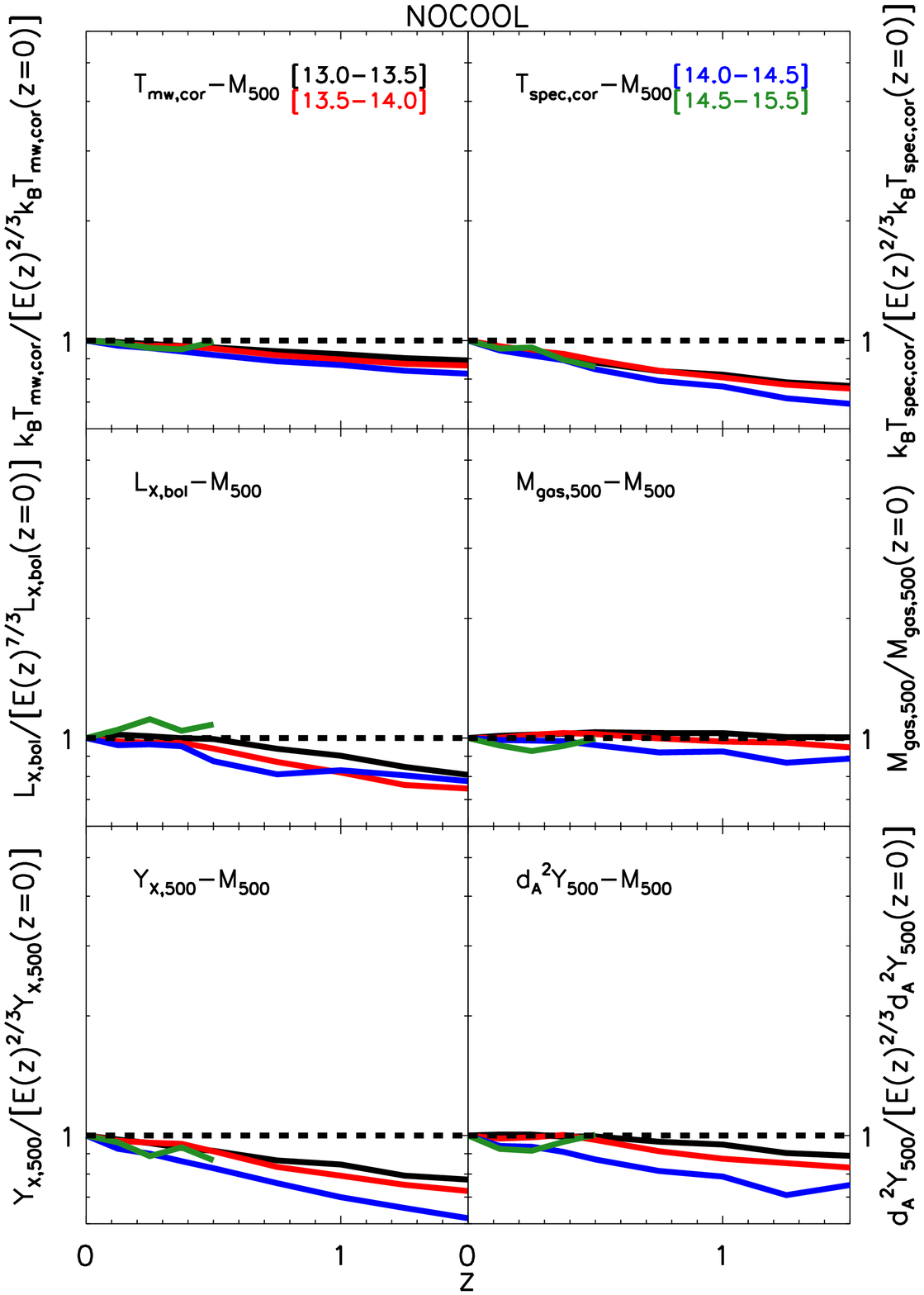}
\includegraphics[width=0.49\hsize]{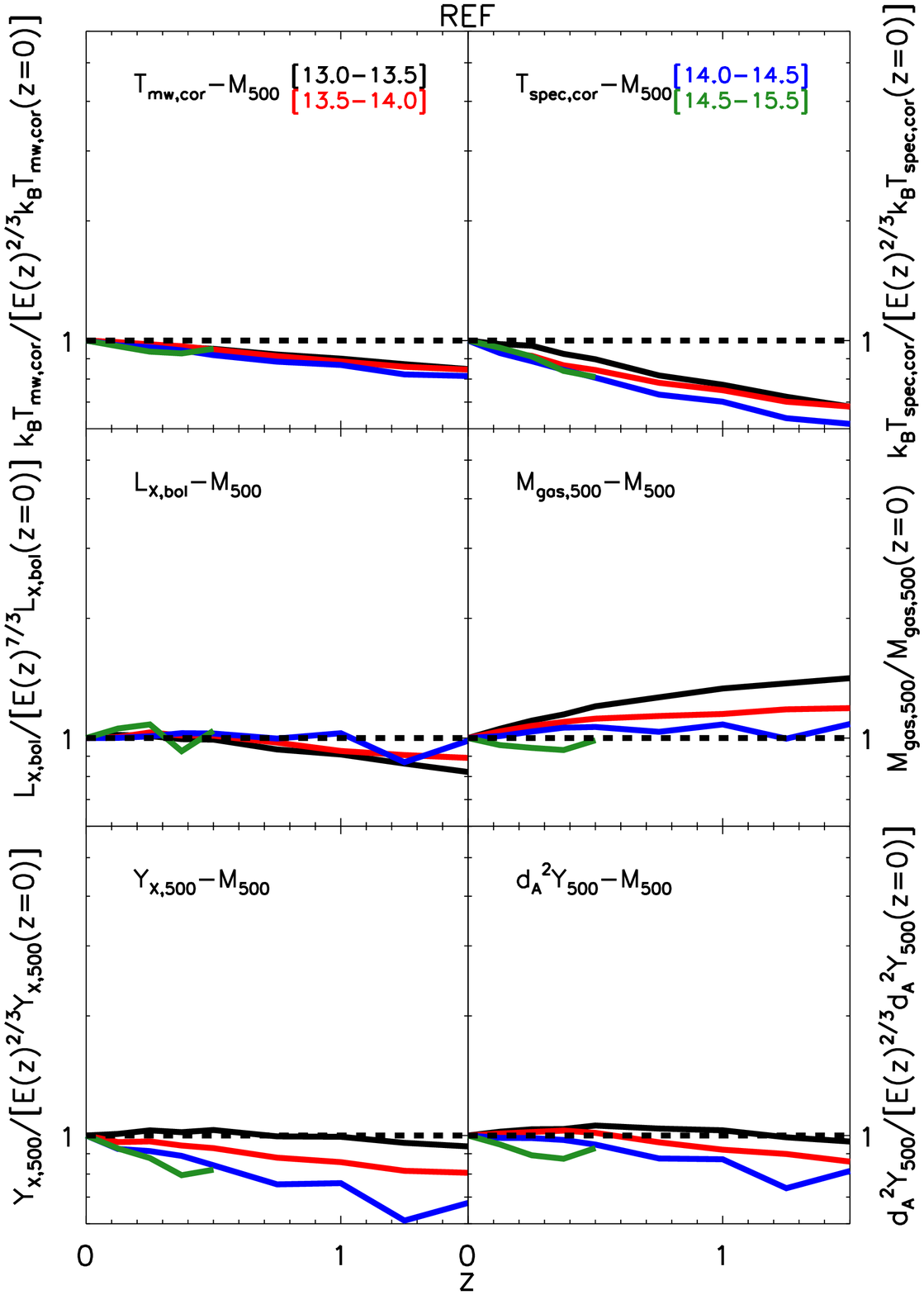}
\caption{Evolution of the normalisations of the scaling relations between total halo mass and core-excised temperature (for both mass-weighted and X-ray spectroscopic temperature), bolometric X-ray luminosity, gas mass, $Y_X$ and the integrated Sunyaev--Zel'dovich signal (from the \emph{top left} subpanel to the \emph{bottom right} subpanel).  The amplitude of each scaling relation in the four $\log_{10}[M_{500}(\textrm{M}_\odot)]$ bins (denoted by solid lines of different colours) has been normalised by the self-similar expectation for the redshift evolution at fixed mass (shown as an horizontal dashed line). The different panels (continued over the page) correspond to the different physical models.  The mass--temperature relation evolves in a negative fashion with respect to the self-similar model (i.e.\ slower), independently of the included ICM physics.  The gas mass evolves approximately self-similarly for the non-radiative simulation but shows a positive evolution (i.e.\ faster than self-similar) when radiative cooling and particularly AGN feedback is included, a result which is strongly mass dependent.  The SZ flux and $Y_X$ evolve negatively with respect to the self-similar expectation for models which neglect efficient feedback, driven by the negative evolution of the temperature.  When AGN feedback is included, the {\it sign} of evolution of the SZ flux and $Y_X$ with respect to self-similar depends on halo mass, driven by the strong halo mass-dependence of the gas mass evolution combined with the negative evolution of the temperature.}
\label{fig:normevo}
\end{center}
\end{figure*}

\begin{figure*}
\begin{center}
\includegraphics[width=0.49\hsize]{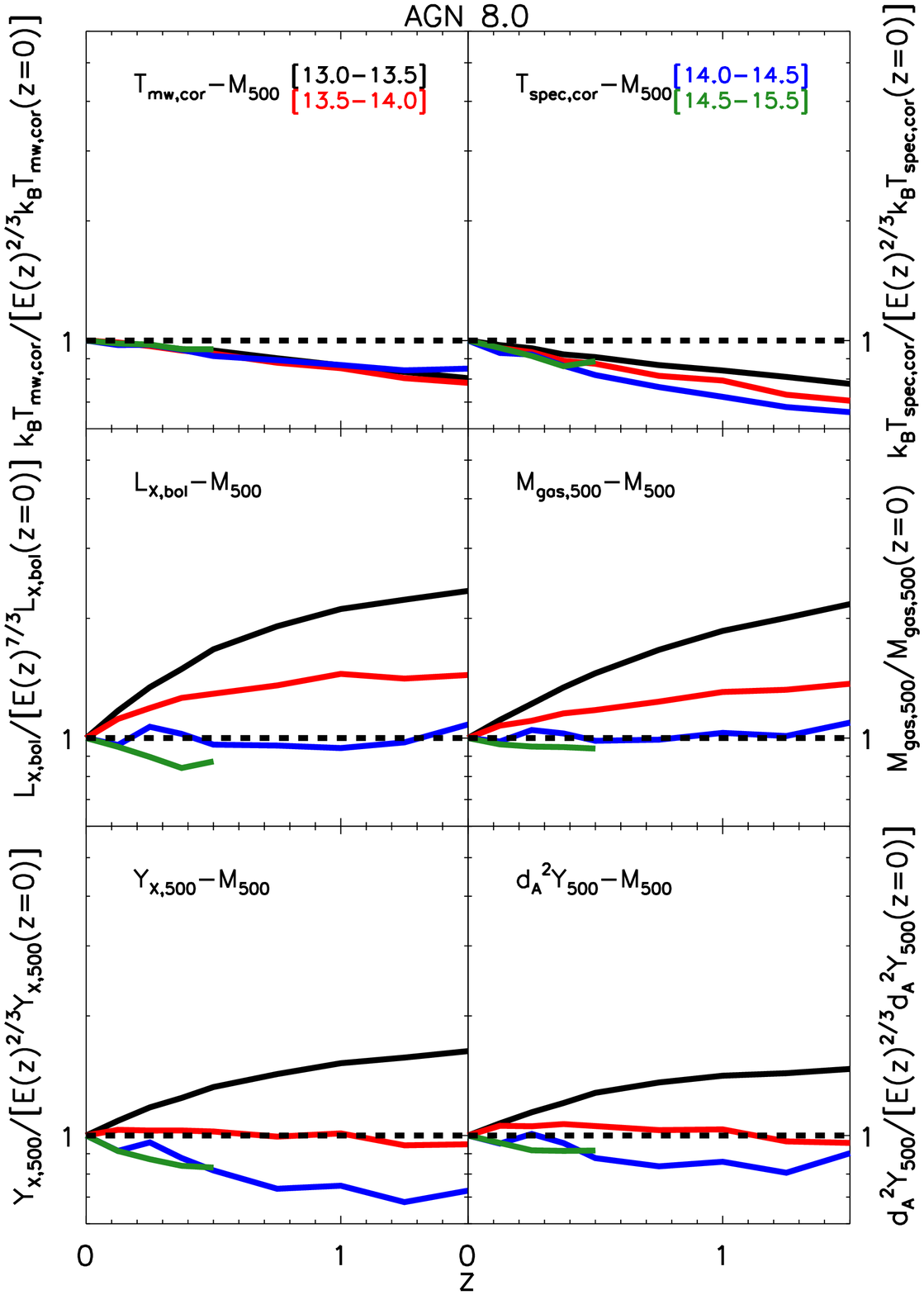}
\includegraphics[width=0.49\hsize]{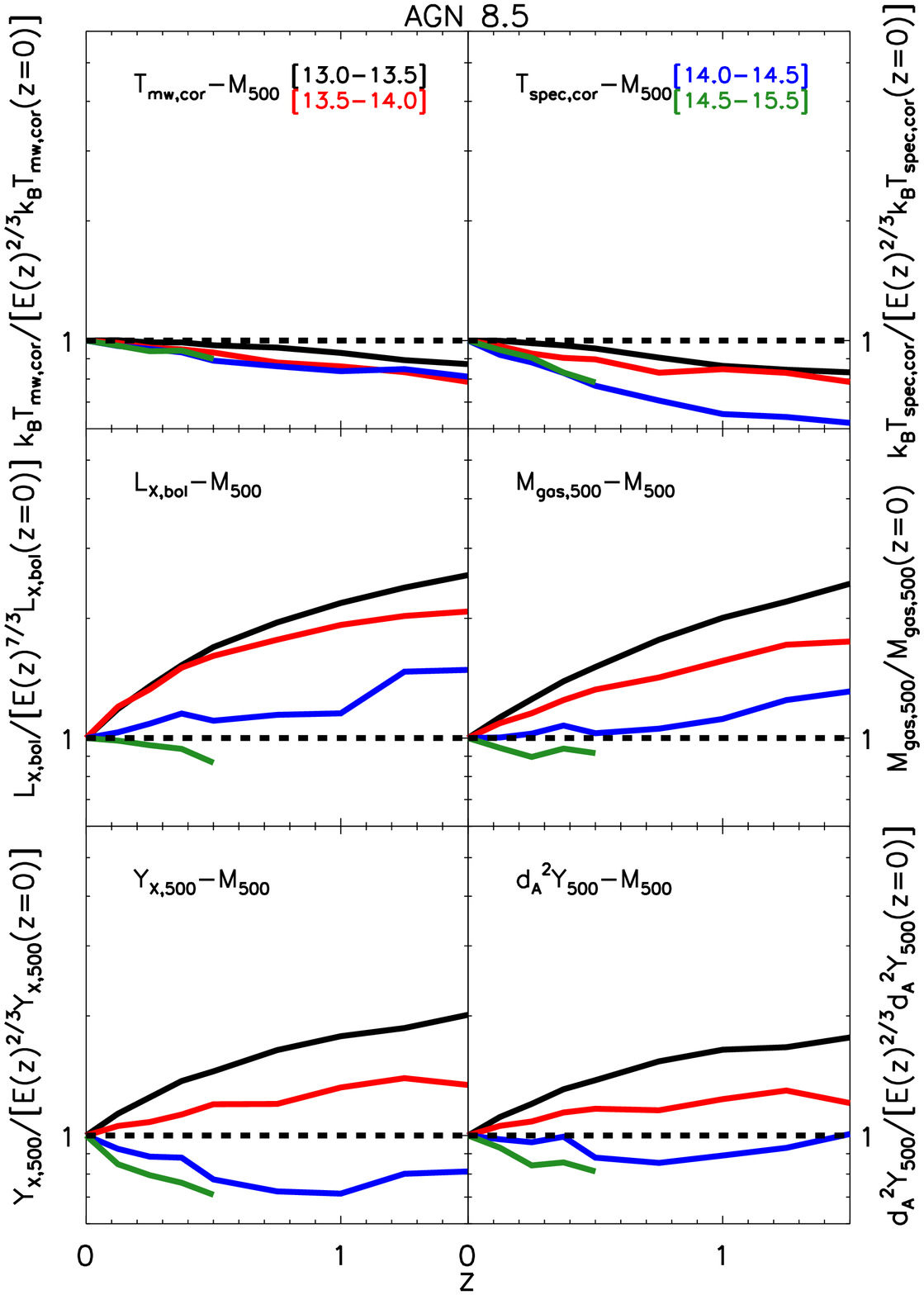}
\contcaption{Same as Fig.~\ref{fig:normevo} but for the \agn~8.0 and \agn~8.5 simulations.}
\label{fig:normevocont}
\end{center}
\end{figure*}

In the bottom panels of Figs.~\ref{fig:slopeevo} and~\ref{fig:slopeevobroken}, we examine the slopes of the relations between total mass and the integrated Sunyaev--Zel'dovich signal and its X-ray analogue, $Y_X$.  The models that neglect feedback (\nocool~and \refsim) exhibit approximately self-similar behaviour over the full range of redshifts we explore.  The inclusion of AGN feedback, however, results in mass slopes that are significantly steeper than the self-similar expectation of 5/3, but move closer to the self-similar result with increasing mass and redshift, in analogy to the gas mass trends discussed above. Taken at face value, the results obtained here for the Sunyaev--Zel'dovich signal in the \agn~models conflict with the observational results of \citeauthor{PlanckIntXI} and \citet{Greco2015}. But, as discussed in detail in \citet{LeBrun2015} (see especially their fig. 5 and also fig. 20 of \citealt{McCarthy2016}), due to its limited resolution ($\sim5$~arcminutes beam), \planck~does not measure the Sunyaev--Zel'dovich flux within $r_{500}$ but within a much larger aperture of size $5r_{500}$. This effectively diminishes the sensitivity of the Sunyaev--Zel'dovich signal to the impact of the non-gravitational physics of galaxy formation.    

Finally, it is worth pointing out that all the above results are independent of the choice of cosmology. 

In short, we find that our most realistic models that include efficient feedback from AGN predict significant deviations in the mass slope from the self-similar expectation for all of the scaling relations we have examined.  The one exception to this is the mass--temperature relation (especially in the mass-weighted case), where only a small deviation from the self-similar expectation is found (this is generally true, independent of the details of the included ICM physics).  For the other scaling relations, all of which depend directly on the gas density/mass, the deviations from the self-similar prediction are strongest at low redshifts and low halo masses.  The models that neglect efficient feedback (\nocool~and \refsim), on the other hand, have mass slopes that are approximately consistent with self-similar expectations.

\begin{figure*}
\begin{center}
\includegraphics[width=0.497\hsize]{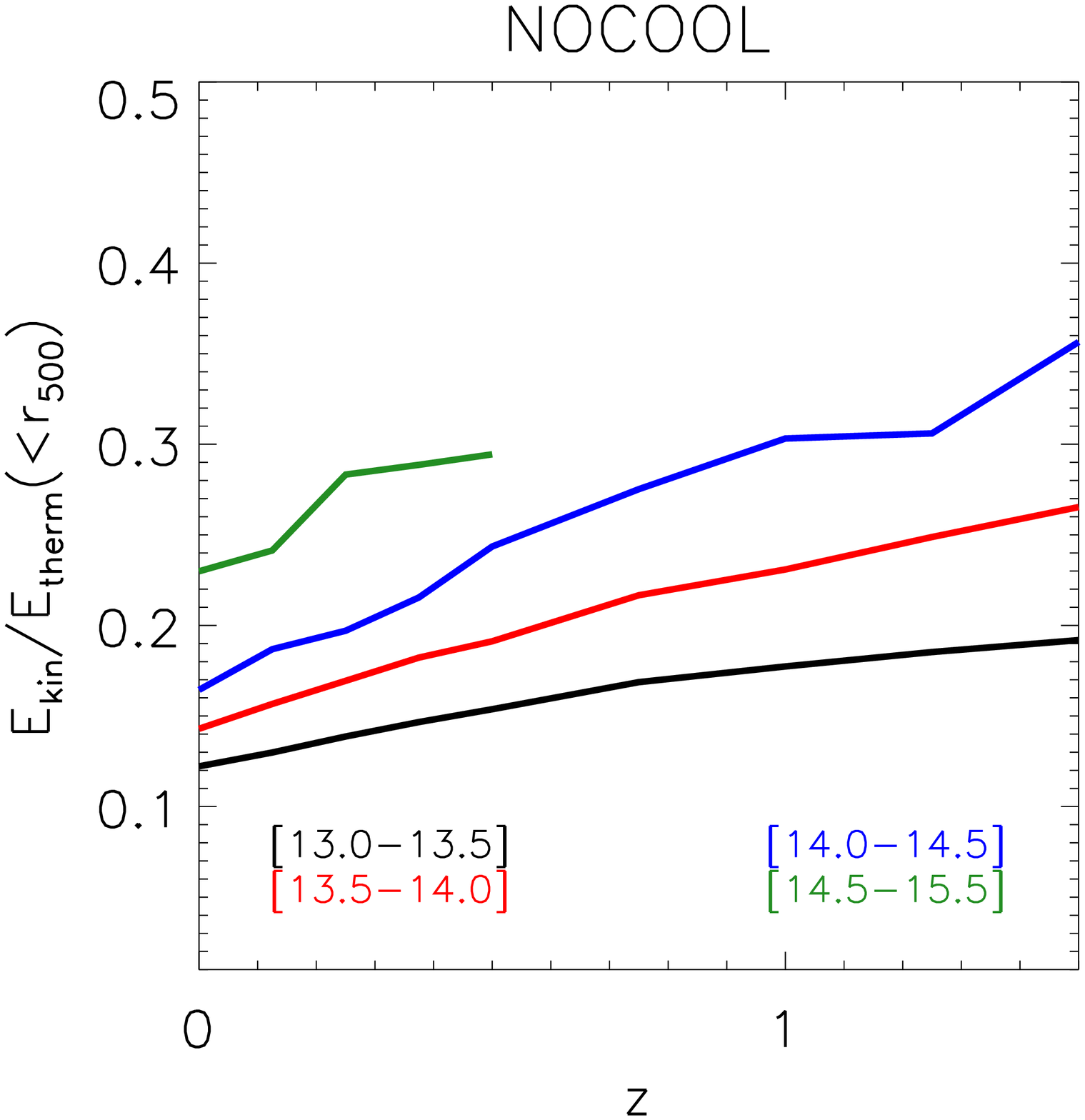}
\includegraphics[width=0.497\hsize]{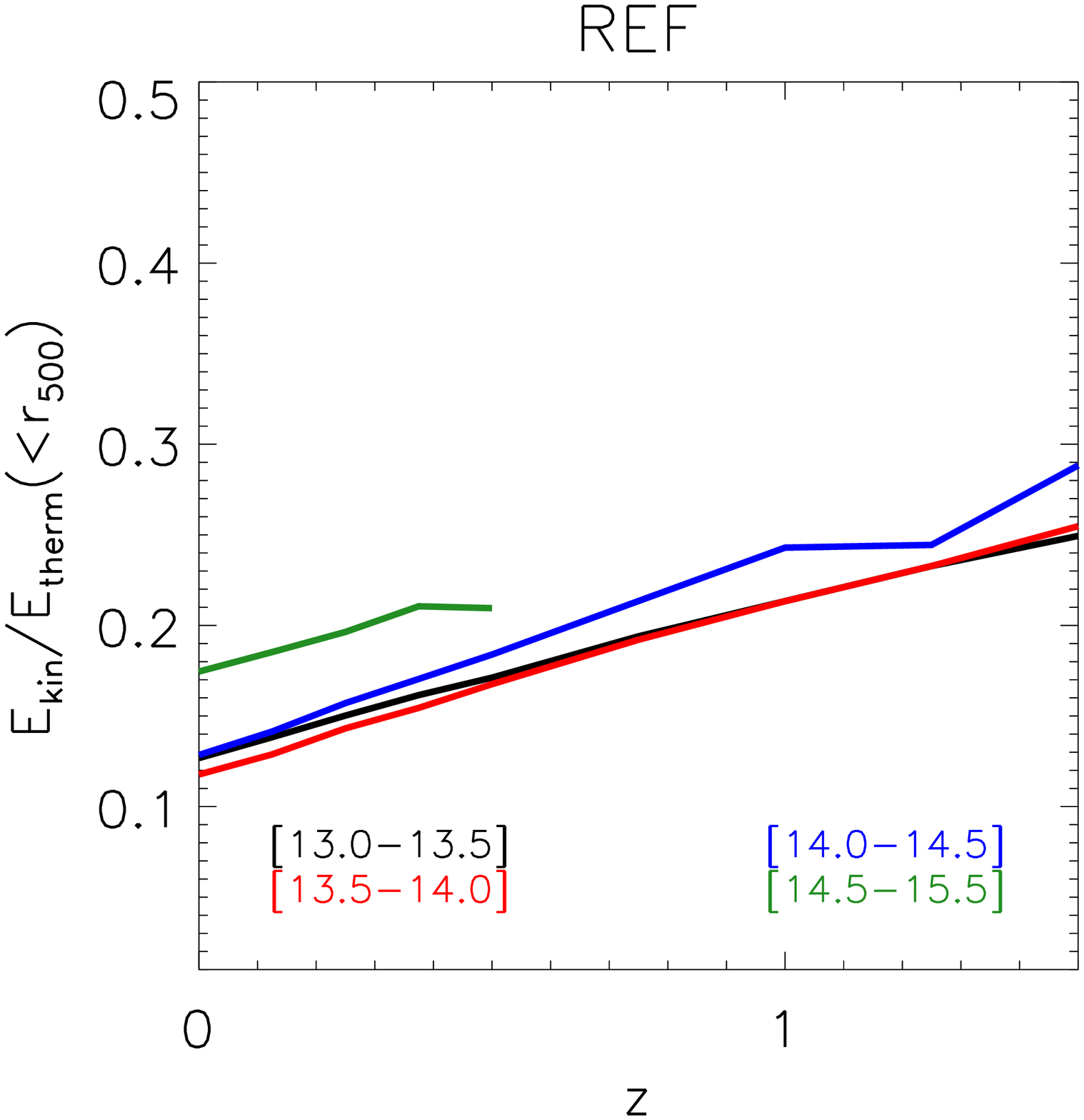}
\includegraphics[width=0.497\hsize]{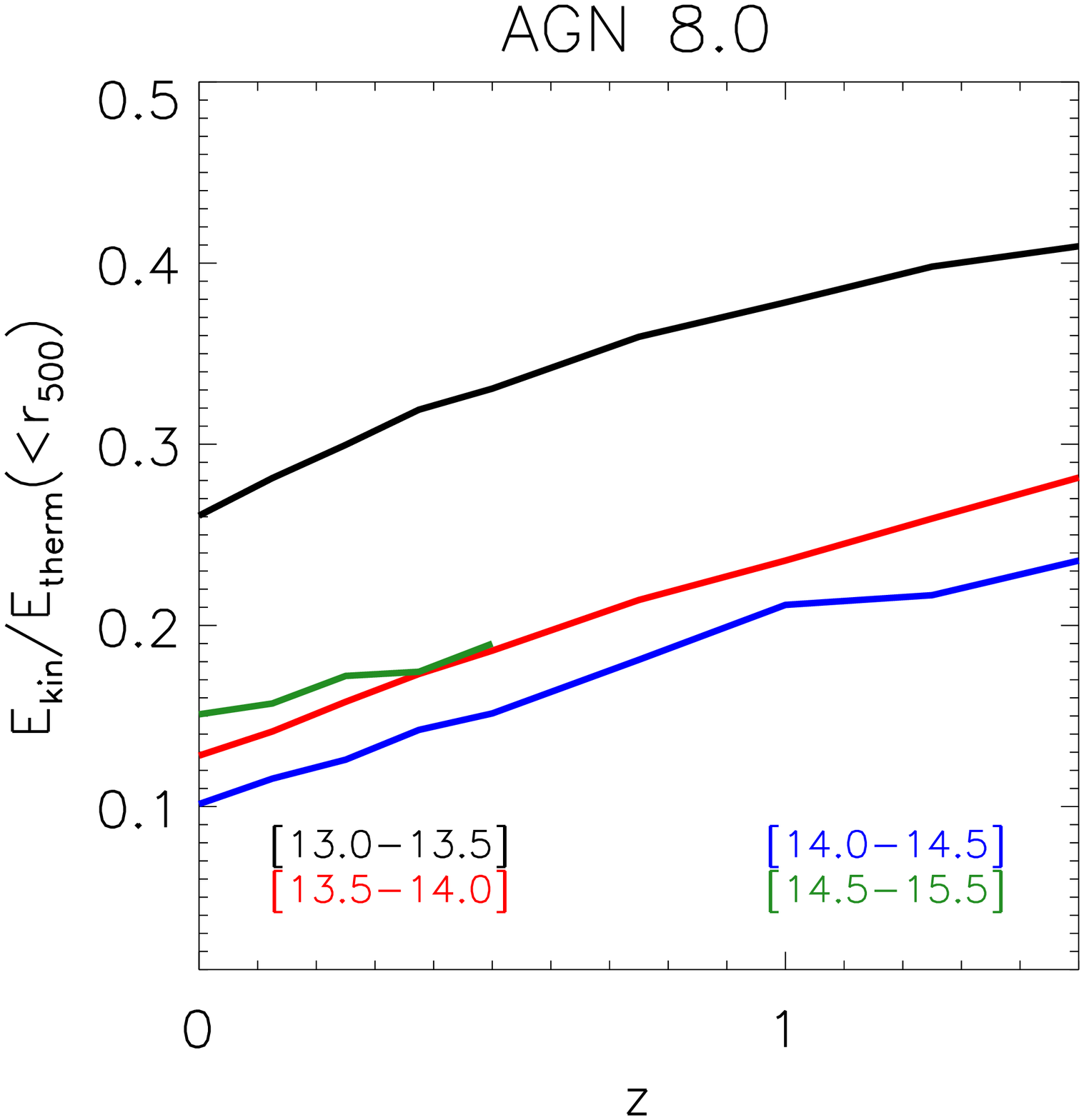}
\includegraphics[width=0.497\hsize]{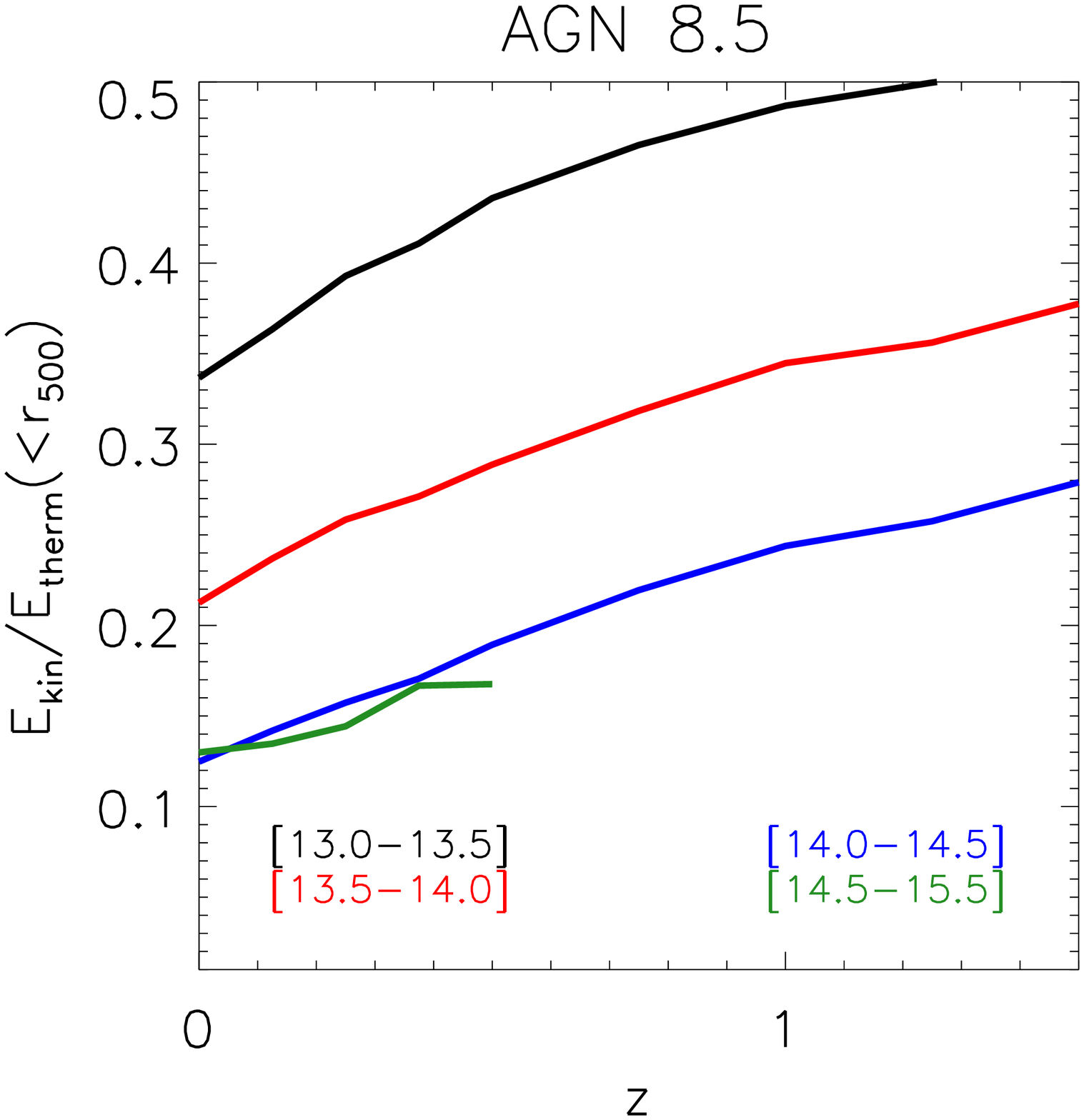}
\caption{Evolution of the kinetic to thermal ratio from $z=0$ to $z=1.5$ in each of the four $\log_{10}[M_{500}(\textrm{M}_\odot)]$ bins (denoted by solid lines of different colours). The kinetic-to-thermal ratio increases with increasing redshift and that independently of mass and physical models. The introduction of AGN feedback notably increases the importance of the kinetic motions in the galaxy groups regime. These results are independent of cosmology.}
\label{fig:KtoTevo}
\end{center}
\end{figure*}


\section{Evolution of the normalisation}
\label{sec:normevo}

In Fig.~\ref{fig:normevo}, we show the evolution of the normalisation of the scaling relations between total mass and core-excised temperature (for both mass-weighted and X-ray spectroscopic temperature), bolometric X-ray luminosity, gas mass, $Y_X$ and the integrated Sunyaev--Zel'dovich signal (from \emph{top left} to \emph{bottom right}). The amplitude of each scaling relation in the four $\log_{10}[M_{500}(\textrm{M}_\odot)]$ bins (denoted by solid lines of different colours) has been normalised by the self-similar expectation for the redshift evolution at fixed mass (shown as an horizontal dashed line). In the remainder of the paper, a scaling relation whose $E(z)$ exponent is smaller (larger) than the self-similar expectation listed in Section~\ref{sec:sstheory} will be referred to as having negative (positive) evolution.

We first consider the evolution of the mass--temperature relation.  Interestingly, the normalisation of this relation evolves in a negative sense with respect to the self-similar expectation, such that the spectroscopic (mass-weighted) temperature is predicted to be $\approx30\%$ ($\approx15\%$) lower than predicted using self-similar arguments by $z=1$.  This negative evolution is approximately independent of halo mass and only weakly dependent on the included gas physics.  That the negative evolution is approximately independent of the included physics implies that it is mainly a consequence of the merger history of clusters.  Two possible causes are: a change in the mass structure of haloes with redshift (i.e.\ evolution of the mass--concentration relation) and/or a change in the degree of virialisation of the gas with redshift.  We expect the former effect to be quite weak, as it only affects the mean temperature through a slight change in the weighting of the particles when calculating the mean.

We have examined the degree of virialisation of the gas as a function of redshift, by tracking the evolution of the ratio of kinetic to thermal energy of the hot gas in different fixed mass bins (see Fig.~\ref{fig:KtoTevo}).  We find that, independently of the included gas physics, haloes of fixed mass have strongly increasing kinetic-to-thermal energy ratios with increasing redshift, evolving from a typical value of 10-15 per cent at $z=0$ up to 20-30 per cent by $z=1$.  These kinetic motions (in both bulk flows and turbulence) act as a source of non-thermal pressure support that increases with increasing redshift, implying that a lower temperature (with respect to low-redshift clusters of the same mass) is required to achieve equilibrium within the overall potential well at high redshift and therefore likely drive the negative evolution (with respect to self-similarity) of the mass--temperature relation. We therefore warn against the `simplistic' interpretation of deviations of any scaling relation from self-similar evolution as indication of the effects of non-gravitational physics.

The amplitude of the bolometric X-ray luminosity--total mass relation evolves positively for all the models which include efficient feedback (i.e.\ the \agn~models). The amplitude of the evolution is strongly mass dependent, slightly redshift dependent (it flattens out as redshift increases) and is strongly sensitive to the non-gravitational physics of galaxy formation (it becomes more positive as the feedback intensity is increased with a reversal from mostly negative to mostly positive when AGN feedback is included). The slight negative evolution in the models without AGN feedback is most likely due to the negative evolution of the mass--temperature relation whereas the positive evolution is linked to `ease' of gas ejection (see the discussion about the evolution of the total mass--gas mass relation below).  

The total mass--gas mass relation is approximately consistent with self-similar evolution for the non-radiative model, but exhibits positive evolution when non-gravitational physics is included, particularly when the feedback `intensity' is increased (going from \refsim~to \agn~8.0 to \agn~8.5).   For our most realistic models (i.e.\ the \agn~models), the evolution is strongly mass dependent and somewhat redshift dependent.  

A likely explanation for the strong positive evolution of the gas mass (and X-ray luminosity) is that, since haloes of fixed mass are denser at higher redshifts, more energy is required to eject gas from these higher redshift haloes.  More precisely, the binding energy can be approximated by 
\begin{equation}
E_{bind}\propto \frac{GM_{\Delta}^{2}}{r_{\Delta}},
\label{eq:Ebind}
\end{equation}
which combined with equation~\eqref{eq:r-M} gives
\begin{equation}
E_{bind}(z)\propto M_{\Delta}^{5/3}E(z)^{2/3}.
\label{eq:Ebindevo}
\end{equation}
Hence, the binding energy increases with redshift making expulsion of gas due to outflows more difficult. Note that the supermassive black holes that power the AGN could `know' about the evolution of the binding energy of their host dark matter halo in the sense that the black hole masses are determined by their halo binding energy through their self-regulation (see \citealt{Booth2010,Booth2011}). However, the growth of black holes in massive galaxies  may be governed by black hole mergers rather than by self-regulated gas accretion \citep[e.g.][]{Peng2007}.

The integrated SZ flux and $Y_X$ show perhaps the most interesting behaviour, in terms of the evolution of the amplitude of their relations with halo mass.  For models with inefficient feedback, there is a mild negative evolution with respect to the self-similar expectation, which is driven by the negative evolution of the temperature combined with the nearly self-similar evolution of the gas mass (note that $Y_{X,SZ,\Delta} \equiv M_{gas,\Delta} T_{\Delta}$).  Things get more interesting when AGN feedback is included.  In particular, low-mass haloes display a positive evolution with respect to the self-similar result, whereas high-mass haloes show negative evolution.  This change in the sign of the effect with respect to the self-similar model is driven by the strong halo mass-dependence of the gas mass evolution.  In particular, low-mass haloes show a strong positive evolution in the gas mass that more than compensates for the negative evolution in the temperature, leading to a positive evolution in the SZ flux and $Y_X$.  High-mass haloes, however, show little evolution in the gas mass (their gas fractions are already near the universal fraction $\Omega_b/\Omega_m$) and, when combined with the negative evolution in the temperature, this leads to a negative evolution of the SZ flux and $Y_X$ with respect to the self-similar expectation.

We note that all the results described in this section are approximately independent of the choice of cosmology, in the sense that the general trends are preserved but the exact values of e.g.\  the $E(z)$ exponents are slightly different. The only noteworthy difference is that the highest mass bin evolves somewhat faster for all the scalings considered in the \wmap7 runs compared to the \planck~runs.  The calibrated \calsim~model has an evolution which is bracketed by that of \agn~8.0 and \agn~8.5 for all the scalings but the mass--temperature one (for both mass-weighted and X-ray temperature).

To summarise, the mass--temperature relation evolves in a negative fashion with respect to the self-similar model, independently of the included ICM physics.  The gas mass evolves approximately self-similarly for the non-radiative simulation but shows a positive evolution when radiative cooling and particularly AGN feedback is included, a result which is strongly mass dependent.  The SZ flux and $Y_X$ evolve negatively with respect to the self-similar expectation for models without efficient feedback, driven by the negative evolution of the temperature.  When AGN feedback is included, however, the {\it sign} of evolution of the SZ flux and $Y_X$ with respect to self-similar depends on halo mass (positive evolution for low-mass haloes, negative evolution for high-mass haloes), driven by the strong halo mass-dependence of the gas mass evolution combined with the negative evolution of the temperature.


\section{Evolution of the scatter}
\label{sec:scatter}

\begin{figure*}
\begin{center}
\includegraphics[width=0.43\hsize]{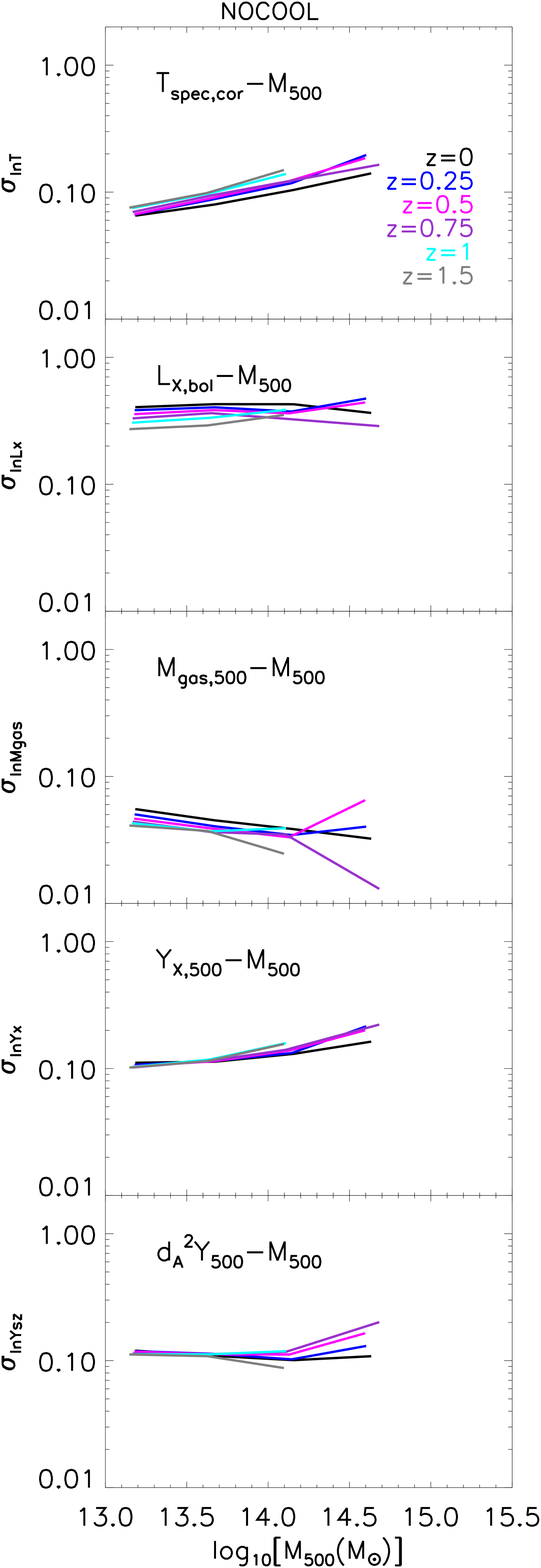}
\includegraphics[width=0.43\hsize]{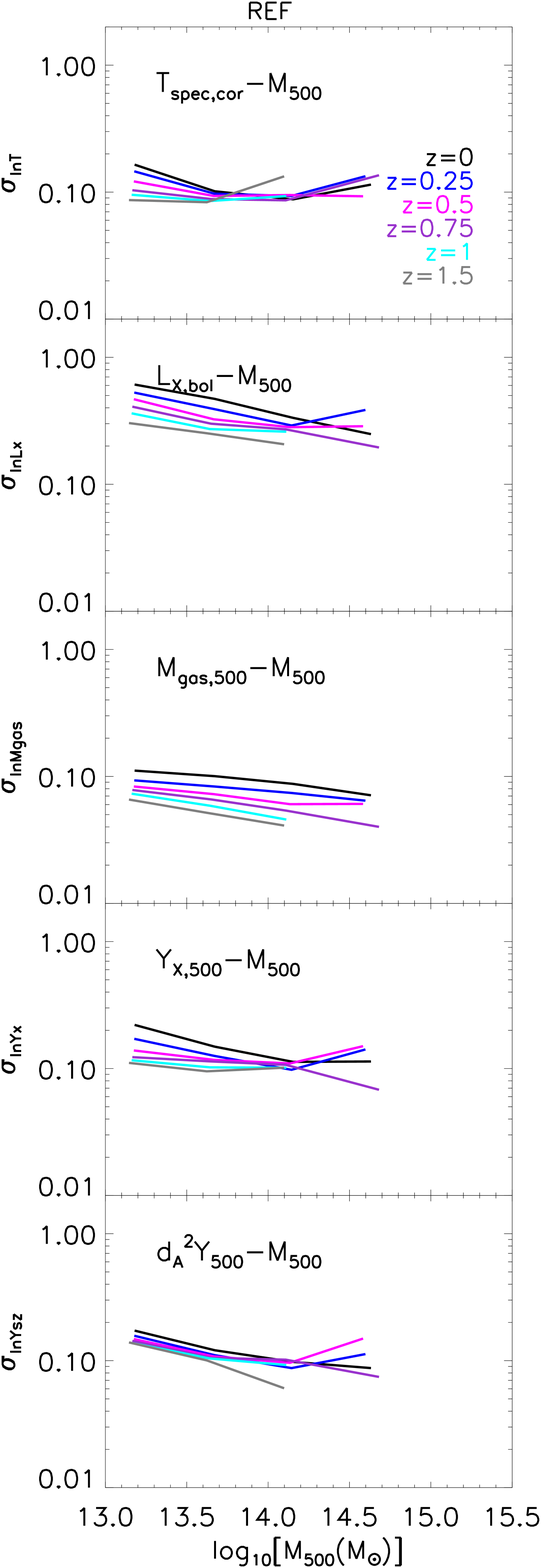}
\caption{Evolution of the log-normal scatter about the scaling relations between total mass and core-excised X-ray spectroscopic temperature, bolometric X-ray luminosity, gas mass, $Y_X$ and the integrated Sunyaev--Zel'dovich signal (from the \emph{top} subpanel to the \emph{bottom} subpanel).  For each simulation and each scaling relation, we plot the log-normal scatter as a function of $M_{500}$ and denote the redshift using lines of different colours. The different columns (continued over the page) correspond to the different physical models. For most scaling relations, the log-normal scatter varies only mildly with mass, is somewhat sensitive to non-gravitational physics, but displays a moderately strong redshift dependence (it tends to decrease with increasing redshift). With the exception of the X-ray luminosity, all the examined hot gas mass proxies have a similar scatter at fixed total mass of about 10 per cent.}
\label{fig:scatter}
\end{center}
\end{figure*}

\begin{figure*}
\begin{center}
\includegraphics[width=0.43\hsize]{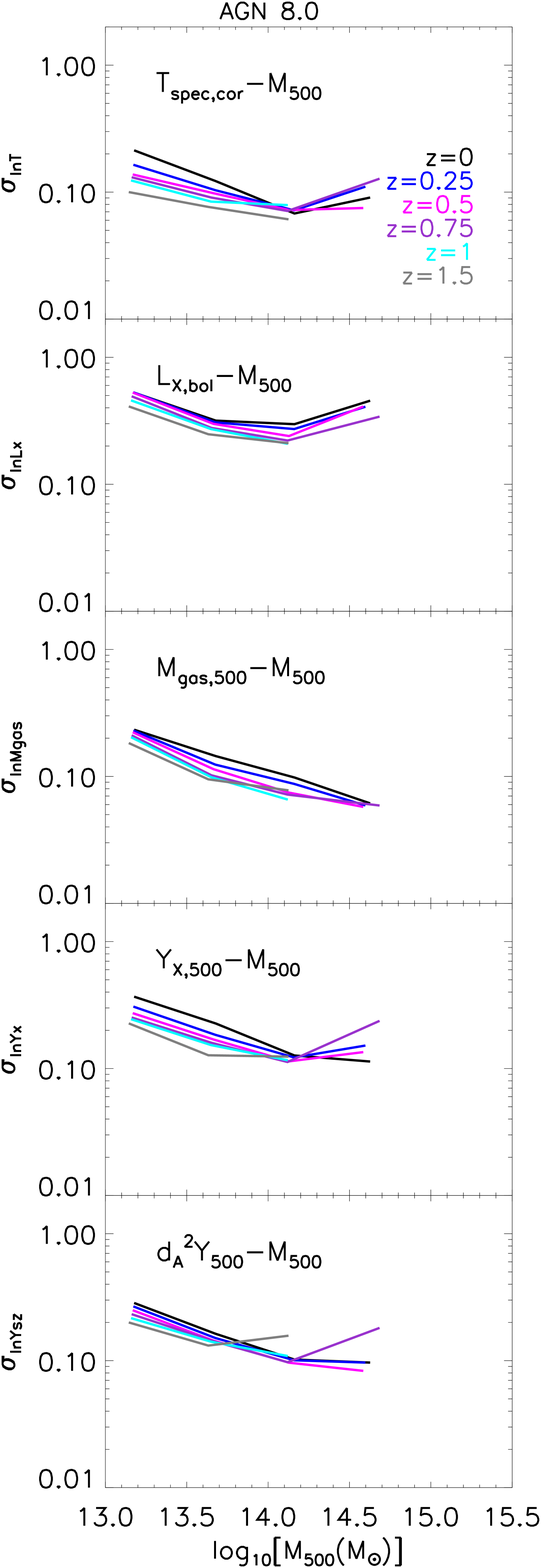}
\includegraphics[width=0.43\hsize]{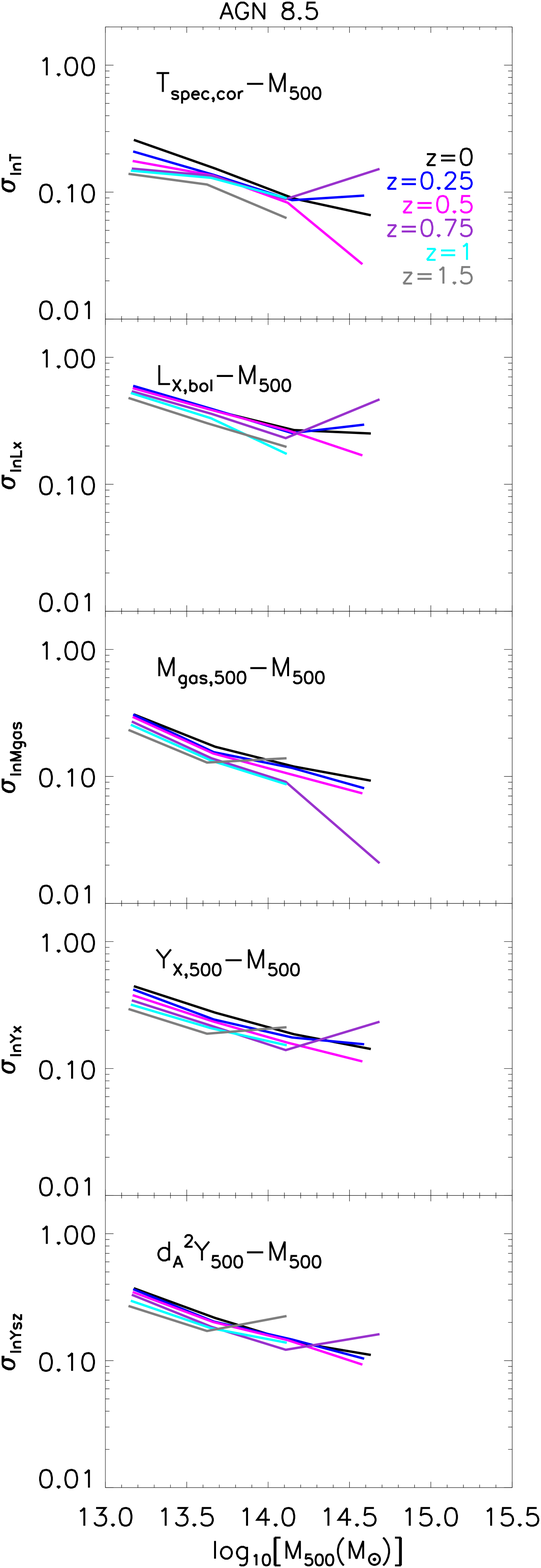}
\contcaption{Same as Fig.~\ref{fig:scatter} but for the \agn~8.0 and \agn~8.5 simulations.}
\label{fig:scattercont}
\end{center}
\end{figure*}

\begin{figure}
\begin{center}
\includegraphics[width=1.0\hsize]{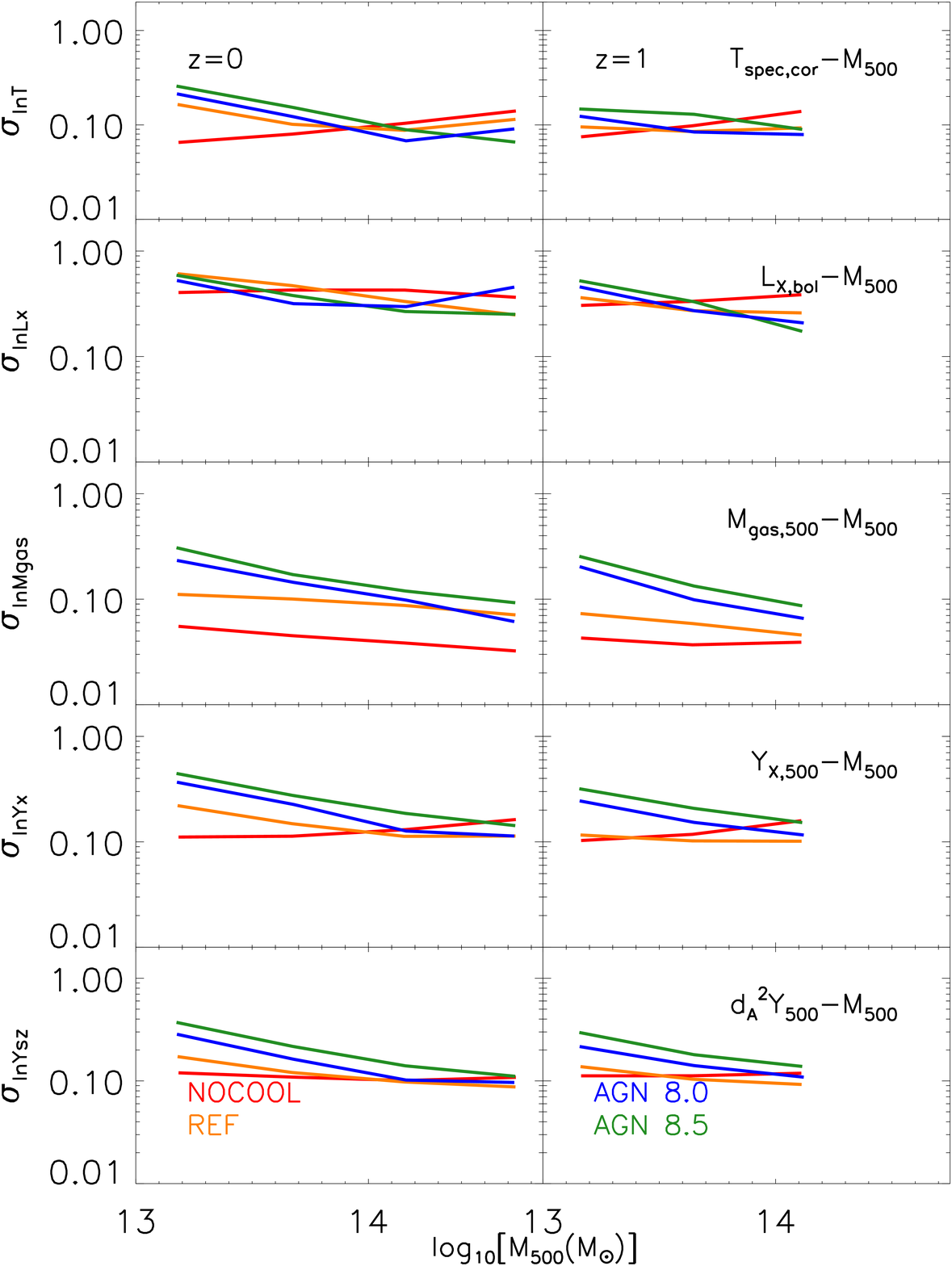}
\caption{Log-normal scatter of the scaling relations between total mass and core-excised X-ray spectroscopic temperature, X-ray luminosity, gas mass, $Y_X$ and the integrated Sunyaev--Zel'dovich signal (from \emph{top} to \emph{bottom}) at $z=0$ (\emph{left}) and $z=1$ (\emph{right}).  For each scaling relation and each redshift, we plot the log-normal scatter as a function of $M_{500}$ and denote the physical model using lines of different colours. The amplitude of the scatter tends to increase with increasing feedback (be it from SN or AGN) intensity at both redshifts.}
\label{fig:scatterbis}
\end{center}
\end{figure}

\begin{table*}
\centering
\caption{Scatter about the median scaling relations at $z=0$. $\sigma_{\ln Y|M}$ was computed using the best-fitting $A''$ values for the log-normal scatter of the \agn~8.0 model listed in Table~\ref{table:brokenpowerlaw}. $\sigma_{\ln M|Y}$ was computed as a sigma-clipped root mean square deviation from the (cubic spline interpolation of) the median $M-Y$ relation (analogous to equation~\ref{eq:rms}) of the \agn~8.0 model. The zero-point uncertainty of $Y$ was estimated as the root mean square dispersion of the normalisation $A$ of the best-fitting evolving power-law obtained for the median relations of the three radiative models (\refsim, \agn~8.0 and 8.5) and by multiplying it by $\ln10$. The observational constraints on $\sigma_{\ln Y|M}$ are summarised in the final column.}
\begin{tabular}{lcccc}
\hline
Scaling relation & $\sigma_{\ln Y|M}$ & $\sigma_{\ln M|Y}$ & Zero-point uncertainty in Y & Observational constraints on $\sigma_{\ln Y|M}$  \\ 
\hline
$T_{spec,cor}-M_{500}$         & $\approx5~\%$   & $\approx20~\%$ & $\approx5~\%$ & $\approx15~\%$ \\ 
$L_{bol}-M_{500}$    	      & $\approx25~\%$ & $\approx25~\%$ & $\approx40~\%$ & $\approx40~\%$ \\ 
$M_{gas,500}-M_{500}$         & $\approx10~\%$ & $\approx10~\%$ & $\approx25~\%$ & $\approx15~\%$ \\ 
$Y_{X,500}-M_{500}$	      & $\approx10~\%$ & $\approx15~\%$  & $\approx25~\%$ & $\approx20~\%$ \\ 
$d_{A}^{2}Y_{500}-M_{500}$  & $\approx10~\%$ & $\approx10~\%$  & $\approx20~\%$ & $\approx20~\%$ \\
$M_{500,hse,spec}-M_{500}$ & $\approx15~\%$ & $\approx25~\%$  & $\approx5~\%$ & ... \\
\hline
\end{tabular}
\label{table:scatter}
\end{table*}

Having quantified the evolution of the median relations between total mass and the various hot gas observables, we now move on to an examination of the scatter about these relations.  Quantifying the degree of scatter (and its possible dependence on halo mass, redshift and ICM physics) is extremely important for cosmological studies using cluster abundances, since the associated Eddington bias can be significant due to the steepness of the halo mass function.

In Fig.~\ref{fig:scatter}, we show the evolution of the log-normal scatter (at fixed total mass) about the scaling relations between total halo mass and core-excised X-ray spectroscopic temperature, bolometric X-ray luminosity, gas mass, $Y_X$ and the integrated Sunyaev--Zel'dovich signal (from \emph{top} to \emph{bottom}). For each simulation and each scaling relation, we plot the log-normal scatter as a function of $M_{500}$ and denote the redshift using lines of different colours.  

For most scaling relations, the log-normal scatter varies only mildly with mass, usually changing by less than a factor of $\approx2-3$ over 1.5 decades in $M_{500}$.  Nevertheless, it is clearly mass dependent in all of the models, whereas virtually all current observational studies adopt mass-invariant scatter.

The amplitude of the scatter at fixed redshift is sensitive to the included non-gravitational physics, with the amplitude tending to increase with increasing complexity (and realism) of the included galaxy formation physics and with increasing AGN feedback intensity, as can be seen more easily in Fig.~\ref{fig:scatterbis} (for both $z=0$ and $z=1$). This is perhaps not so surprising, as we know that AGN feedback is having a large effect on the mean properties.

On the positive side (for observational studies), the amplitude of the scatter tends to decrease with increasing redshift, which is likely due to the lessening importance of non-gravitational physics at higher redshifts when examining the evolution of haloes of fixed mass.  Typically, the scatter is reduced by $\sim50$ per cent from $z=0$ to $z=1$.

We note that the results above are generally robust to changing the cosmology from \planck~to \wmap7.

Table~\ref{table:scatter} summarises the scatter about the median scaling relations at $z=0$. The scatter in the observable at fixed mass, $\sigma_{\ln Y|M}$, is computed using the best-fitting $A''$ values for the log-normal scatter of the \agn~8.0 model listed in Table~\ref{table:brokenpowerlaw}. $\sigma_{\ln M|Y}$ was computed as a sigma-clipped (using a threshold value of 2$\sigma$ for the clipping) root mean square deviation from the (cubic spline interpolation of) the median $M-Y$ relation (analogous to equation~\ref{eq:rms}) of the \agn~8.0 model. The zero-point uncertainty of $Y$ was estimated by computing the root mean square dispersion of the normalisation $A$ of the best-fitting evolving power-law obtained for the median relations of the three radiative models (\refsim, \agn~8.0 and 8.5) and by multiplying it by $\ln10$.  The unphysical non-radiative simulation (\nocool) was excluded from the computation. 

All but one of the hot gas proxies examined here have a similar scatter at fixed total mass of order 10 per cent. Note, however, that the mass dependence is significant (see Fig.~\ref{fig:scatterbis}). The bolometric X-ray luminosity, which is the odd one out, has a significantly larger scatter at fixed total mass (it is about three times higher).  The theoretical relations exhibit values of $\sigma_{\ln Y|M}$ that are slightly smaller than some of the recent observational constraints, which are summarised in the rightmost column of Table~\ref{table:scatter} (see e.g.\ \citealt{Pratt2009,Vikhlinin2009,Mantz2010a,Andersson2011}; \citeauthor{Planck2011c}; \citealt{Lin2012}; \citeauthor{Planck2012b,Planck2013a} and \citealt{Allen2011} for a review).  We caution that this difference is at least partly due to the fact that we are not using full synthetic X-ray observations of the `depth' typical of these observational datasets.  A careful comparison would also require us to fold in the full observational selection functions of these datasets, which is beyond the scope of this work.  We note, however, that the values obtained for $\sigma_{\ln Y|M}$ are similar to those found in several previous simulation studies \citep[see for instance][]{Nagai2006,Kravtsov2006,Stanek2010,Fabjan2011,Kay2012}.


\section{Scatter and evolution of the hydrostatic bias}
\label{sec:HSE}

High-quality X-ray observations that allow for the measurement of spatially-resolved temperature and density profiles can be used to derive a direct estimate of the total mass via the assumption of hydrostatic equilibrium.  However, clusters are not expected to be perfectly in equilibrium, and simulations are often relied upon to calibrate the degree of `hydrostatic bias' that is expected.  While such hydrostatic analysis is generally applicable to local systems only, owing to the rapidly declining X-ray surface brightness with increasing redshift, it is nevertheless interesting to examine the evolution and scatter in the hydrostatic bias and its sensitivity to the included gas physics, as future X-ray missions such as {\it eRosita} and eventually {\it Athena} will be able to extend this kind of analysis to higher redshifts.  (Stacking of large numbers of high-redshift systems in current data, to derive mean hydrostatic halo mass estimates, is also possible.)  We note that there has already been considerable attention devoted to quantifying the hydrostatic bias in simulated clusters (e.g.\ \citealt{Evrard1996,Kay2007}; \citealt*{Nagai2007b}; \citealt{Piffaretti2008,Meneghetti2010,Battaglia2012,Kay2012,Rasia2012,Rasia2014}). However, most of these studies examined only relatively small samples of clusters in zoomed simulations and could not explore the system-to-system scatter in the hydrostatic bias and its evolution, which we do below.

\begin{figure}
\begin{center}
\includegraphics[width=1.0\hsize]{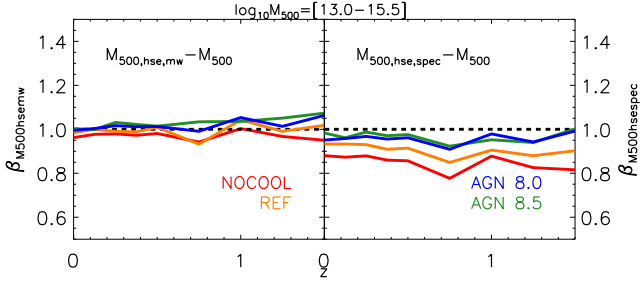}
\caption{Evolution of the mass slope of the total mass--hydrostatic mass scaling relations obtained when using either the density and temperature profiles derived from spectral fitting (\emph{right}) or the mass-weighted profiles (\emph{left}) to compute the hydrostatic masses. In each panel, we plot the redshift evolution of the best-fitting power-law indices obtained by fitting the broken power-law given by equation~\eqref{eq:power}. The solid curves (red, orange, blue and green) correspond to the different simulations and the horizontal dashed lines to the self-similar expectation, respectively.  The mass-weighted hydrostatic mass tracks the absolute mass well (i.e.\ $M_{\rm 500,hse,mw} \propto M_{500}$), whereas the spectroscopically-derived hydrostatic mass tracks the true halo mass slightly less well, with $M_{\rm 500,hse,spec} \propto M_{500}^{0.9}$ typically.}
\label{fig:HSEslopeevo}
\end{center}
\end{figure}

\begin{figure}
\begin{center}
\includegraphics[width=1.0\hsize]{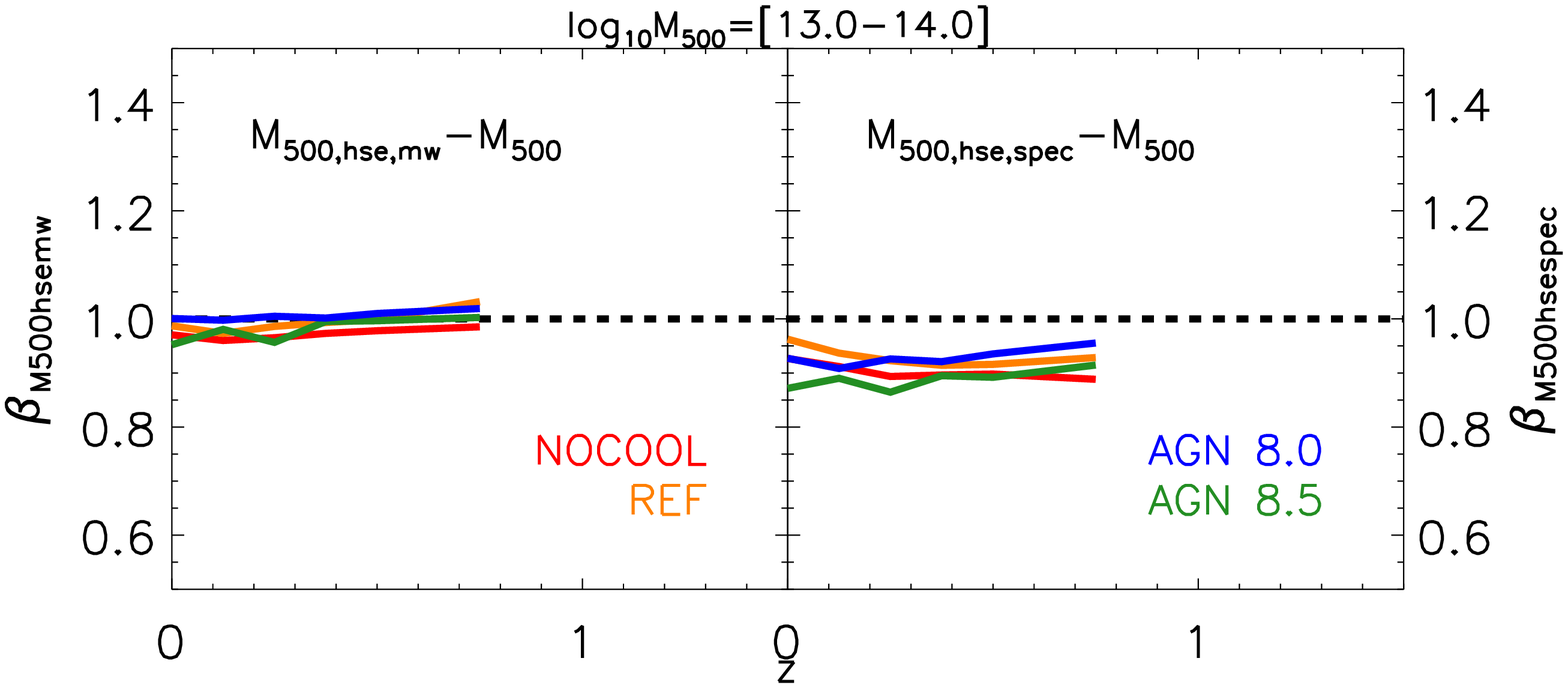}
\includegraphics[width=1.0\hsize]{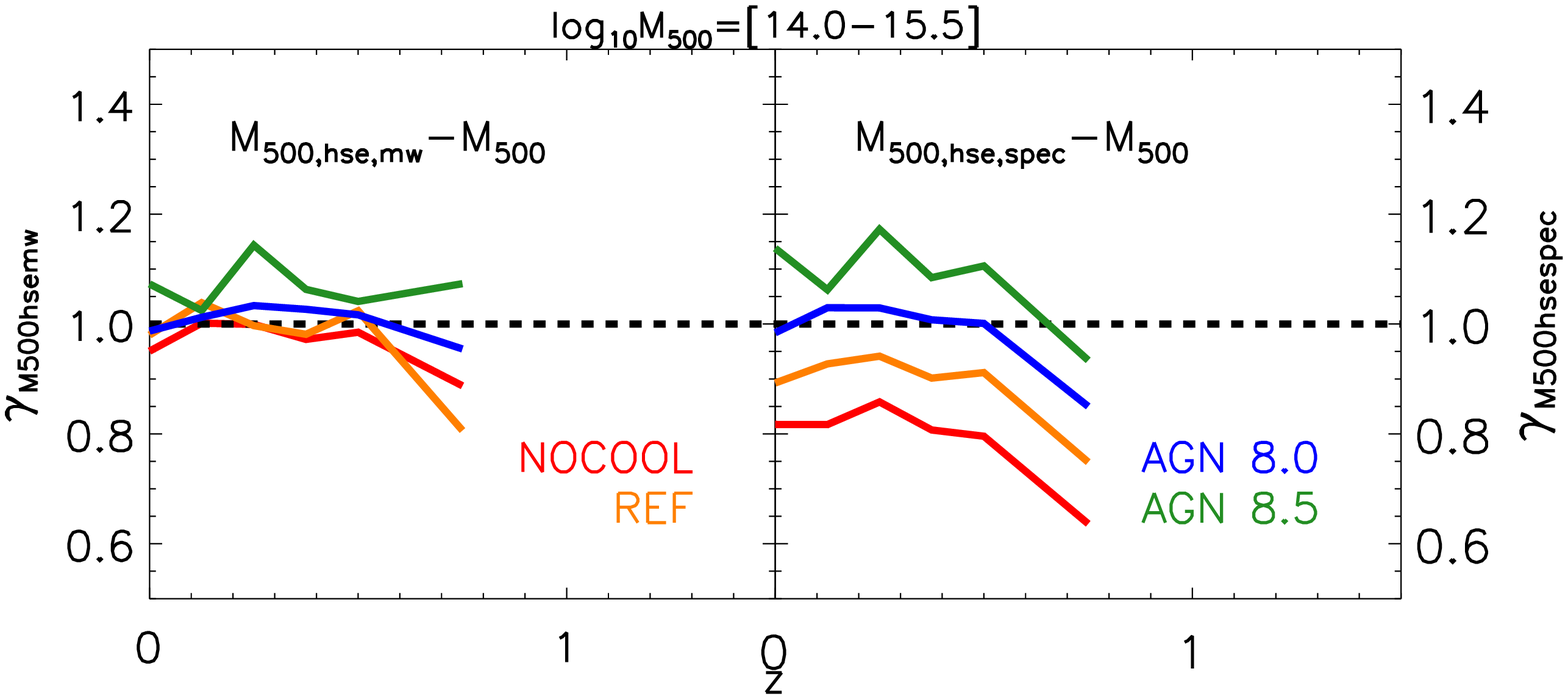}
\caption{Evolution of the mass slope of the total mass--hydrostatic mass scaling relations obtained when using either the density and temperature profiles derived from spectral fitting (\emph{right} subpanels) or the mass-weighted profiles (\emph{left} subpanels) to compute the hydrostatic masses.  In each subpanel, we plot the redshift evolution of the low-mass (\emph{top} panel) and high-mass (\emph{bottom} panel) best-fitting power-law indices obtained by fitting the broken power-law given by equation~\eqref{eq:broken}. The solid curves (red, orange, blue and green) correspond to the different simulations and the horizontal dashed lines to the self-similar expectation, respectively.  The mass-weighted hydrostatic mass tracks the absolute mass well (i.e.\ $M_{\rm 500,hse,mw} \propto M_{500}$), whereas the spectroscopically-derived hydrostatic mass tracks the true halo mass less well, with the exact scaling depending on the included gas physics and halo mass range.}
\label{fig:HSEslopeevobroken}
\end{center}
\end{figure} 

\begin{figure}
\begin{center}
\includegraphics[width=1.0\hsize]{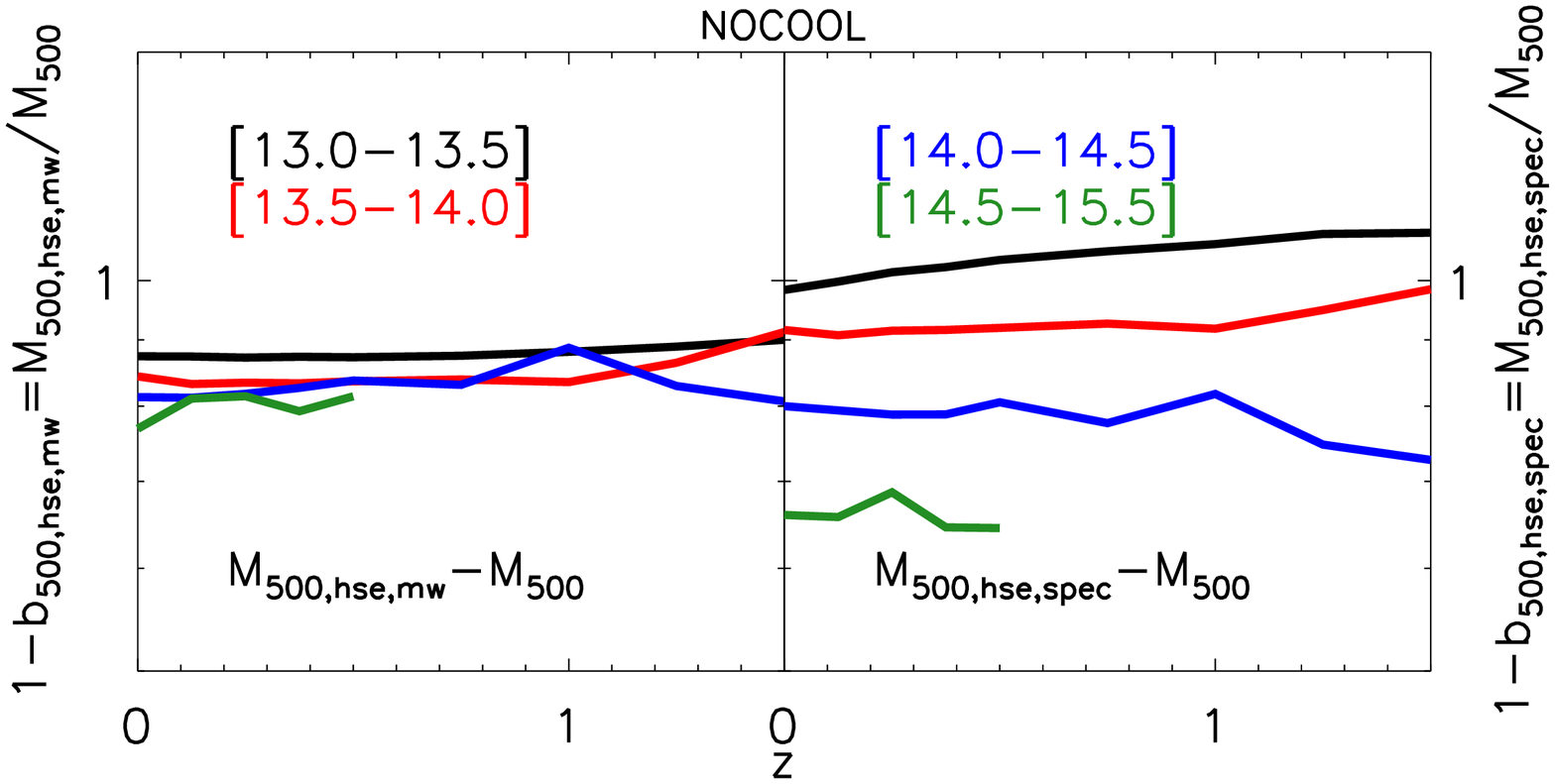}
\includegraphics[width=1.0\hsize]{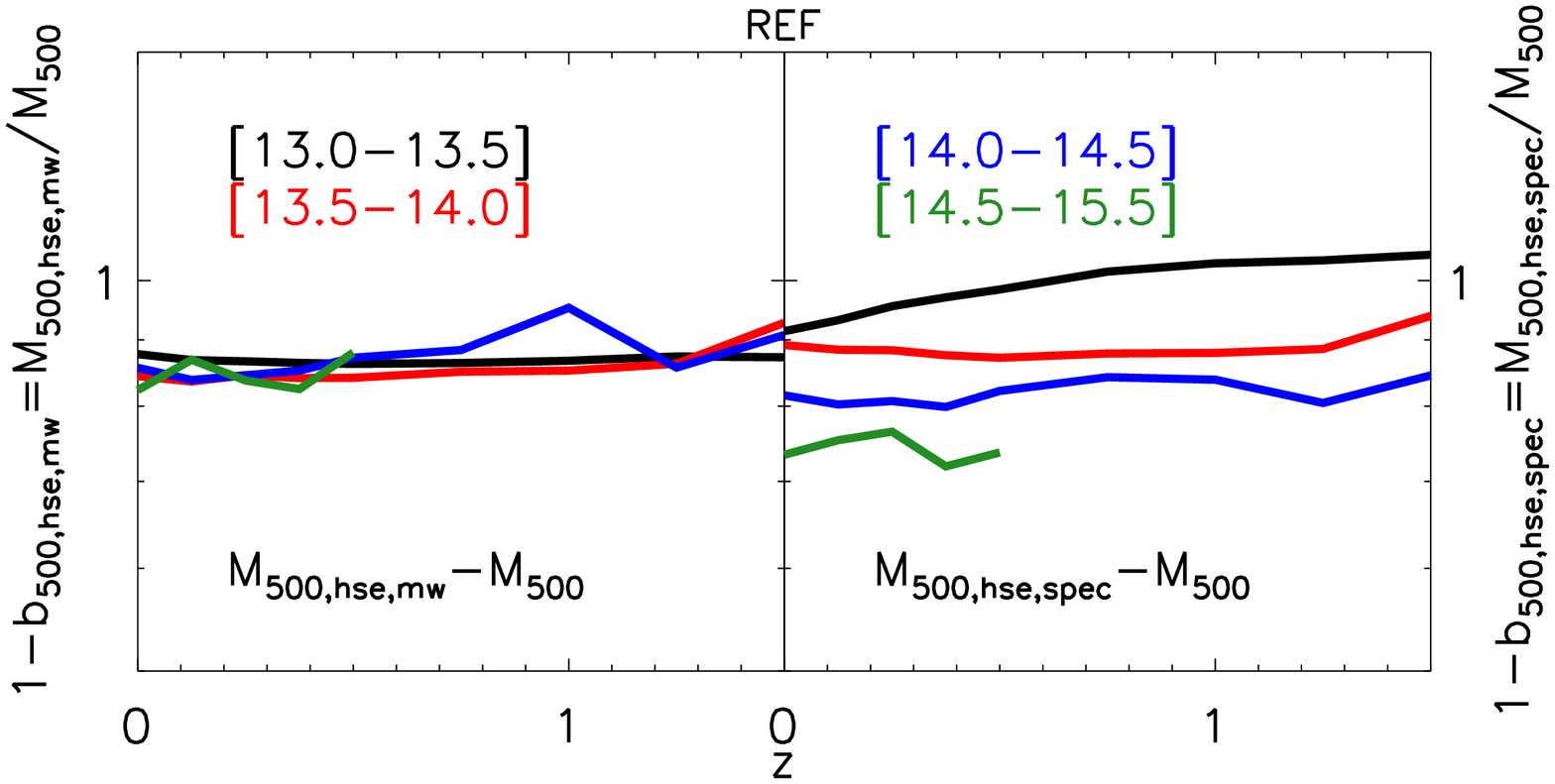}
\includegraphics[width=1.0\hsize]{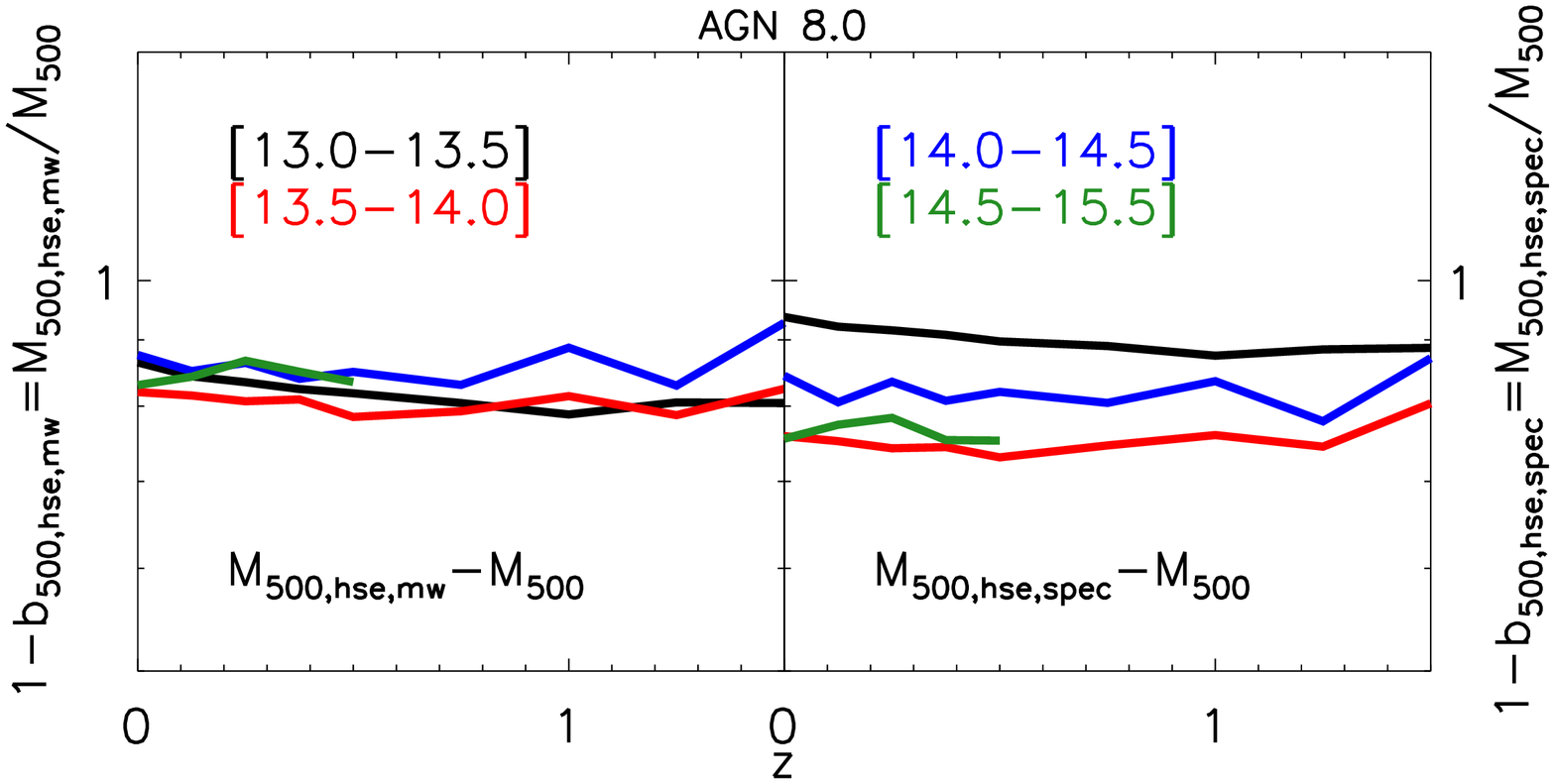}
\includegraphics[width=1.0\hsize]{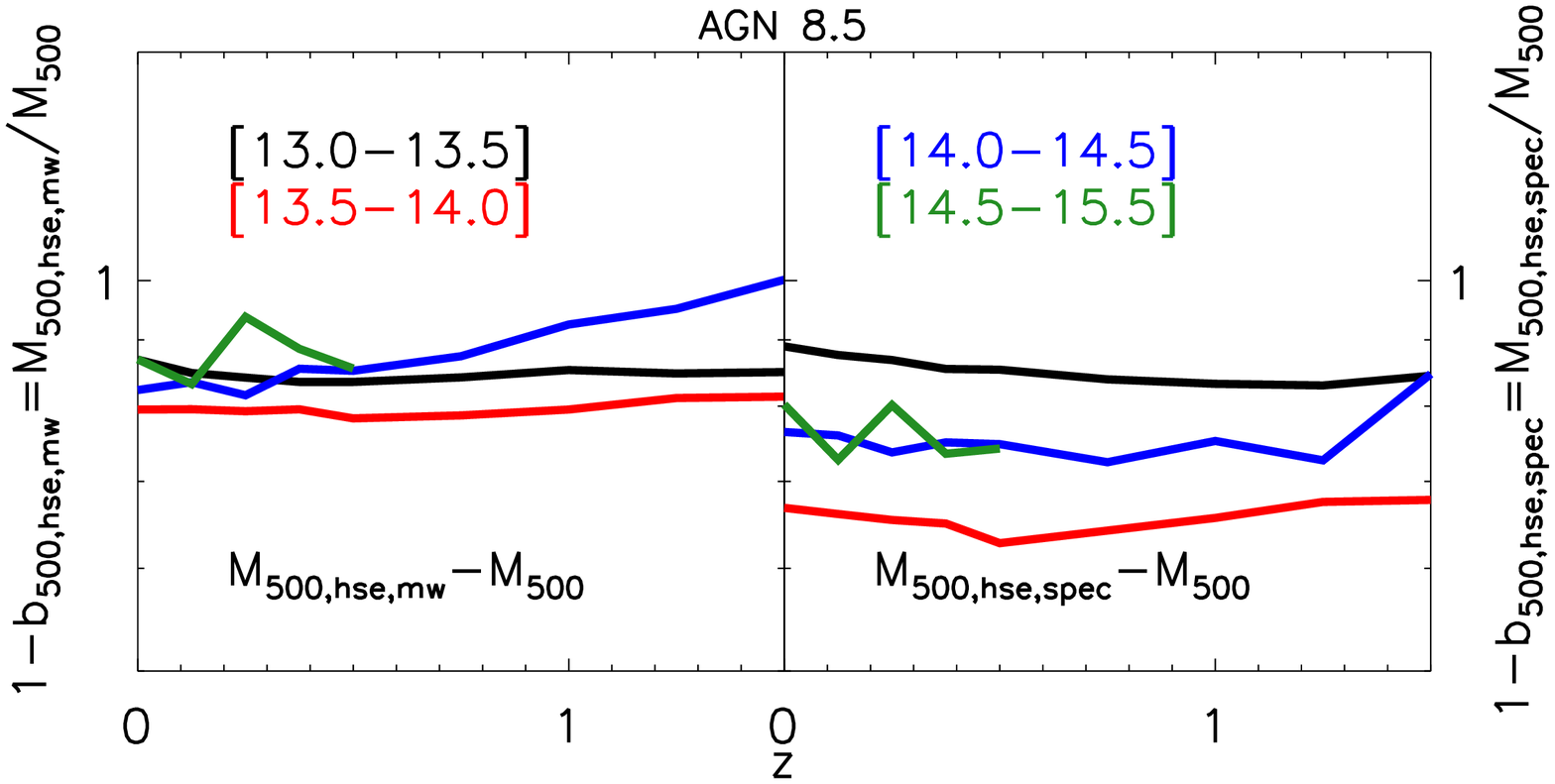}
\caption{Evolution of the normalisation of the total mass--hydrostatic equilibrium bias scaling relations obtained when using either the density and temperature profiles derived from from spectral fitting (\emph{right} subpanels) or the mass-weighted profiles (\emph{left} subpanels) to compute the hydrostatic masses. The four $\log_{10}[M_{500}(\textrm{M}_\odot)]$ bins are denoted by solid lines of different colours and the different panels correspond to the different physical models. A mass and non-gravitational physics dependence is introduced when the profiles coming from the spectral fitting are used. Galaxy groups and clusters have different sensitivity to galaxy formation physics: their hydrostatic equilibrium biases vary in opposite directions with increased feedback intensity.}
\label{fig:HSEnormevo}
\end{center}
\end{figure}

\begin{figure*}
\begin{center}
\includegraphics[width=0.497\hsize]{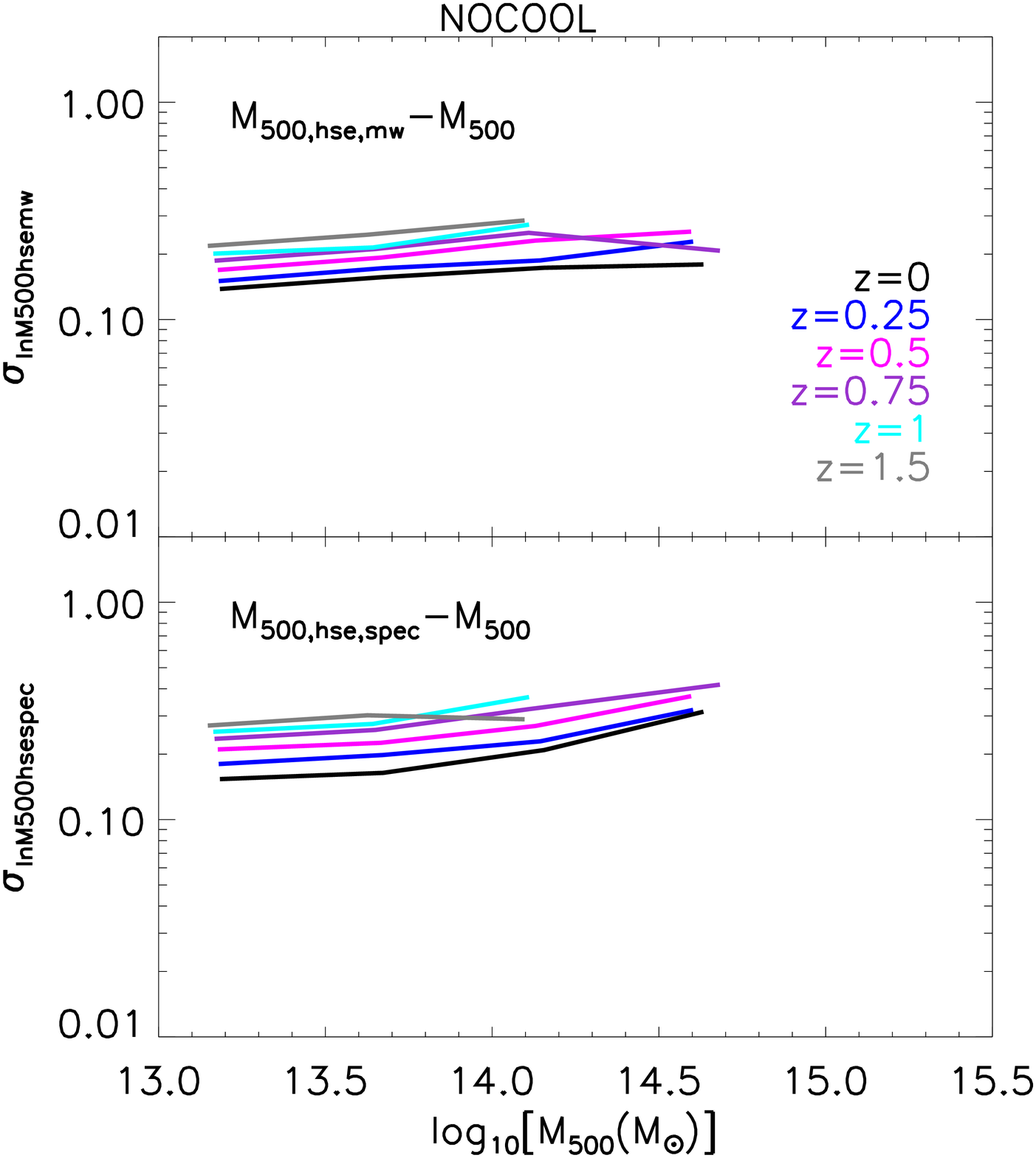}
\includegraphics[width=0.497\hsize]{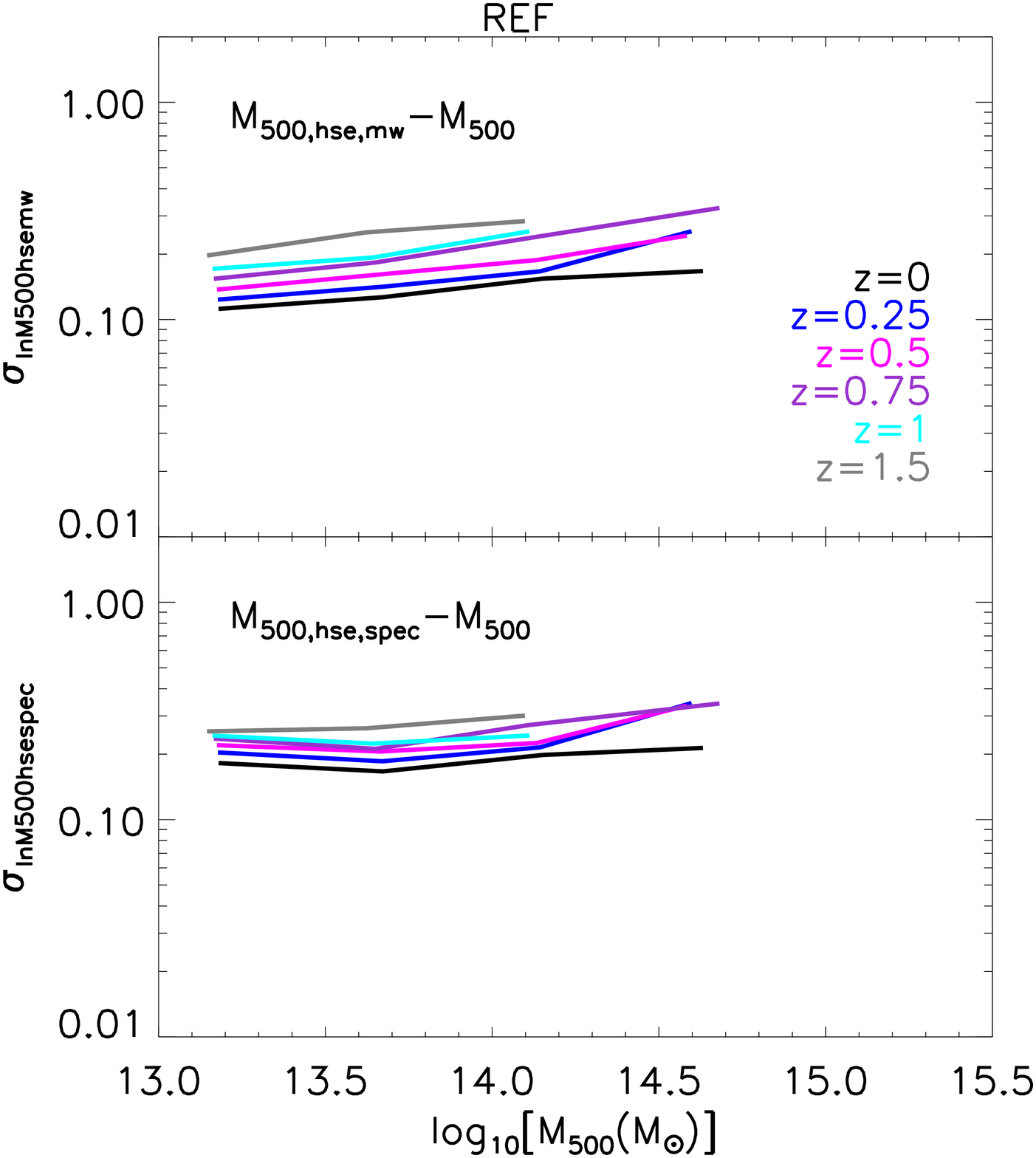}
\includegraphics[width=0.497\hsize]{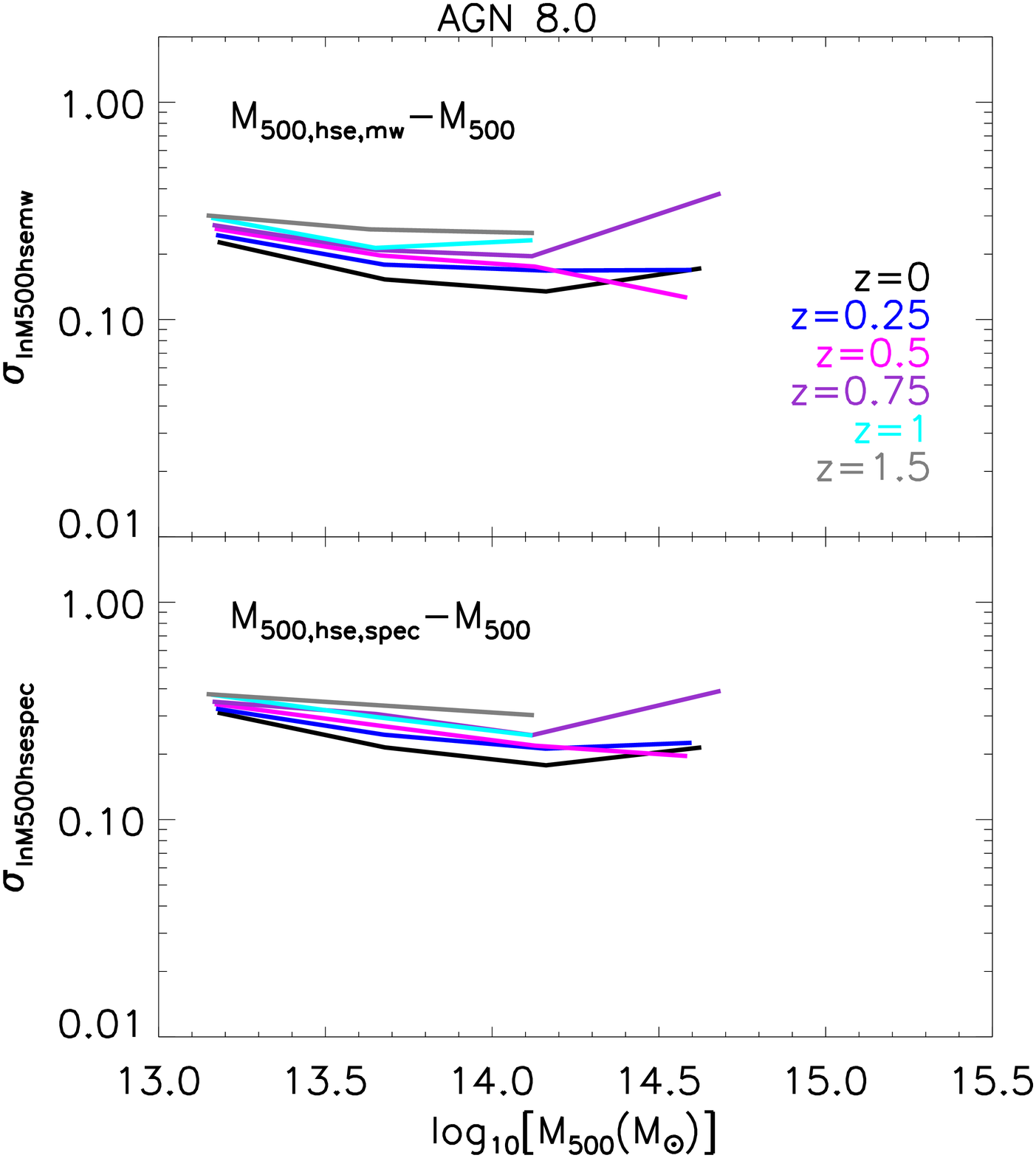}
\includegraphics[width=0.497\hsize]{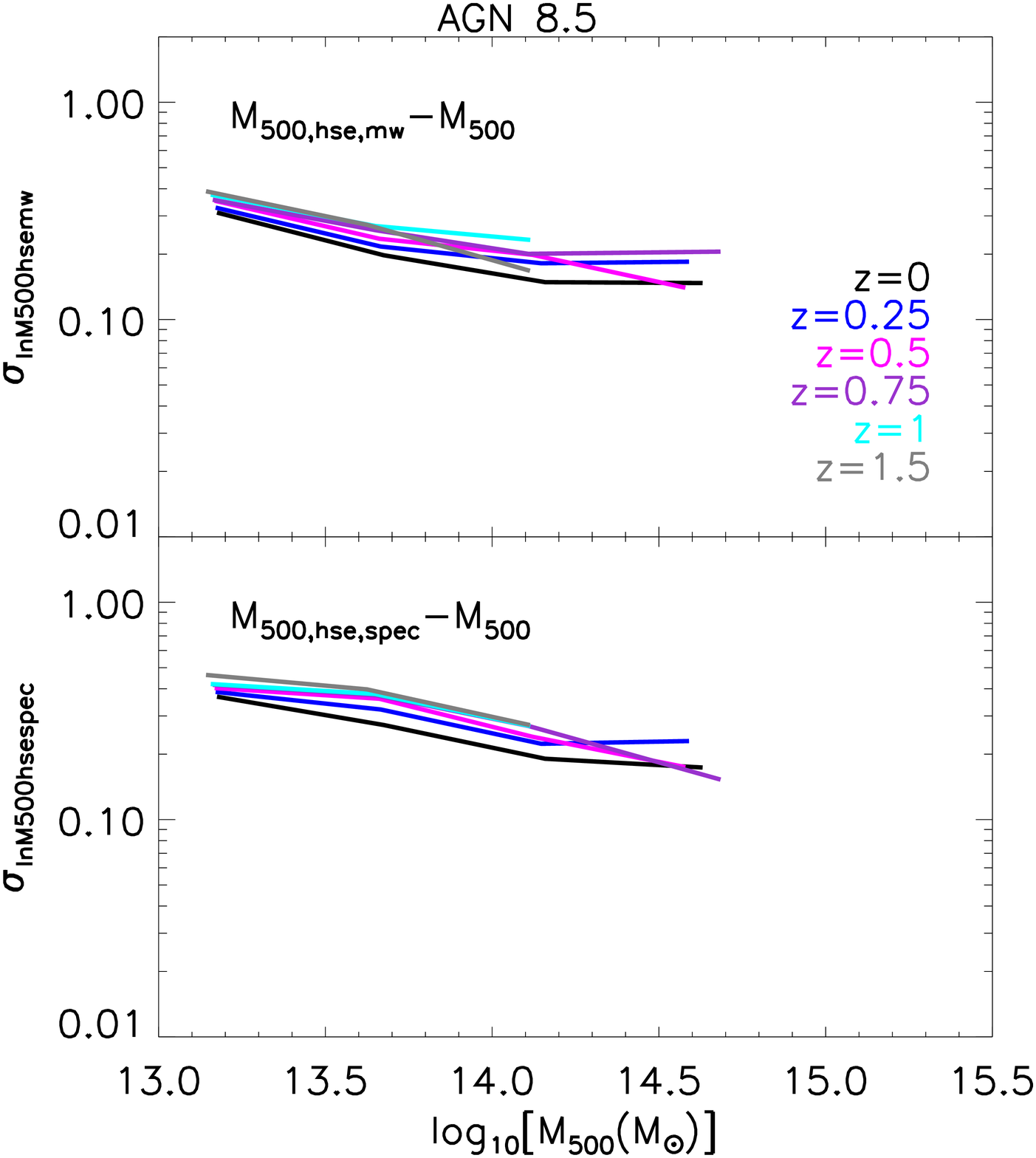}
\caption{Evolution of the log-normal scatter about the total mass--hydrostatic mass scaling relations obtained when using either the density and temperature profiles coming from the results of the spectral fitting (\emph{bottom} subpanels) or the mass-weighted profiles (\emph{top} subpanels) to compute the hydrostatic masses. For each simulation and each scaling relation, we plot the log-normal scatter as a function of $M_{500}$ and denote the redshift using lines of different colours. For both types of hydrostatic masses, the log-normal scatter varies only mildly with mass, but displays slightly stronger redshift (it tends to increase with increasing redshift contrarily to most of the hot gas scalings; see Fig.~\ref{fig:scatter}) and gravitational physics dependencies. About $0.04$ dex additional scatter is introduced in the relation by the use of spectral fitting to obtain the density and temperature profiles.}
\label{fig:HSEscatter}
\end{center}
\end{figure*}

The hydrostatic masses were computed as in \citetalias{LeBrun2014}. Briefly, the parametric models of \citet{Vikhlinin2006} were fitted to either the temperature and density profiles coming from the results of the spectral fitting or the mass-weighted profiles over the radial range $[0.1-1.1]r_{500}$ and hydrostatic equilibrium was assumed to compute the mass profiles.

In Figs.~\ref{fig:HSEslopeevo} and~\ref{fig:HSEslopeevobroken}, we show the evolution of the mass slopes from $z=0$ to $z=1.5$ for the total mass--hydrostatic mass scaling relations for each of the four physical models. In Fig.~\ref{fig:HSEslopeevo}, we plot the redshift evolution of the best-fitting power-law indices obtained by fitting the power-law given by equation~\eqref{eq:power}, whereas in Fig.~\ref{fig:HSEslopeevobroken}, we plot the redshift evolutions of the low-mass (\emph{top} panel) and high-mass (\emph{bottom} panel) best-fitting power-law indices obtained by fitting the broken power-law given by equation~\eqref{eq:broken}.  The solid curves (red, orange, blue and green) represent the various simulations.  The horizontal dashed lines correspond to the self-similar expectation (i.e.\ hydrostatic mass tracks the true mass).  We consider two varieties of hydrostatic mass: one where we use the mass-weighted density and temperature profiles directly from the simulations (\emph{left} panels) and one where we use the spectroscopically-derived density and temperature profiles in the hydrostatic equation (\emph{right} panels).  The former is a better indicator of the true degree of hydrostatic bias, whereas the latter is also sensitive to observational biases in the derived gas density and temperature profiles (e.g.\ due to gas clumping and multiphase structure).

When the hydrostatic masses are computed using the mass-weighted profiles, the mass slopes are close to the self-similar expectation of 1, independent of mass, redshift and the nature of the included ICM physics.  The use of the spectroscopic profiles, however, breaks the self-similarity, by making the relation shallower than the self-similar expectation of 1 in the vast majority of cases (the only exceptions being the high-mass end slope of the \agn~8.0 model for $z<0.6$).  This is generally true for all redshifts, masses and physical models.  It is worth mentioning that the model that displays the largest deviation from self-similarity is the unphysical non-radiative model \nocool, likely due to the excessive gas clumping and high degree of multiphase structure present in this simulation. Because the gas is unable to radiate, large quantities of gas can exist at very high densities/short cooling times in the \nocool~model.  Such gas is thermally unstable when radiative cooling is switched on and it quickly cools down to very low (non-X-ray-emitting) temperatures, where it is either converted into stars or heated by the central AGN.  Thus, inclusion of radiative cooling and feedback generally leads to a {\it less} clumpy X-ray emitting ICM. Additionally, the mass slope is shallower for clusters than for groups and, in the clusters' case, the slope gets steeper with increasing feedback intensity. This is probably due to the fact that clusters, being less relaxed, tend to have more mass in substructures (be clumpier; see e.g.\ \citealt{Neto2007}) and to the degree of clumpiness being dependent on feedback intensity \citep[see e.g.][]{Battaglia2015}. Note that both are only apparent dependencies which are due to the use of the spectroscopic profiles since they disappear in the left panels of Fig.~\ref{fig:HSEslopeevobroken} when the mass-weighted profiles are used.

In Fig.~\ref{fig:HSEnormevo}, we show the evolution of the normalisation of the total mass--hydrostatic equilibrium bias scaling relations obtained when using both the spectroscopic (\emph{right} subpanels) and mass-weighted (\emph{left} subpanels) density and temperature profiles to compute the hydrostatic masses.  The results are shown for the four $\log_{10}[M_{500}(\textrm{M}_\odot)]$ bins (denoted by solid lines of different colours). 

When the hydrostatic masses are computed using the true mass-weighted profiles, the bias $1-b\equiv \frac{M_{500,hse,mw/spec}}{M_{500}}$ is nearly independent of redshift and only weakly dependent on mass and included gas physics, with a typical value of $\approx0.8$; i.e.\ $M_{\rm 500,hse,mw} \approx 0.8 M_{500}$.  The use of spectroscopic profiles changes the situation, however, introducing a strong mass dependence in the bias for the \nocool~and \refsim~models.  As already noted, this is likely due to the effects of gas clumping and the high degree of multiphase temperature structure present in these models, particularly the non-radiative one.  Inclusion of energetic AGN feedback reduces the degree of gas clumping, which removes some (but not all) of the mass dependence of the spectroscopic `hydrostatic' bias parameter.  Typically, the spectroscopic bias parameter ranges between 0.7 to 1.0 depending on the exact mass range under consideration for the AGN feedback models, and is approximately independent of redshift. This is similar to the range ($1-b\approx0.7-0.95$) suggested by recent measurements of the scaling relation between weak lensing \citep[see e.g.][]{vonderLinden2014,Hoekstra2015,Penna2016,Smith2016} or CMB lensing \citep[see e.g.][]{Melin2015} and hydrostatic masses. 

Finally, in Fig.~\ref{fig:HSEscatter}, we show the evolution of the log-normal scatter (at fixed total mass) about the total mass--hydrostatic mass scaling relations obtained when using either the spectroscopic (\emph{bottom} subpanels) or mass-weighted (\emph{top} subpanels) density and temperature profiles to compute the hydrostatic masses.  For each simulation and each scaling relation, we plot the log-normal scatter as a function of $M_{500}$ and denote the redshift using lines of different colours. 

For both types of hydrostatic masses, the log-normal scatter varies only mildly with mass, but displays a slightly stronger redshift dependence, tending to increase with increasing redshift.  This trend is contrary to that of the scaling relations we investigated in Section~\ref{sec:scatter}, which show decreasing scatter with increasing redshift.   Only a modest amount of additional scatter ($\approx0.04$ dex, typically) is introduced when we use the spectroscopically-derived bias as opposed to the mass-weighted hydrostatic bias.  

It is noteworthy that, on average, the hydrostatic bias has a larger overall scatter than the other mass proxies (apart from the X-ray luminosity) we have considered (see Table~\ref{table:scatter}) and becomes increasingly less competitive with the other proxies with increasing redshift. However, together with the core-excised X-ray temperature, it has the smallest zero-point uncertainty due to the uncertain non-gravitational physics (see Table 4 and discussion in Section~\ref{sec:disc}).

All these results are robust to changing the cosmology from \planck~to \wmap7.


\section{Comparison with previous work}
\label{sec:comp}

Hot gas scaling relations and their (potential) deviations from self-similar expectations have been explored in a number of previous studies using cosmological hydrodynamical simulations, some of which also included AGN feedback \citep[or preheating;][]{Short2010,Stanek2010,Battaglia2010,Battaglia2012,Fabjan2011,Planelles2013,Pike2014}.  Here we compare our findings with previous studies.

In terms of the mass slopes, \citet{Stanek2010}, \citet{Fabjan2011}, \citet{Biffi2014} and \citet{Pike2014} all found that the slope of the mass--temperature scaling is shallower than the self-similar expectation, in agreement with the results of this study (but see \citealt{Kravtsov2006} and \citealt{Nagai2007a} who found no deviation from self-similarity). It is worth noting though that Pike et al.\ found that the slope decreases with increasing redshift whereas we found a steepening and that Biffi et al.\ also concluded that some of the deviation is due to X-ray spectral fitting. \citet{Kay2012} and \citet{Battaglia2012} similarly saw a steepening of the SZ flux--mass relation when AGN feedback is included (but see \citealt{Pike2014}), whereas \citet{Nagai2006} concluded that galaxy formation physics has a small impact by comparing a non-radiative zoom simulation with one that included cooling and star formation but no AGN feedback.  Our results on the SZ flux--mass relation are consistent with these studies, in that we find that inclusion of cooling, star formation, and stellar feedback (as in the \refsim~model) does not significantly affect the slope of the relation, but the further inclusion of energetic AGN feedback, which lowers the gas fractions of groups, steepens the relation.

\citet{Short2010}, \citet{Stanek2010} and \citet{Fabjan2011}, in agreement with the results discussed here, concluded that the mass--temperature relation evolves negatively (but see \citealt{Nagai2007a}). Stanek et al.\ similarly attributed it to the fact that the kinetic-to-thermal ratio increases with redshift.  In terms of the gas mass--total mass scaling, \citet{Planelles2013} found that it evolves self-similarly even when AGN feedback is included and \citet{Fabjan2011} found a mild positive evolution even in the adiabatic case and in a model similar in spirit to \refsim~(see also \citealt{Kravtsov2006}), both in clear disagreement with the present study.  Note that for Planelles et al., this is due to the fact that the baryon fraction nearly tracks the universal baryon fraction $\Omega_b/\Omega_m$ even in their simulations with AGN feedback (i.e.\ the feedback was not sufficiently energetic to lower the gas fractions of groups). \citet{Short2010} and \citet{Battaglia2012} reported noticeable deviations from self-similarity in the evolution of $Y_{X}$ and SZ flux, respectively, in agreement with the results presented here, whereas \citet{Nagai2006}, \citet{Kay2012}, \citet{Pike2014}, \citet{Kravtsov2006}, \citet{Nagai2007a} and \citet{Fabjan2011} saw no deviations from the self-similar expectation for the normalisation of the SZ flux (for the first three references) and of $Y_{X}$ (for the last three) even when non-gravitational physics is included.  We again attribute these differences to inefficient feedback in these studies, that we argue, in \citet{McCarthy2011} and \citetalias{LeBrun2014}, is necessary to reproduce the observed low baryon fractions of groups.

\citet{Nagai2007b}, in reasonable agreement with the present study, found a $5-20$ per cent hydrostatic bias (see also e.g.\ \citealt{Meneghetti2010,Battaglia2012,Rasia2012}). Additionally, \citet{Kay2012} reported a $20-30$ per cent hydrostatic bias for the non-radiative runs and a $\sim10$ per cent decrease when non-gravitational physics is included, in agreement with the results presented here.


\section{Summary and Discussion}
\label{sec:disc}

We have used the cosmo-OWLS suite of large-volume hydrodynamical simulations \citep{LeBrun2014,McCarthy2014} to investigate the scatter and evolution of the global hot gas properties of large populations of simulated galaxy groups and clusters.  The simulations use $2\times1024^3$ particles in a 400 $h^{-1}$ Mpc on a side box, and assume either a \wmap~7-year or a \planck~2013 cosmology. We have studied all the haloes with $M_{500}\ge10^{13}~\textrm{M}_{\odot}$ over the redshift range $z=0-1.5$. For instance, there are over $25,000$ such systems in the \planck~cosmology version of the \nocool~model. The most realistic models, which include efficient feedback from AGN, have been shown by Le Brun et al.\ to reproduce a wide range of observable properties of local systems.  Note that cosmo-OWLS forms an extension of the OWLS project (\citealt{Schaye2010}) and was designed to help quantify the importance of uncertainties in `sub-grid' physics for cluster cosmology efforts.

From the study presented here, we conclude the following:
\begin{enumerate}
\item The median relations between (true) halo mass and the (core-excised) mass-weighted and X-ray spectroscopic temperature, bolometric X-ray luminosity, gas mass, Sunyaev-Zel'dovich (SZ) flux, its X-ray analogue and the hydrostatic mass, and the scatter about them are reasonably well modelled by evolving broken power-laws (equation~\ref{eq:brokenextra}), whereas single power-laws of the form given by equation~\eqref{eq:power} perform less well (Fig.~\ref{fig:reconstructed}).  In the \agn~models, the physical origin of this result is tied to the break of the self-similarity in the relation between gas mass fraction and halo mass (which is also observed): massive clusters have approximately constant gas fractions near the universal value of $\Omega_b/\Omega_m$, while for lower mass systems ($M_{500} \la 10^{14} M_\odot$), the gas fraction is a steadily declining function of halo mass.
\item In terms of the mass slopes (i.e.\ the logarithmic slopes of the various mass--observable relations), we find that our most realistic (\agn) models predict large deviations from the self-similar expectation for all of the scaling relations we have examined.  The one exception is the mass--temperature relation, where only a weak, though still significant, deviation from self-similarity is found (independently of the details of the included ICM physics).  For the other scaling relations, all of which depend directly on the gas density/mass, the deviation from the self-similar prediction is strongest at low redshifts and low halo masses.  Models without efficient feedback (\nocool~and \refsim), on the other hand, have mass slopes that are approximately consistent with self-similar expectations (Figs.~\ref{fig:slopeevo}-\ref{fig:slopeevobroken}).
\item In terms of the evolution of the normalisation of the scaling relations, we find that the mass--temperature relation evolves in a negative fashion with respect to the self-similar model (i.e.\ slower than predicted), independently of the included ICM physics.  This is likely due to the increasing contribution of non-thermal pressure support with increasing redshift (the kinetic-to-thermal energy ratio of the ICM increases from a typical value of 10-15 per cent at $z=0$ up to 20-30 per cent by $z=1$; see Fig.~\ref{fig:KtoTevo}).  The gas mass evolves approximately self-similarly (i.e.\ remains constant) for the non-radiative simulation, but shows a positive evolution when radiative cooling and particularly AGN feedback is included, a result which is strongly mass dependent.  We hypothesize that, in the case of the \agn~models,  the positive evolution is due to the increasing binding energy of haloes of {\it fixed total mass} with increasing redshift (making gas expulsion more difficult at high redshift).  The SZ flux and $Y_X$ evolve negatively with respect to the self-similar expectation for models without efficient feedback, driven by the negative evolution of the temperature.  When AGN feedback is included, however, the {\it sign} of the evolution of the SZ flux and $Y_X$ with respect to self-similar depends on halo mass (positive evolution for low-mass haloes, negative evolution for high-mass haloes), driven by the strong halo mass-dependence of the gas mass evolution combined with the negative evolution of the temperature (Fig.~\ref{fig:normevo}).
\item The scatter about the various mass--observable relations is mass dependent, varying by a factor of $2-3$ over 1.5 decades in $M_{500}$ (Fig.~\ref{fig:scatterbis}).  The overall amplitude of the scatter at fixed redshift is somewhat sensitive to the included non-gravitational physics, with the amplitude tending to increase with increasing complexity (and realism) of the included galaxy formation physics and with increasing AGN feedback intensity.  Encouragingly, the overall amplitude of the scatter tends to decrease with increasing redshift, which is likely due to the lessening importance of non-gravitational physics at higher redshifts when examining the evolution of haloes of fixed mass.  Typically, the scatter is reduced by $\sim50$ per cent from $z=0$ to $z=1$ (Fig.~\ref{fig:scatter}). At $z=0$, the X-ray temperature, gas mass, Sunyaev--Zel'dovich signal and $Y_X$ all have similar amounts of scatter at fixed total mass (typically 10 per cent), whereas the bolometric X-ray luminosity has considerably larger scatter (Table~\ref{table:scatter}).  
\item We have also separately analysed the scatter and evolution in the hydrostatic mass--true mass relation (see Section~\ref{sec:HSE} and Figs.~\ref{fig:HSEslopeevo}-\ref{fig:HSEscatter}).  We find that, overall, the hydrostatic mass tracks the true mass reasonably well (such that $M_{500,hse} \propto M_{500}$).  The typical value of the {\it true} bias is $1-b\equiv\frac{M_{500,hse,mw}}{M_{500}}\approx0.8$ and is insensitive to redshift, true halo mass, and included ICM physics.  When we derive hydrostatic masses using synthetic X-ray spectroscopy to infer the density and temperature profiles, however, the bias becomes more dependent on the true halo mass range and the included ICM physics, with typical median values ranging between $1-b\equiv\frac{M_{500,hse,spec}}{M_{500}}\approx0.7$ and 1.0 with little evolution.  At $z=0$, the typical level of scatter in the hydrostatic mass is 15 per cent (i.e.\ slightly larger than for most of the other proxies we have considered, barring the bolometric X-ray luminosity). However, contrary to what was found for the other proxies, the scatter in the hydrostatic mass {\it increases} with increasing redshift, making it an increasingly less competitive proxy at higher redshifts.
\item The bolometric X-ray luminosity--X-ray temperature relation qualitatively behaves in approximately the same way as the bolometric X-ray luminosity--mass relation (compare Figs.~\ref{fig:slopeevo}-\ref{fig:scatter} to Figs.~\ref{fig:LTslopeevo}-\ref{fig:LTscatter}). The differences can be explained by the behaviour of the mass--temperature relation.
\item The vast majority of these results are robust to changing the cosmology from \planck~to \wmap7. The only noteworthy exceptions are the high-mass ends of all the studied mass-observable relations and the bolometric X-ray luminosity--temperature relation which evolve slightly faster in the \wmap7 cosmology. 
\end{enumerate}

Our results have highlighted that, while some of the trends are sensitive to the sub-grid modelling, there are many robust trends that are directly relevant for ongoing and upcoming surveys.  For example, for samples including both galaxy groups and clusters, {\it none} of the mass--observable relations are well-fit by a single power-law for any of the physical models (apart from the unphysical non-radiative model).  We advocate the use of broken power-laws as an alternative.  In terms of amplitude evolution, with the exception of the gas mass--halo mass relation restricted to high halo masses, {\it none} of the scaling relations we have examined evolves self-similarly in any of the models: we therefore warn against the `simplistic' interpretation of deviations of any scaling relation from self-similar evolution as an indication of the effects of feedback processes. The precise evolution generally depends on the included ICM physics, for which we advocate using the predictions of the \agn~models (which at least reproduce the $z=0$ cluster population well). In addition, we find that the scatter about the relations generally decreases with increasing redshift, independently of the included ICM physics.  The one important exception is the hydrostatic mass, whose performance as a mass proxy worsens (scatter increases) with increasing redshift.

Our detailed analysis of the various mass--observable relations also allows us to comment on what is the `best' total mass proxy.  We argue that an ideal mass proxy should: (i) be easy to measure (including with relatively shallow data); (ii) be strongly correlated with mass; (iii) have a small intrinsic scatter; (iv) be insensitive to the choice of cosmology; (v) be insensitive to the cluster dynamical state; (vi) be insensitive to uncertain baryon physics (i.e.\ the zero point is well-known); and (vii) evolve in a manner that is easy to characterise.  On the basis of the above criteria, we argue that the X-ray temperature performs the best.  Its main strength relative to the other mass proxies is that it is least sensitive to uncertain baryon physics (in terms of its zero point, evolution and intrinsic scatter) while having amongst the smallest intrinsic scatter in mass at fixed proxy (see Table~\ref{table:scatter}).  

If one's sole criterion for judging the best mass proxy is instead based on the intrinsic scatter (for example, because one plans to use a high quality sub-sample with mass proxy information in combination with `self-calibration' to remove any zero point uncertainties; e.g.\ \citealt{Levine2002,Hu2003,Majumdar2003,Majumdar2004,Mantz2010b,Wu2010}), one might instead conclude from inspection of Table~\ref{table:scatter} that the integrated Sunyaev--Zel'dovich signal, $Y_X$ and the gas mass have a slight advantage over X-ray temperature (see \citealt{Allen2011} and references therein).  We caution, however, that our predictions for the intrinsic scatter of these mass proxies (and their dependencies on halo mass and redshift) are sensitive to unresolved feedback physics, whereas the scatter in the mass--temperature relation is generally insensitive to baryonic physics (see Fig.~\ref{fig:scatterbis}).
 
In the present study, we have examined the scaling relations between various global hot gas properties and halo mass.  In a future study, we plan to examine the evolution and scatter in the structure of the hot gas (i.e.\ profiles).

\section*{Acknowledgements}

The authors would like to thank the members of the OWLS team for their contributions to the development of the simulation code used here and the anonymous referee for a constructive report. AMCLB is grateful to Monique Arnaud, Andrey Kravtsov, Daisuke Nagai and Gabriel Pratt for constructive discussions and to Remco van der Burg for useful comments on an earlier version of the manuscript. AMCLB acknowledges support from an internally funded PhD studentship at the Astrophysics Research Institute of Liverpool John Moores University, from the French Agence Nationale de la Recherche under grant ANR-11-BS56-015 and from the European Research Council under the European Union's Seventh Framework Programme (FP7/2007-2013) / ERC grant agreement number 340519. IGM is supported by an STFC Advanced Fellowship. JS is sponsored by the European Research Council under the European Union's Seventh Framework Programme (FP7/2007-2013)/ERC Grant agreement 278594-GasAroundGalaxies. 
This work used the DiRAC Data Centric system at Durham University, operated by the Institute for Computational Cosmology on behalf of the STFC DiRAC HPC Facility (www.dirac.ac.uk). This equipment was funded by BIS National E-infrastructure capital grant ST/K00042X/1, STFC capital grant ST/H008519/1, and STFC DiRAC Operations grant ST/K003267/1 and Durham University. DiRAC is part of the National E-Infrastructure. 

\bibliographystyle{mn2e}
\bibliography{scatter}

\appendix

\section{Results for the mass--non-core-excised temperatures scaling relations}
\label{app:Tnocor}

As it is sometimes impossible to compute core-excised temperatures for observed groups and clusters (especially for instance at high redshift or for short exposure times), we also present the results for the mass--temperature relation in the case of non-core excised temperatures (be it mass-weighted or X-ray) and conclude that whether one excises the core or not when computing the temperature does not noticeably affect the results presented in the main part of the paper, with a few noteworthy exceptions: (i) the high-mass end of the mass--X-ray temperature relation of the \agn~8.5 model is steeper in the absence of core excision and it even goes from being shallower than the self-similar expectation when the core is excised to being steeper (compare the left panel of Fig.~\ref{fig:slopeevobroken} to the bottom panel of Fig.~\ref{fig:Tnocorslopeevobroken}), (ii) the evolution of the normalisation being slightly more negative for the X-ray temperature when the core is excised (compare the top right subpanels of Fig.~\ref{fig:normevo} to the panels of Fig.~\ref{fig:Tnocornormevo}) and feedback is included (be it from SNe or AGN) and (iii) as might be expected, a global increase (at all masses and redhsift) of the scatter for all the physical models (compare the top panels of Fig.~\ref{fig:scatter} to the panels of Fig.~\ref{fig:Tnocorscatter}) in the absence of core excision. Finally, it is worth noting that the fact that, for the highest-mass bin, the finding that the mass--temperature scalings have a slightly faster evolution in the \wmap7 cosmology than in the \planck~one is also unaffected by core excision.

\begin{figure}
\begin{center}
\includegraphics[width=1.0\hsize]{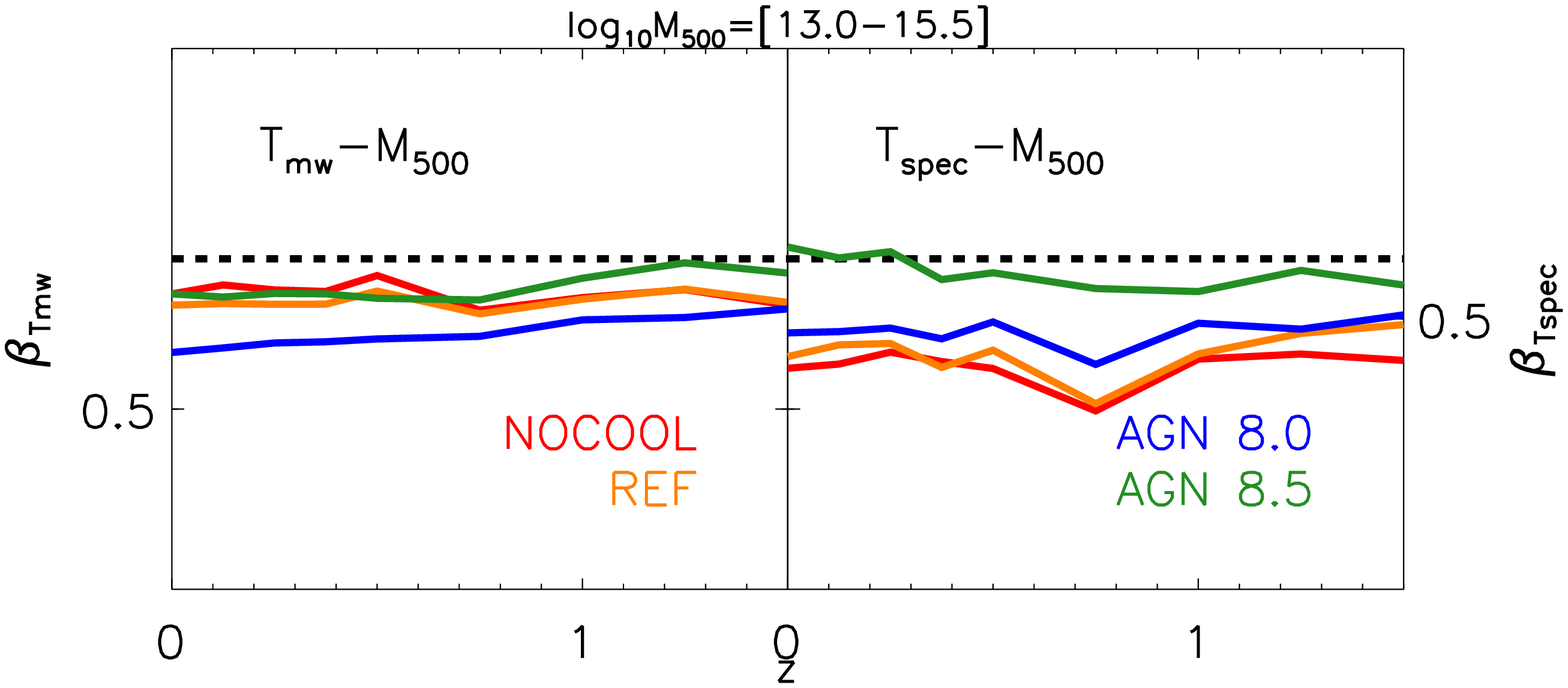}
\caption{Evolution of the mass slope from $z=0$ to $z=1.5$ for the total mass--temperature scaling relations (for both non-core-excised mass-weighted and X-ray spectroscopic temperatures). In each panel, we plot the redshift evolution of the best-fitting power-law indices obtained by fitting the broken power-law given by equation~\eqref{eq:power} at each redshift independently. The solid curves (red, orange, blue and green) correspond to the different simulations and the horizontal dashed lines to the self-similar expectation, respectively. Whether one excises the core or not when computing the temperature does not noticeably affect the results presented in the main part of the paper (compare with Fig.~\ref{fig:slopeevo}).}
\end{center}
\end{figure}

\begin{figure}
\begin{center}
\includegraphics[width=1.0\hsize]{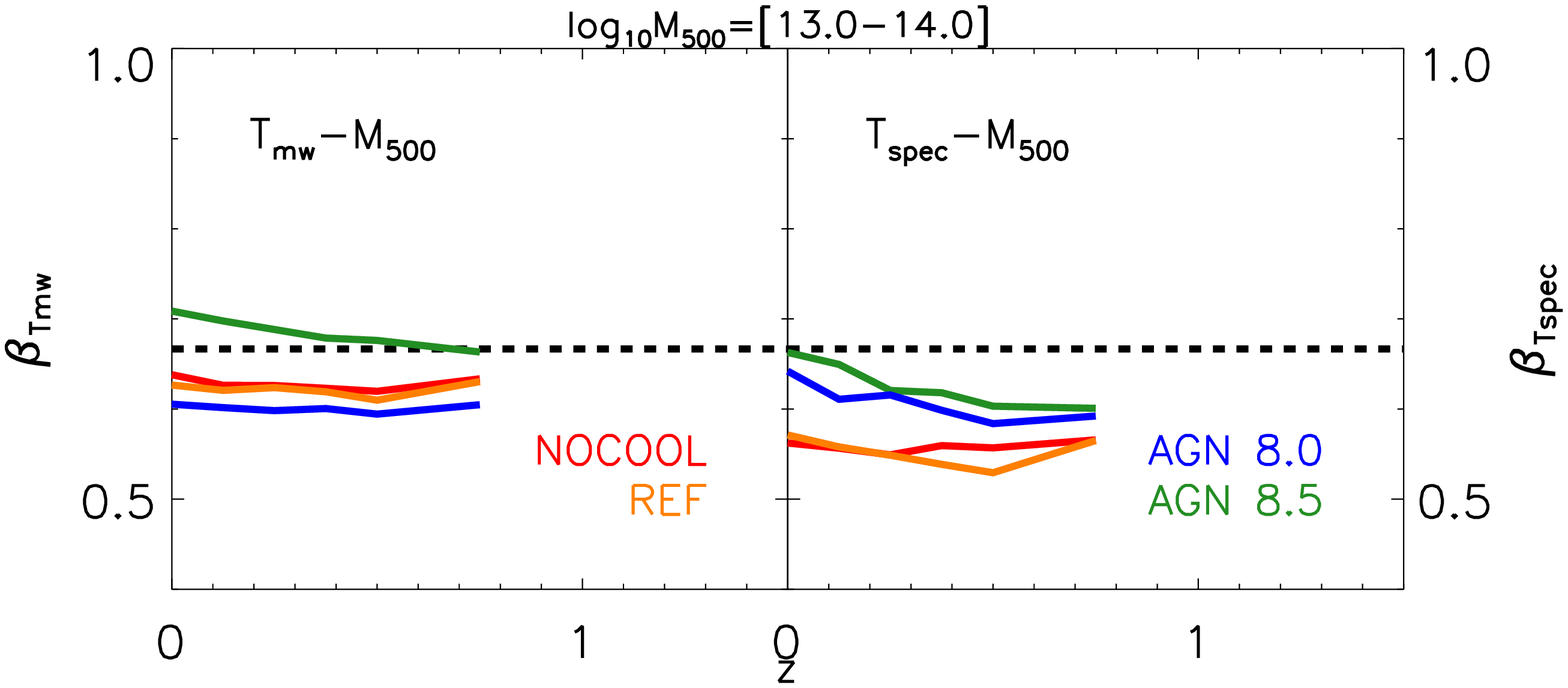}
\includegraphics[width=1.0\hsize]{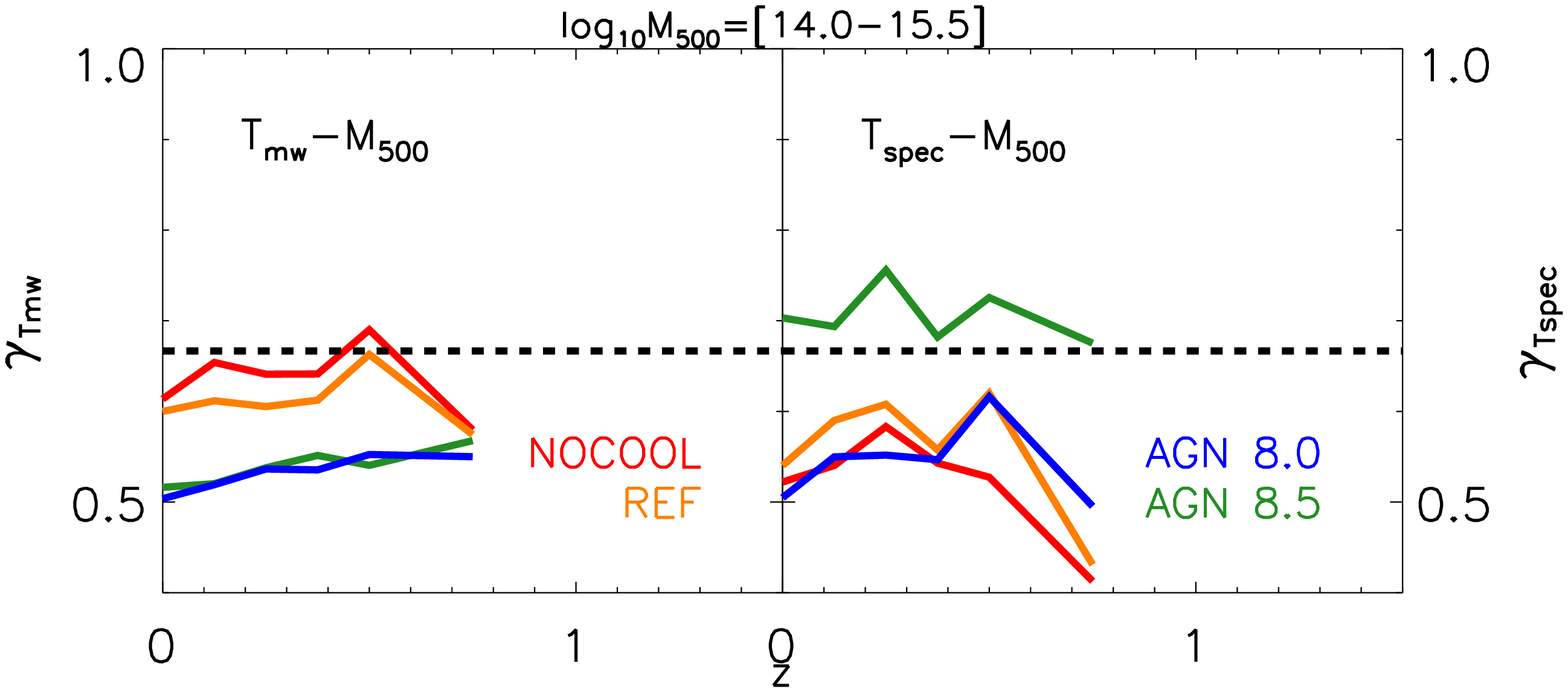}
\caption{Evolution of the mass slope from $z=0$ to $z=1.5$ for the total mass--temperature scaling relations (for both non-core-excised mass-weighted and X-ray spectroscopic temperatures). In each subpanel, we plot the redshift evolution of the low-mass (\emph{top} panel) and high-mass (\emph{bottom} panel) best-fitting power-law indices obtained by fitting the broken power-law given by equation~\eqref{eq:broken} at each individual redshift. The solid curves (red, orange, blue and green) correspond to the different simulations and the horizontal dashed lines to the self-similar expectation, respectively. The high-mass end of the mass--X-ray temperature relation of the \agn~8.5 model is steeper in the absence of core excision (compare with Fig.~\ref{fig:slopeevobroken}).}
\label{fig:Tnocorslopeevobroken}
\end{center}
\end{figure} 

\begin{figure}
\begin{center}
\includegraphics[width=1.0\hsize]{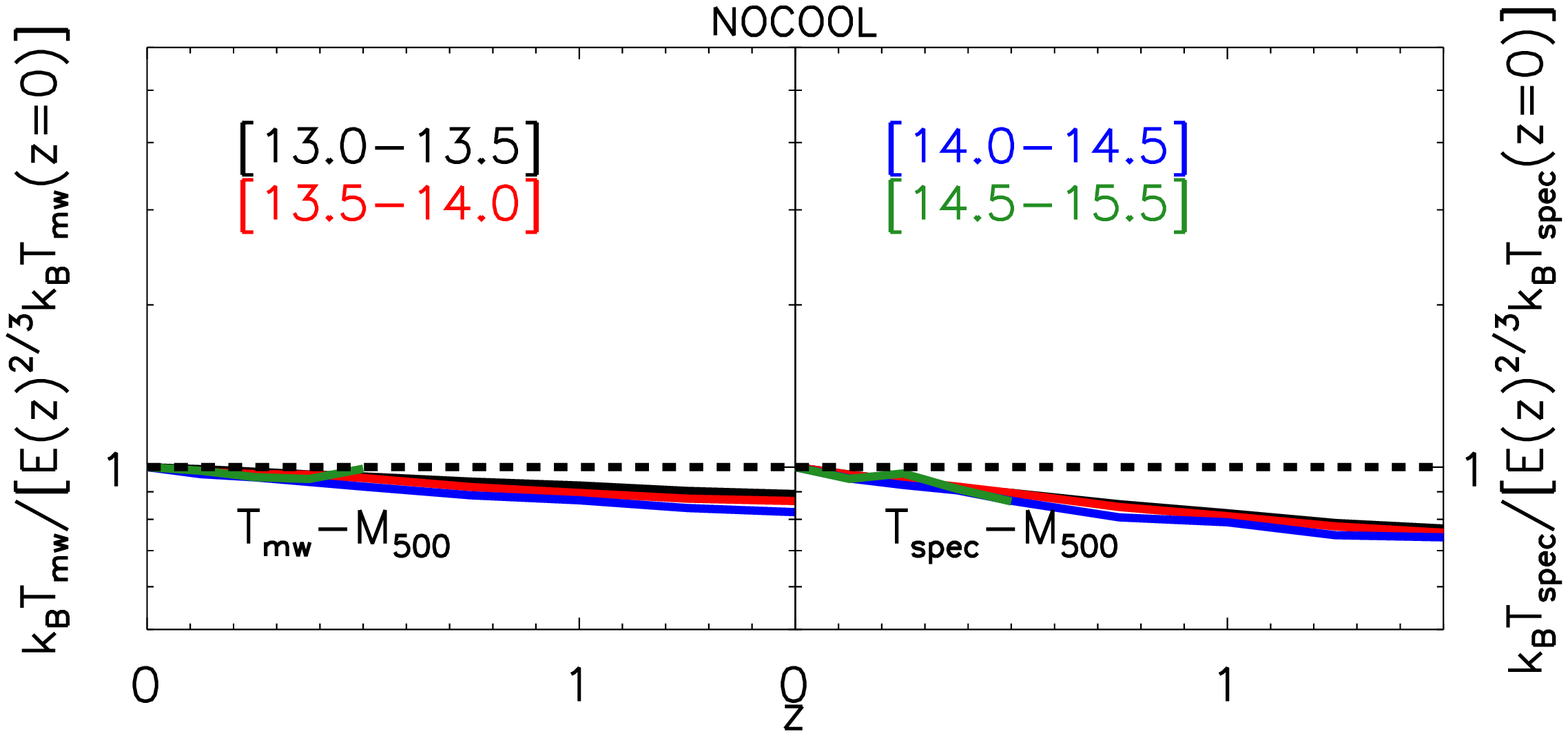}
\includegraphics[width=1.0\hsize]{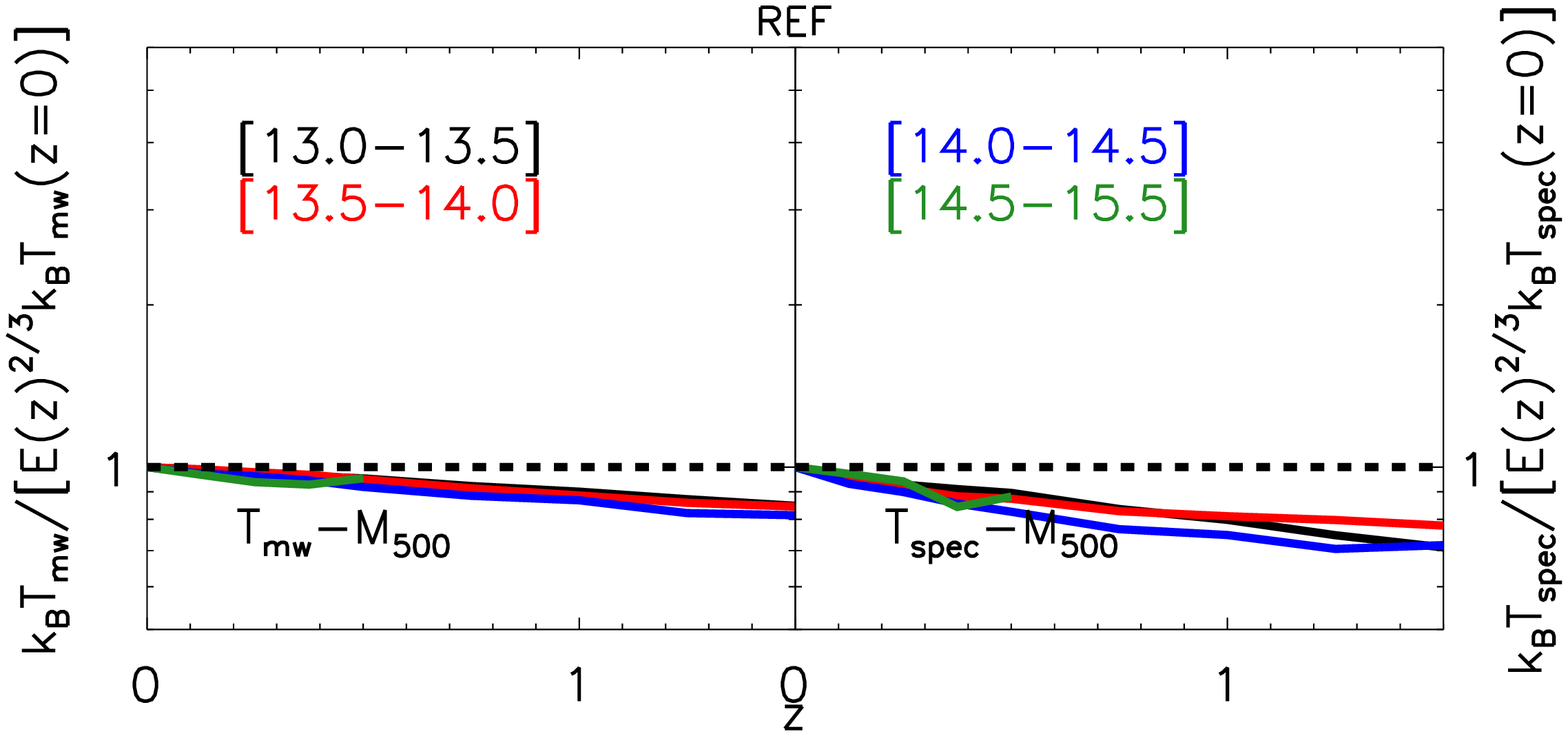}
\includegraphics[width=1.0\hsize]{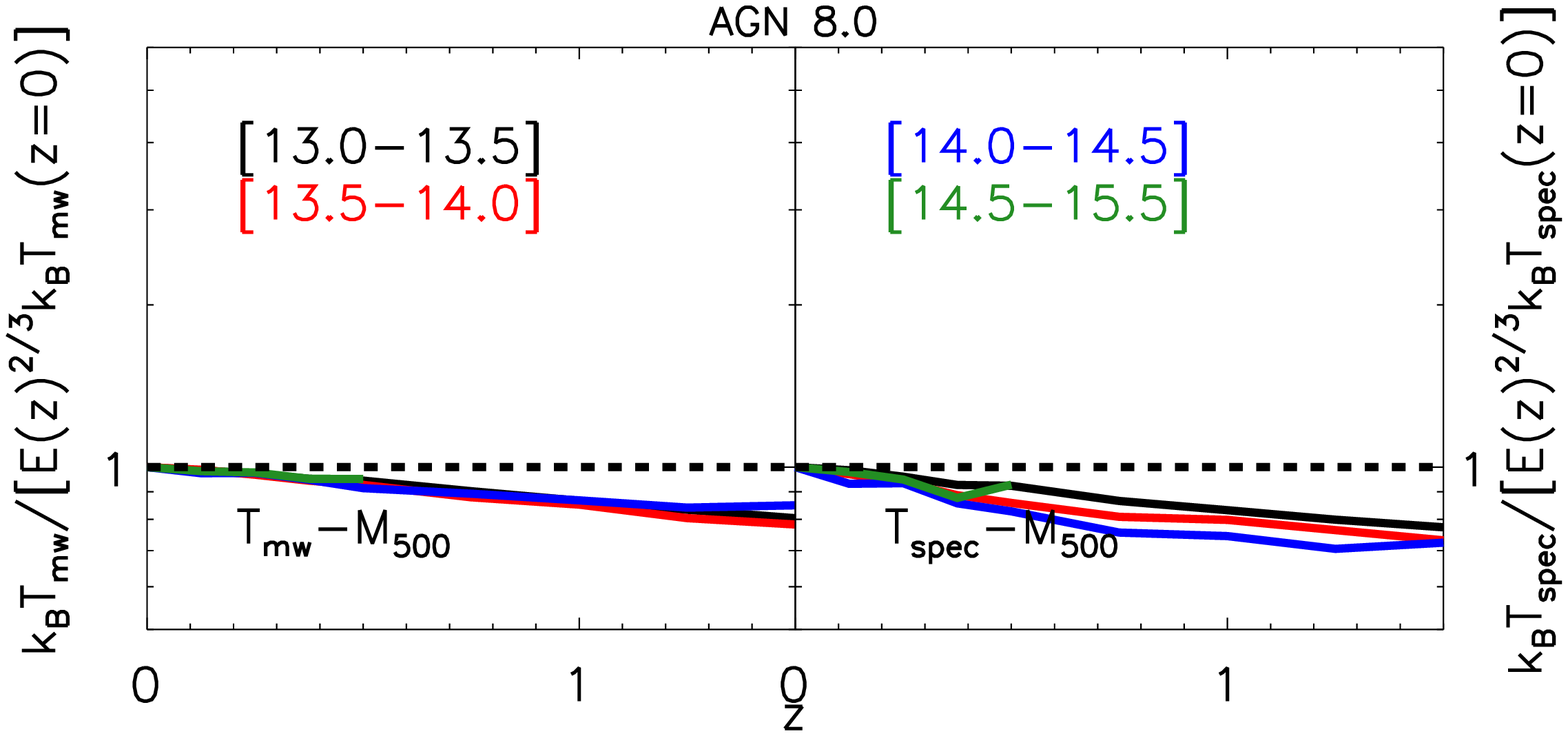}
\includegraphics[width=1.0\hsize]{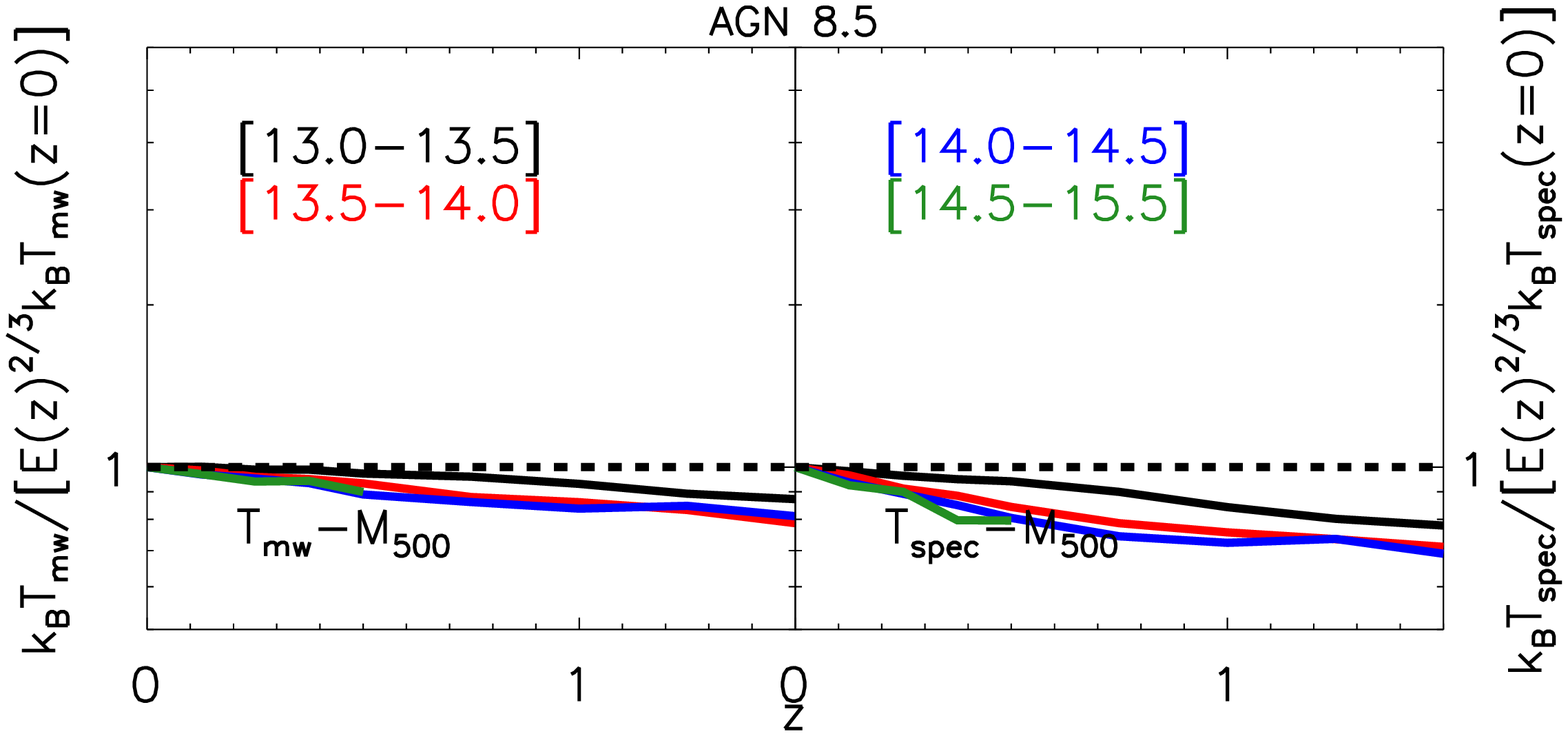}
\caption{Evolution of the normalisation from $z=0$ to $z=1.5$  for the total mass--temperature scaling relations (for both non-core-excised mass-weighted and X-ray spectroscopic temperatures) for each simulation. The normalisations of each scaling relation in the four $\log_{10}[M_{500}(\textrm{M}_\odot)]$ bins (denoted by solid lines of different colours) have been normalised by the self-similar expectation for the redshift evolution at fixed mass (shown as an horizontal dashed line). The evolution of the normalisation is slightly more negative for the X-ray temperature when the core is excised (compare with Fig.~\ref{fig:normevo}). The fact that, for the highest mass bin, the mass--temperature scalings have a slightly faster evolution in the \wmap7 cosmology than in the \planck~one is also unaffected by core excision.}
\label{fig:Tnocornormevo}
\end{center}
\end{figure}

\begin{figure}
\begin{center}
\includegraphics[width=1.0\hsize]{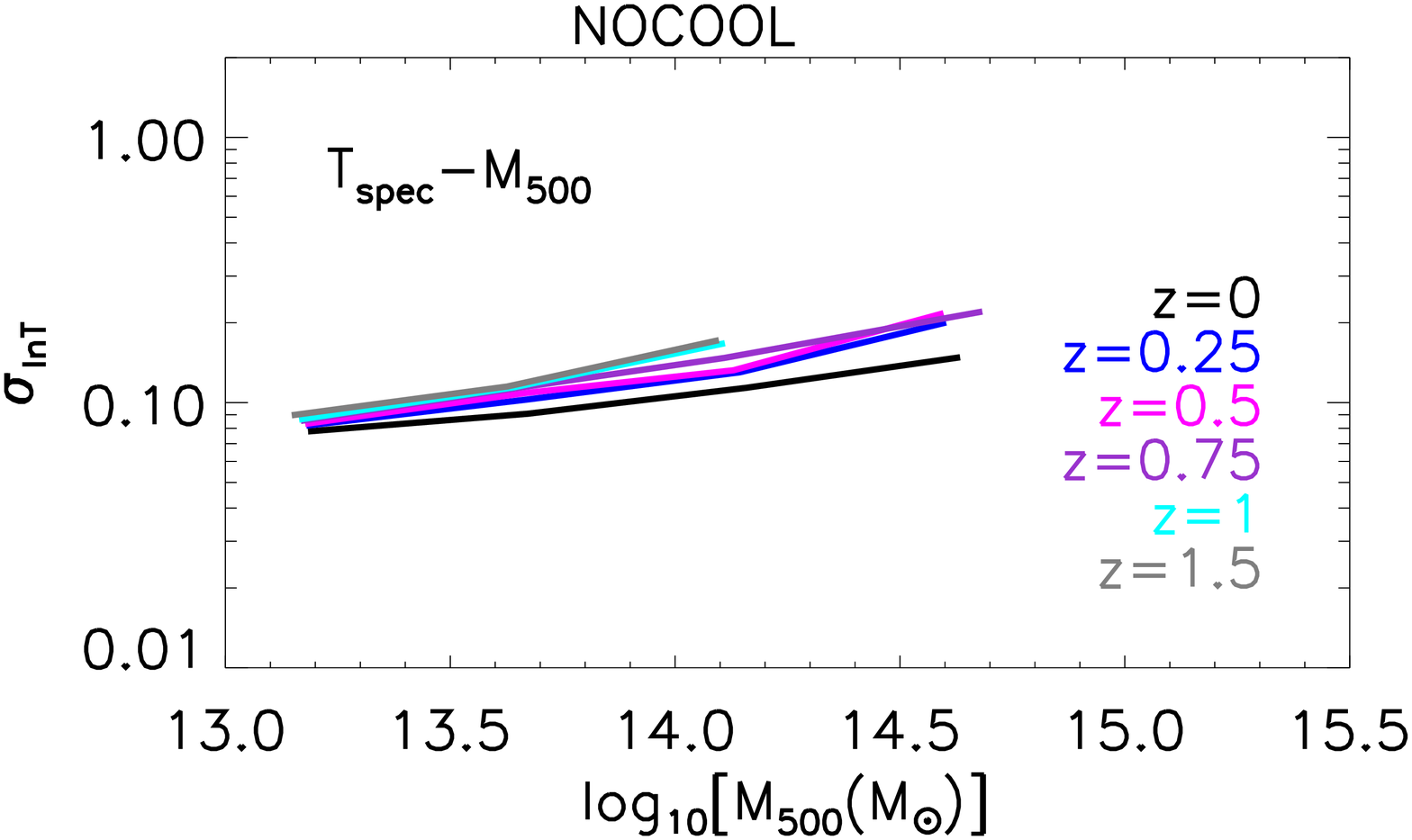}
\includegraphics[width=1.0\hsize]{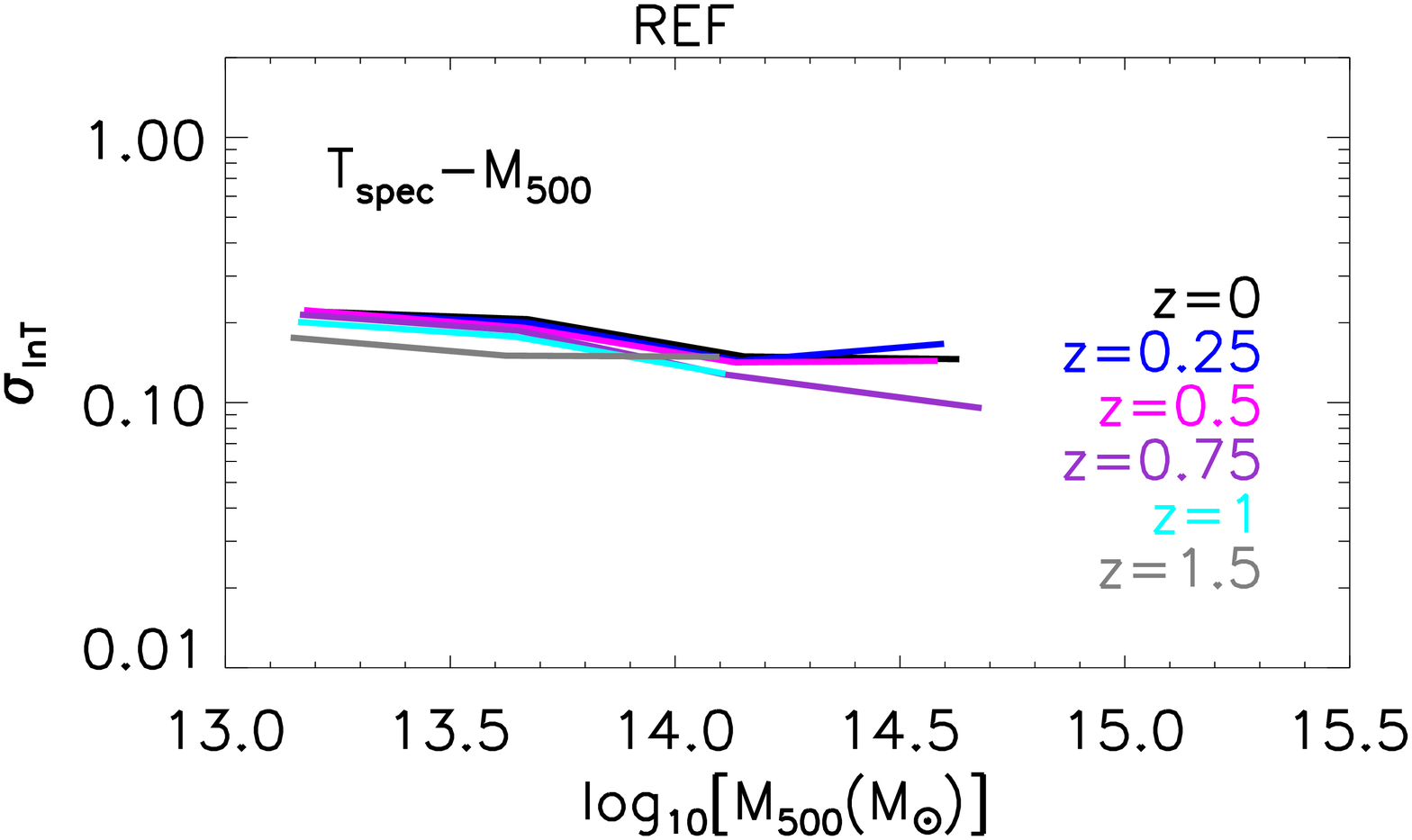}
\includegraphics[width=1.0\hsize]{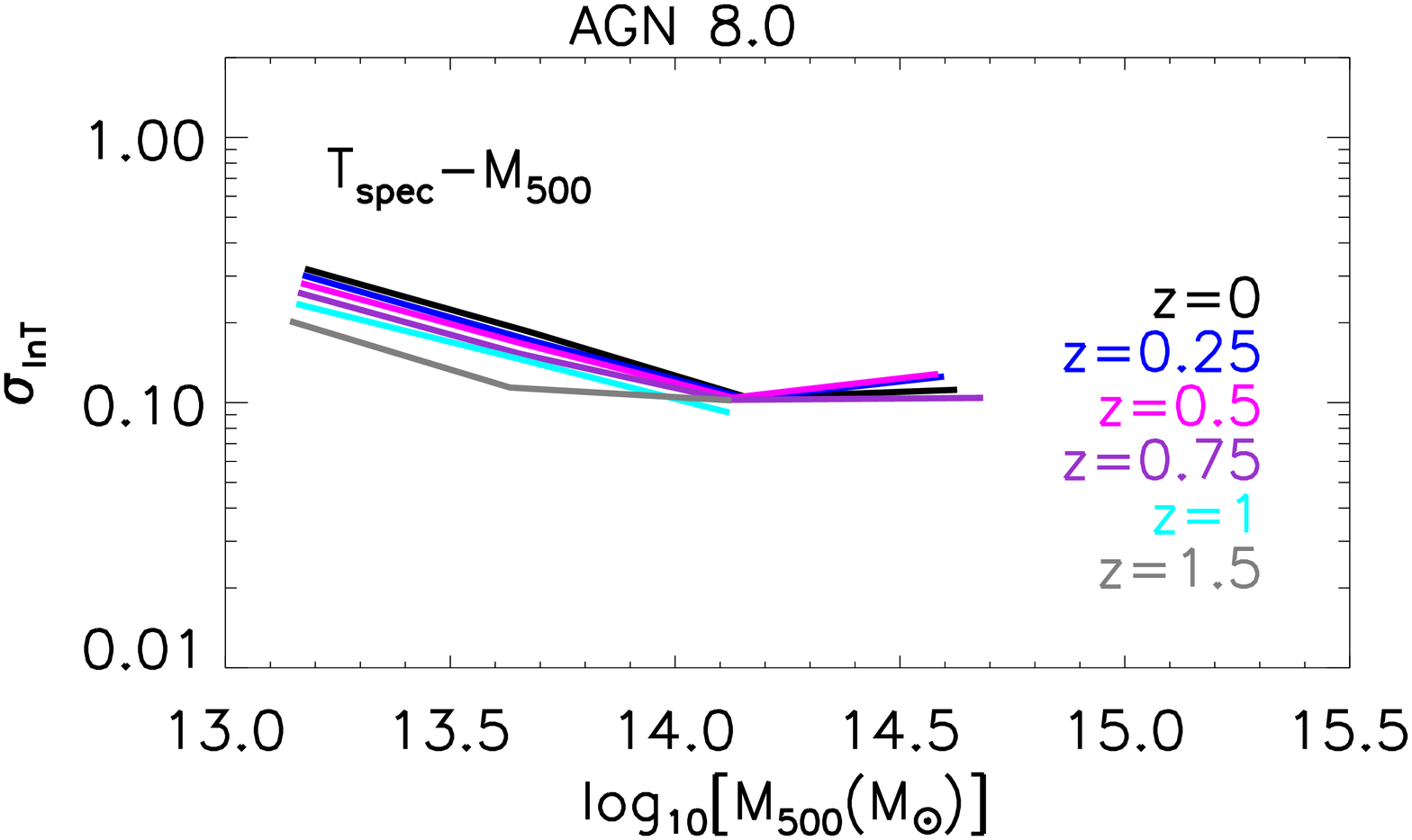}
\includegraphics[width=1.0\hsize]{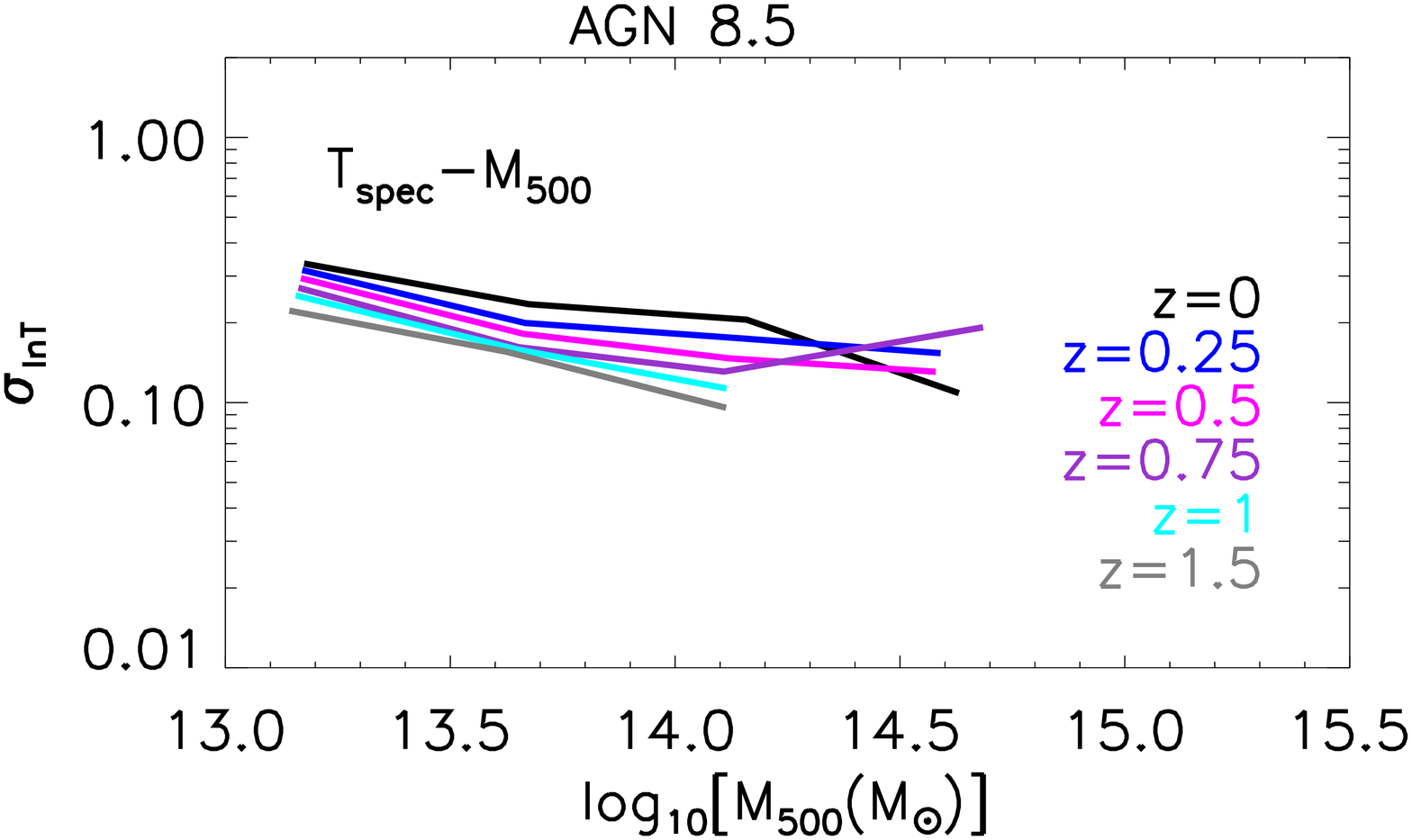}
\caption{Evolution of the log-normal scatter from $z=0$ to $z=1.5$ for the total mass--non-core-excised X-ray spectroscopic temperature scaling relation. For each simulation, we plot the log-normal scatter as a function of $M_{500}$ and denote the redshift using lines of different colours. There is a  global increase (at all masses and redhsift) of the scatter for all the physical models (compare to the top panels of Fig.~\ref{fig:scatter}) in the absence of core excision.} 
\label{fig:Tnocorscatter}
\end{center}
\end{figure}

\section{Results for the other physical models}
\label{app:fits}

In Tables~\ref{table:powernocool}-\ref{table:brokenpowerlawagn8.5}, we present the results of the fitting procedures described in Section~\ref{sec:fitting} for three of the other cosmo-OWLS physical models in \planck~cosmology.

\begin{table*}
\centering
\caption{Results of fitting the evolving power-law (equation~\ref{eq:power}) to both the median relation and the log-normal scatter about it for the \nocool~simulation. The scatter is the log-normal scatter in the natural logarithm of the $Y$ variable.}
\begin{tabular}{lp{1.217cm}ccccc}
\hline
Scaling relation & Median or scatter & $A$ & $\alpha$ & $\beta$ & $\chi^2$ & d.o.f. \\ 
\hline
$T_{spec,cor}-M_{500}$        	    & Median & \phantom{$--$}0.173$\pm$0.006     & \phantom{$-$}0.377$\pm$0.032  & \phantom{$-$}0.547$\pm$0.007 & 0.074 & 30 \\  
$L_{bol}-M_{500}$   & Median & \phantom{$-$-}44.129$\pm$0.009    & \phantom{$-$}2.267$\pm$0.051 & \phantom{$-$}1.411$\pm$0.012  & 0.190 & 30 \\ 
$M_{gas,500}-M_{500}$        & Median & \phantom{$-$-}13.136$\pm$0.002   & \phantom{$-$}0.066$\pm$0.011  & \phantom{$-$}1.020$\pm$0.002  & 0.008 & 30 \\ 
$Y_{X,500}-M_{500}$	     & Median & \phantom{$-$-}13.311$\pm$0.004   & \phantom{$-$}0.471$\pm$0.021  & \phantom{$-$}1.569$\pm$0.005  & 0.033 & 30 \\ 
$d_{A}^{2}Y_{500}-M_{500}$ & Median & \phantom{$-$}$-$5.573$\pm$0.003 & \phantom{$-$}0.647$\pm$0.017  & \phantom{$-$}1.645$\pm$0.004  & 0.021 & 30 \\ 
$M_{500,hse,spec}-M_{500}$        & Median & \phantom{$-$-}13.904$\pm$0.008   & $-$0.006$\pm$0.041 & \phantom{$-$}0.849$\pm$0.010  & 0.124 & 30 \\ 
\hline
$T_{spec,cor}-M_{500}$        	    & Scatter & \phantom{$-$}$-$0.983$\pm$0.008 & \phantom{$-$}0.303$\pm$0.045  & \phantom{$-$}0.280$\pm$0.010 & 0.147 & 30 \\
$L_{bol}-M_{500}$   & Scatter & \phantom{$-$}$-$0.381$\pm$0.011  & $-$0.325$\pm$0.060 		       & \phantom{$-$}0.025$\pm$0.014 		  & 0.259 & 30 \\
$M_{gas,500}-M_{500}$        & Scatter & \phantom{$-$}$-$1.381$\pm$0.026  & $-$0.360$\pm$0.139		       & $-$0.124$\pm$0.032		  & 1.400 & 30 \\ 
$Y_{X,500}-M_{500}$	     & Scatter & \phantom{$-$}$-$0.863$\pm$0.012  & \phantom{$-$}0.111$\pm$0.064 & \phantom{$-$}0.192$\pm$0.015  & 0.301  & 30 \\ 
$d_{A}^{2}Y_{500}-M_{500}$ & Scatter & \phantom{$-$}$-$0.930$\pm$0.016  & \phantom{$-$}0.010$\pm$0.089 & \phantom{$-$}0.035$\pm$0.021  & 0.581 & 30 \\
$M_{500,hse,spec}-M_{500}$        & Scatter & \phantom{$-$}$-$0.657$\pm$0.012  & \phantom{$-$}0.607$\pm$0.065		       & \phantom{$-$}0.165$\pm$0.015		  & 0.307 & 30 \\ 
\hline
\end{tabular}
\label{table:powernocool}
\end{table*}

\begin{table*}
\caption{Results of fitting the evolving broken power-law (equation~\ref{eq:brokenextra}) to both the median relation and the log-normal scatter about it for the \nocool~simulation. The scatter is the log-normal scatter in the natural logarithm of the $Y$ variable.}
\centering
\begin{tabular}{lp{1.217cm}ccccccc}
\hline
Scaling relation & Median or scatter & $A''$ & $\alpha''$ & $\beta''$ & $\gamma''$ & $\delta''$ & $\chi^2$ & d.o.f. \\ 
\hline
$T_{spec,cor}-M_{500}$         & Median & \phantom{$--$}0.200$\pm$0.010     & \phantom{$-$}0.290$\pm$0.042 & \phantom{$-$}0.656$\pm$0.037  & \phantom{$-$}0.499$\pm$0.019 & $-$0.056$\pm$0.023		       & 0.054 & 28 \\ 
$L_{bol}-M_{500}$    & Median & \phantom{$-$-}44.178$\pm$0.009    & \phantom{$-$}2.260$\pm$0.039 & \phantom{$-$}1.457$\pm$0.035  & \phantom{$-$}1.269$\pm$0.018 & \phantom{$-$}0.026$\pm$0.021		      & 0.047 & 28 \\  
$M_{gas,500}-M_{500}$         & Median & \phantom{$-$-}13.143$\pm$0.003   & \phantom{$-$}0.038$\pm$0.015 & \phantom{$-$}1.054$\pm$0.013  & \phantom{$-$}1.008$\pm$0.007 & $-$0.019$\pm$0.008 		      & 0.007 & 28 \\  
$Y_{X,500}-M_{500}$	      & Median & \phantom{$-$-}13.328$\pm$0.007   & \phantom{$-$}0.406$\pm$0.028 & \phantom{$-$}1.648$\pm$0.025  & \phantom{$-$}1.543$\pm$0.013 &$-$0.044$\pm$0.015		       & 0.024 & 28 \\ 
$d_{A}^{2}Y_{500}-M_{500}$  & Median & \phantom{$-$}$-$5.577$\pm$0.006 & \phantom{$-$}0.633$\pm$0.024 & \phantom{$-$}1.654$\pm$0.021 & \phantom{$-$}1.663$\pm$0.011 & $-$0.014$\pm$0.013 		       & 0.017 & 28 \\ 
$M_{500,hse,spec}-M_{500}$ & Median & \phantom{$-$-}13.951$\pm$0.025  & $-$0.156$\pm$0.046 		       & \phantom{$-$}1.035$\pm$0.041 & \phantom{$-$}0.765$\pm$0.021 & $-$0.097$\pm$0.025 		       & 0.064 & 28 \\ 
\hline
$T_{spec,cor}-M_{500}$         & Scatter & \phantom{$-$}$-$1.011$\pm$0.015  & \phantom{$-$}0.393$\pm$0.064 & \phantom{$-$}0.168$\pm$0.057  & \phantom{$-$}0.330$\pm$0.029 & \phantom{$-$}0.058$\pm$0.034 & 0.125 & 28 \\ 
$L_{bol}-M_{500}$    & Scatter & \phantom{$-$}$-$0.374$\pm$0.019  & $-$0.231$\pm$0.080		       & $-$0.062$\pm$0.071 		  & $-$0.030$\pm$0.036		    & \phantom{$-$}0.084$\pm$0.043 & 0.196 & 28 \\
$M_{gas,500}-M_{500}$         & Scatter & \phantom{$-$}$-$1.378$\pm$0.050  & $-$0.427$\pm$0.214		       & $-$0.055$\pm$0.188		  	  & $-$0.108$\pm$0.097		    & $-$0.055$\pm$0.115 		      & 1.383 & 28 \\
$Y_{X,500}-M_{500}$	      & Scatter & \phantom{$-$}$-$0.943$\pm$0.015  & \phantom{$-$}0.348$\pm$0.066 & $-$0.107$\pm$0.058 	    	  & \phantom{$-$}0.341$\pm$0.030 & \phantom{$-$}0.149$\pm$0.035 & 0.131 & 28 \\ 
$d_{A}^{2}Y_{500}-M_{500}$  & Scatter & \phantom{$-$}$-$1.018$\pm$0.026 & \phantom{$-$}0.207$\pm$0.109 & $-$0.231$\pm$0.096  		 & \phantom{$-$}0.225$\pm$0.049 & \phantom{$-$}0.109$\pm$0.058  & 0.358 & 28 \\
$M_{500,hse,spec}-M_{500}$ & Scatter & \phantom{$-$}$-$0.705$\pm$0.019 & \phantom{$-$}0.629$\pm$0.079 & \phantom{$-$}0.105$\pm$0.070	 & \phantom{$-$}0.297$\pm$0.036 & $-$0.012$\pm$0.042 		       & 0.189 & 28 \\ 
\hline
\end{tabular}
\end{table*}

\begin{table*}
\caption{Results of fitting the evolving power-law (equation~\ref{eq:power}) to both the median relation and the log-normal scatter about it for the \refsim~simulation. The scatter is the log-normal scatter in the natural logarithm of the $Y$ variable.}
\centering
\begin{tabular}{lp{1.217cm}ccccc}
\hline
Scaling relation & Median or scatter & $A$ & $\alpha$ & $\beta$ & $\chi^2$ & d.o.f. \\ 
\hline
$T_{spec,cor}-M_{500}$    	    & Median & \phantom{$--$}0.314$\pm$0.003     & \phantom{$-$}0.248$\pm$0.019  & \phantom{$-$}0.530$\pm$0.004 & 0.027 & 30 \\  
$L_{bol}-M_{500}$   & Median & \phantom{$-$-}43.546$\pm$0.007    & \phantom{$-$}2.349$\pm$0.038 & \phantom{$-$}1.365$\pm$0.009  & 0.106 & 30 \\ 
$M_{gas,500}-M_{500}$        & Median & \phantom{$-$-}12.917$\pm$0.004  & \phantom{$-$}0.352$\pm$0.021  & \phantom{$-$}1.091$\pm$0.005  & 0.032 & 30 \\ 
$Y_{X,500}-M_{500}$	     & Median & \phantom{$-$-}13.229$\pm$0.006   & \phantom{$-$}0.588$\pm$0.034  & \phantom{$-$}1.609$\pm$0.008  & 0.085 & 30 \\  
$d_{A}^{2}Y_{500}-M_{500}$ & Median & \phantom{$-$}$-$5.720$\pm$0.006 & \phantom{$-$}0.723$\pm$0.034  & \phantom{$-$}1.745$\pm$0.008  & 0.082 & 30 \\
$M_{500,hse,spec}-M_{500}$        & Median & \phantom{$-$-}13.912$\pm$0.006   & \phantom{$-$}0.041$\pm$0.031 & \phantom{$-$}0.907$\pm$0.007  & 0.069 & 30 \\ 
\hline
$T_{spec,cor}-M_{500}$     	    & Scatter & \phantom{$-$}$-$0.950$\pm$0.022  & $-$0.285$\pm$0.118  		      & $-$0.025$\pm$0.028  		 & 1.014 & 30 \\
$L_{bol}-M_{500}$   & Scatter & \phantom{$-$}$-$0.423$\pm$0.014  & $-$0.654$\pm$0.076		       & $-$0.191$\pm$0.018 		  & 0.415 & 30 \\ 
$M_{gas,500}-M_{500}$        & Scatter & \phantom{$-$}$-$1.069$\pm$0.009  & $-$0.805$\pm$0.051		       & $-$0.137$\pm$0.012			  & 0.185 & 30 \\ 
$Y_{X,500}-M_{500}$	     & Scatter & \phantom{$-$}$-$0.866$\pm$0.017  & $-$0.538$\pm$0.092 		       & $-$0.109$\pm$0.022  		  & 0.619 & 30 \\
$d_{A}^{2}Y_{500}-M_{500}$ & Scatter & \phantom{$-$}$-$0.937$\pm$0.018  & $-$0.290$\pm$0.100 		       & $-$0.150$\pm$0.023 		  & 0.728 & 30 \\ 
$M_{500,hse,spec}-M_{500}$        & Scatter & \phantom{$-$}$-$0.690$\pm$0.015  & \phantom{$-$}0.426$\pm$0.082		       & \phantom{$-$}0.080$\pm$0.019	  & 0.485 & 30 \\ 
\hline
\end{tabular}
\end{table*}

\begin{table*}
\caption{Results of fitting the evolving broken power-law (equation~\ref{eq:brokenextra}) to both the median relation and the log-normal scatter about it for the \refsim~simulation. The scatter is the log-normal scatter in the natural logarithm of the $Y$ variable.}
\centering
\begin{tabular}{lp{1.217cm}ccccccc}
\hline
Scaling relation & Median or scatter & $A''$ & $\alpha''$ & $\beta''$ & $\gamma''$ & $\delta''$ & $\chi^2$ & d.o.f. \\ 
\hline
$T_{spec,cor}-M_{500}$    	    & Median & \phantom{$--$}0.317$\pm$0.007      & \phantom{$-$}0.221$\pm$0.029 & \phantom{$-$}0.559$\pm$0.025  & \phantom{$-$}0.531$\pm$0.013 & $-$0.021$\pm$0.015		        & 0.025 & 28 \\  
$L_{bol}-M_{500}$   & Median & \phantom{$-$-}43.505$\pm$0.010    & \phantom{$-$}2.495$\pm$0.043 & \phantom{$-$}1.187$\pm$0.038  & \phantom{$-$}1.435$\pm$0.019 & \phantom{$-$}0.097$\pm$0.023  & 0.056 & 28 \\ 
$M_{gas,500}-M_{500}$        & Median & \phantom{$-$-}12.934$\pm$0.005    & \phantom{$-$}0.256$\pm$0.021 & \phantom{$-$}1.199$\pm$0.019  & \phantom{$-$}1.077$\pm$0.010 & $-$0.070$\pm$0.011 		       & 0.014 & 28 \\ 
$Y_{X,500}-M_{500}$	     & Median & \phantom{$-$-}13.254$\pm$0.009    & \phantom{$-$}0.454$\pm$0.040 & \phantom{$-$}1.762$\pm$0.035 & \phantom{$-$}1.585$\pm$0.018  & $-$0.097$\pm$0.021		       & 0.049 & 28 \\  
$d_{A}^{2}Y_{500}-M_{500}$ & Median & \phantom{$-$}$-$5.684$\pm$0.009 & \phantom{$-$}0.643$\pm$0.039  & \phantom{$-$}1.853$\pm$0.034 & \phantom{$-$}1.669$\pm$0.018  & $-$0.045$\pm$0.021		       & 0.047 & 28 \\ 
$M_{500,hse,spec}-M_{500}$        & Median & \phantom{$-$-}13.935$\pm$0.010    & $-$0.052$\pm$0.041 & \phantom{$-$}1.018$\pm$0.036  & \phantom{$-$}0.871$\pm$0.019 & $-$0.063$\pm$0.022 		       & 0.050 & 28 \\ 
\hline
$T_{spec,cor}-M_{500}$    	    & Scatter &  \phantom{$-$}$-$1.111$\pm$0.024 & \phantom{$-$}0.233$\pm$0.101 		      & $-$0.665$\pm$0.088  		  & \phantom{$-$}0.264$\pm$0.046 		    & \phantom{$-$}0.332$\pm$0.054  & 0.305 & 28 \\
$L_{bol}-M_{500}$   & Scatter & \phantom{$-$}$-$0.467$\pm$0.025  & $-$0.482$\pm$0.108 		       & $-$0.397$\pm$0.095			  & $-$0.120$\pm$0.049 		    & \phantom{$-$}0.116$\pm$0.058  & 0.352 & 28 \\
$M_{gas,500}-M_{500}$        & Scatter & \phantom{$-$}$-$1.048$\pm$0.014  & $-$0.987$\pm$0.058		       & \phantom{$-$}0.060	$\pm$0.051 & $-$0.133$\pm$0.026 		    & $-$0.140$\pm$0.031  		       & 0.103 & 28 \\ 
$Y_{X,500}-M_{500}$	     & Scatter &  \phantom{$-$}$-$0.936$\pm$0.030 & $-$0.331$\pm$0.127		       & $-$0.369$\pm$0.112		 & \phantom{$-$}0.024$\pm$0.058	             & \phantom{$-$}0.129$\pm$0.068 & 0.489 & 28 \\ 
$d_{A}^{2}Y_{500}-M_{500}$ & Scatter & \phantom{$-$}$-$1.011$\pm$0.030 & $-$0.232$\pm$0.126  & $-$0.266$\pm$0.111 		 & \phantom{$-$}0.046$\pm$0.057  & \phantom{$-$}0.001$\pm$0.067 & 0.480 & 28 \\
$M_{500,hse,spec}-M_{500}$        & Scatter & \phantom{$-$}$-$0.763$\pm$0.024  & \phantom{$-$}0.560$\pm$0.104		       & $-$0.109$\pm$0.091	 & \phantom{$-$}0.245$\pm$0.047 		    & \phantom{$-$}0.066$\pm$0.055  		       & 0.324 & 28 \\ 
\hline
\end{tabular}
\end{table*}

\begin{table*}
\caption{Results of fitting the evolving power-law (equation~\ref{eq:power}) to both the median relation and the log-normal scatter about it for the \agn~8.5 simulation. The scatter is the log-normal scatter in the natural logarithm of the $Y$ variable.}
\centering
\begin{tabular}{lp{1.217cm}ccccc}
\hline
Scaling relation & Median or scatter & $A$ & $\alpha$ & $\beta$ & $\chi^2$ & d.o.f. \\ 
\hline
$T_{spec,cor}-M_{500}$        	    & Median & \phantom{$--$}0.265$\pm$0.008    & \phantom{$-$}0.390$\pm$0.042  & \phantom{$-$}0.654$\pm$0.010  & 0.127 & 30 \\ 
$L_{bol}-M_{500}$   & Median & \phantom{$-$-}43.130$\pm$0.019   & \phantom{$-$}3.191$\pm$0.103  & \phantom{$-$}1.767$\pm$0.024   & 0.770 & 30 \\ 
$M_{gas,500}-M_{500}$        & Median & \phantom{$-$-}12.684$\pm$0.012   & \phantom{$-$}0.753$\pm$0.064  & \phantom{$-$}1.364$\pm$0.015  & 0.295 & 30 \\ 
$Y_{X,500}-M_{500}$	     & Median & \phantom{$-$-}12.951$\pm$0.018   & \phantom{$-$}1.132$\pm$0.096  & \phantom{$-$}2.010$\pm$0.022  & 0.667 & 30 \\ 
$d_{A}^{2}Y_{500}-M_{500}$ & Median & \phantom{$-$}$-$5.905$\pm$0.011 & \phantom{$-$}1.166$\pm$0.061  & \phantom{$-$}2.055$\pm$0.014  & 0.269 & 30 \\ 
$M_{500,hse,spec}-M_{500}$        & Median & \phantom{$-$-}13.874$\pm$0.013   & $-$0.043$\pm$0.068  & \phantom{$-$}0.964$\pm$0.016  & 0.340 & 30 \\ 
\hline
$T_{spec,cor}-M_{500}$        	    & Scatter & \phantom{$-$}$-$0.951$\pm$0.032 & $-$0.323$\pm$0.175		       & $-$0.310$\pm$0.041 	  	  & 2.232 & 30 \\ 
$L_{bol}-M_{500}$   & Scatter & \phantom{$-$}$-$0.470$\pm$0.025 & $-$0.321$\pm$0.136 		       & $-$0.268$\pm$0.032 & 1.351 & 30 \\ 
$M_{gas,500}-M_{500}$        & Scatter & \phantom{$-$}$-$0.875$\pm$0.027  & $-$0.489$\pm$0.146		       & $-$0.457$\pm$0.034			  & 1.558 & 30 \\ 
$Y_{X,500}-M_{500}$	     & Scatter & \phantom{$-$}$-$0.661$\pm$0.020  & $-$0.309$\pm$0.109 		       & $-$0.289$\pm$0.025  		  & 0.862 & 30 \\ 
$d_{A}^{2}Y_{500}-M_{500}$ & Scatter & \phantom{$-$}$-$0.764$\pm$0.020  & $-$0.145$\pm$0.107 &$-$0.328$\pm$0.025  		  & 0.830 & 30 \\
$M_{500,hse,spec}-M_{500}$        & Scatter & \phantom{$-$}$-$0.672$\pm$0.056  & \phantom{$-$}0.469$\pm$0.302		       & $-$0.201$\pm$0.070		  & 6.646 & 30 \\ 
\hline
\end{tabular}
\end{table*}

\begin{table*}
\caption{Results of fitting the evolving broken power-law (equation~\ref{eq:brokenextra}) to both the median relation and the log-normal scatter about it for the \agn~8.5 simulation. The scatter is the log-normal scatter in the natural logarithm of the $Y$ variable.}
\centering
\begin{tabular}{lp{1.217cm}ccccccc}
\hline
Scaling relation & Median or scatter & $A''$ & $\alpha''$ & $\beta''$ & $\gamma''$ & $\delta''$ & $\chi^2$ & d.o.f. \\ 
\hline
$T_{spec,cor}-M_{500}$         & Median & \phantom{$--$}0.298$\pm$0.013       & \phantom{$-$}0.254$\pm$0.053  & \phantom{$-$}0.816$\pm$0.047 & \phantom{$-$}0.604$\pm$0.025 & $-$0.094$\pm$0.029	           	& 0.088 & 28 \\ 
$L_{bol}-M_{500}$    & Median & \phantom{$-$-}43.062$\pm$0.022    & \phantom{$-$}3.084$\pm$0.092  & \phantom{$-$}1.816$\pm$0.080 & \phantom{$-$}2.005$\pm$0.042 & $-$0.130$\pm$0.049  & 0.258 & 28 \\  
$M_{gas,500}-M_{500}$         & Median & \phantom{$-$-}12.713$\pm$0.015     & \phantom{$-$}0.489$\pm$0.062 & \phantom{$-$}1.649$\pm$0.054 & \phantom{$-$}1.374$\pm$0.028  & $-$0.204$\pm$0.033 		         & 0.117 & 28 \\ 
$Y_{X,500}-M_{500}$	      & Median & \phantom{$-$-}13.004$\pm$0.025     & \phantom{$-$}0.752$\pm$0.106 & \phantom{$-$}2.429$\pm$0.092 & \phantom{$-$}1.991$\pm$0.048  &$-$0.286$\pm$0.056		         & 0.342 & 28 \\  
$d_{A}^{2}Y_{500}-M_{500}$  & Median & \phantom{$-$}$-$5.827$\pm$0.012  & \phantom{$-$}0.887$\pm$0.053 & \phantom{$-$}2.395$\pm$0.047  & \phantom{$-$}1.927$\pm$0.024  & $-$0.187$\pm$0.028 		 & 0.087 & 28 \\
$M_{500,hse,spec}-M_{500}$ & Median & \phantom{$-$-}13.833$\pm$0.023   & \phantom{$-$}0.013$\pm$0.096 & \phantom{$-$}0.875$\pm$0.084   & \phantom{$-$}1.064$\pm$0.044  & \phantom{$-$}0.022$\pm$0.051 & 0.281 & 28 \\ 
\hline
$T_{spec,cor}-M_{500}$         & Scatter & \phantom{$-$}$-$1.024$\pm$0.061   & $-$0.016$\pm$0.258  & $-$0.673$\pm$0.226  		    & $-$0.205$\pm$0.118 & \phantom{$-$}0.213$\pm$0.138 & 2.036 & 28 \\ 
$L_{bol}-M_{500}$    & Scatter & \phantom{$-$}$-$0.589$\pm$0.039    & $-$0.156$\pm$0.165		         & $-$0.532$\pm$0.145 		   & \phantom{$-$}0.027$\pm$0.076  & \phantom{$-$}0.066$\pm$0.088  & 0.835 & 28 \\
$M_{gas,500}-M_{500}$         & Scatter & \phantom{$-$}$-$0.848$\pm$0.053   & $-$0.600$\pm$0.223		         & $-$0.325$\pm$0.196	   	   & $-$0.496$\pm$0.103 		      & $-$0.076$\pm$0.119		         & 1.533 & 28 \\ 
$Y_{X,500}-M_{500}$	      & Scatter & \phantom{$-$}$-$0.773$\pm$0.031   & \phantom{$-$}0.031$\pm$0.130 & $-$0.717$\pm$0.114		   & $-$0.081$\pm$0.060	               & \phantom{$-$}0.216$\pm$0.069 & 0.518 & 28 \\
$d_{A}^{2}Y_{500}-M_{500}$  & Scatter & \phantom{$-$}$-$0.868$\pm$0.030  & \phantom{$-$}0.237$\pm$0.127 & $-$0.791$\pm$0.112  		   & $-$0.160$\pm$0.058 		       & \phantom{$-$}0.256$\pm$0.068 & 0.497 & 28 \\
$M_{500,hse,spec}-M_{500}$ & Scatter & \phantom{$-$}$-$0.733$\pm$0.109  & \phantom{$-$}0.576$\pm$0.461 & $-$0.356$\pm$0.404	 	    & $-$0.062$\pm$0.212 		       & \phantom{$-$}0.053$\pm$0.246 & 6.532 & 28 \\ 
\hline
\end{tabular}
\label{table:brokenpowerlawagn8.5}
\end{table*}


\section{Scatter and evolution of the bolometric X-ray luminosity--temperature relation}
\label{app:LT}

\begin{table*}
\caption{Results for all the physical models of fitting the evolving power-law (equation~\ref{eq:LTpower}) to both the median bolometric X-ray luminosity--temperature relation and the log-normal scatter about it (for both core-excised and non-core-excised X-ray spectroscopic temperatures). The scatter is the log-normal scatter in the natural logarithm of the $Y$ variable.}
\centering
\begin{tabular}{llp{1.217cm}ccccc}
\hline
Scaling relation & Simulation &Median or scatter & $A$ & $\alpha$ & $\beta$ & $\chi^2$ & d.o.f. \\ 
\hline 
$L_{bol}-T_{spec}$ & \nocool & Median & \phantom{$-$-}44.413$\pm$0.010    & \phantom{$-$}1.206$\pm$0.049 & \phantom{$-$}2.590$\pm$0.024  & 0.242 & 33 \\  
$L_{bol}-T_{spec}$ & \refsim & Median & \phantom{$-$-}43.519$\pm$0.008    & \phantom{$-$}1.474$\pm$0.041 & \phantom{$-$}2.442$\pm$0.023  & 0.166 & 33 \\ 
$L_{bol}-T_{spec}$ & \agn~8.0 & Median & \phantom{$-$-}43.478$\pm$0.019    & \phantom{$-$}1.909$\pm$0.092 & \phantom{$-$}3.211$\pm$0.049 & 0.832 & 33 \\ 
$L_{bol}-T_{spec}$ & \agn~8.5 & Median & \phantom{$-$-}43.255$\pm$0.018    & \phantom{$-$}2.239$\pm$0.087 & \phantom{$-$}2.834$\pm$0.043  & 0.755 & 33 \\ 
\hline
$L_{bol}-T_{spec}$ & \nocool & Scatter & \phantom{$-$}$-$0.569$\pm$0.011  & $-$0.647$\pm$0.055 		       & $-$0.053$\pm$0.027		 &  0.298 & 33 \\ 
$L_{bol}-T_{spec}$ & \refsim & Scatter & \phantom{$-$}$-$0.377$\pm$0.010  & $-$0.561$\pm$0.051 		       & $-$0.494$\pm$0.028 		 & 0.256 & 33 \\
$L_{bol}-T_{spec}$ & \agn~8.0 & Scatter & \phantom{$-$}$-$0.374$\pm$0.018  & $-$0.233$\pm$0.091 		       & $-$0.644$\pm$0.048 		 & 0.814 & 33 \\ 
$L_{bol}-T_{spec}$ & \agn~8.5 & Scatter & \phantom{$-$}$-$0.300$\pm$0.015  & $-$0.429$\pm$0.075		       & $-$0.584$\pm$0.037 		 & 0.552 & 33 \\ 
\hline
$L_{bol}-T_{spec,cor}$ & \nocool & Median & \phantom{$-$-}44.470$\pm$0.011    & \phantom{$-$}1.227$\pm$0.056 & \phantom{$-$}2.557$\pm$0.026  & 0.253 & 31 \\ 
$L_{bol}-T_{spec,cor}$ & \refsim & Median & \phantom{$-$-}43.491$\pm$0.010    & \phantom{$-$}1.755$\pm$0.048 & \phantom{$-$}2.563$\pm$0.027  & 0.225 & 33 \\
$L_{bol}-T_{spec,cor}$ & \agn~8.0 & Median & \phantom{$-$-}43.416$\pm$0.017    & \phantom{$-$}2.021$\pm$0.084 & \phantom{$-$}3.246$\pm$0.044  & 0.696 & 33 \\ 
$L_{bol}-T_{spec,cor}$ & \agn~8.5 & Median & \phantom{$-$-}43.079$\pm$0.018    & \phantom{$-$}2.417$\pm$0.089 & \phantom{$-$}2.730$\pm$0.044  & 0.781 & 33 \\ 
\hline
$L_{bol}-T_{spec,cor}$ & \nocool & Scatter & \phantom{$-$}$-$0.561$\pm$0.018  & $-$0.592$\pm$0.096 		       & $-$0.156$\pm$0.044 		 & 0.732 & 31 \\ 
$L_{bol}-T_{spec,cor}$ & \refsim & Scatter & \phantom{$-$}$-$0.398$\pm$0.011  & $-$0.637$\pm$0.056		       & $-$0.202$\pm$0.032 		 & 0.315 & 33 \\ 
$L_{bol}-T_{spec,cor}$ & \agn~8.0 & Scatter & \phantom{$-$}$-$0.428$\pm$0.015  & $-$0.373$\pm$0.074 		       & $-$0.515$\pm$0.039 		 & 0.542 & 33 \\ 
$L_{bol}-T_{spec,cor}$ & \agn~8.5 & Scatter & \phantom{$-$}$-$0.446$\pm$0.038  & $-$0.092$\pm$0.186 		       & $-$0.544$\pm$0.091 		 & 3.398 & 33 \\
\hline
\end{tabular}
\label{table:LTpowerlaw}
\end{table*}

\begin{table*}
\caption{Results for all the physical models of fitting the evolving broken power-law (equation~\ref{eq:LTbrokenextra}) to both the median bolometric X-ray luminosity--temperature relation and the log-normal scatter about it (for both core-excised and non-core-excised X-ray spectroscopic temperatures). The scatter is the log-normal scatter in the natural logarithm of the $Y$ variable.}
\centering
\begin{tabular}{llp{1.217cm}ccccccc}
\hline
Scaling relation & Simulation & Median or scatter & $A''$ & $\alpha''$ & $\beta''$ & $\gamma''$ & $\delta''$ & $\chi^2$ & d.o.f. \\ 
\hline
$L_{bol}-T_{spec}$    & \nocool & Median & \phantom{$-$-}44.426$\pm$0.013     & \phantom{$-$}1.305$\pm$0.048 & \phantom{$-$}2.353$\pm$0.096  & \phantom{$-$}2.386$\pm$0.066 & \phantom{$-$}0.223$\pm$0.061 		       & 0.138 & 31 \\ 
$L_{bol}-T_{spec}$    & \refsim & Median & \phantom{$-$-}43.495$\pm$0.013     & \phantom{$-$}1.488$\pm$0.049 & \phantom{$-$}2.318$\pm$0.115  & \phantom{$-$}2.586$\pm$0.066 & \phantom{$-$}0.027$\pm$0.073 		       & 0.141 & 31 \\ 
$L_{bol}-T_{spec}$    & \agn~8.0 & Median & \phantom{$-$-}43.490$\pm$0.019     & \phantom{$-$}1.606$\pm$0.070 & \phantom{$-$}4.122$\pm$0.153  & \phantom{$-$}3.450$\pm$0.095 & $-$0.700$\pm$0.097 		       & 0.289 & 31 \\  
$L_{bol}-T_{spec}$    & \agn~8.5 & Median & \phantom{$-$-}43.276$\pm$0.023     & \phantom{$-$}2.011$\pm$0.087 & \phantom{$-$}3.510$\pm$0.175  & \phantom{$-$}2.933$\pm$0.111 & $-$0.483$\pm$0.108 		       & 0.451 & 31 \\
\hline
$L_{bol}-T_{spec}$    & \nocool & Scatter  & \phantom{$-$}$-$0.591$\pm$0.018  & $-$0.663$\pm$0.067		        & $-$0.064$\pm$0.132 		  & \phantom{$-$}0.121$\pm$0.091 & $-$0.047$\pm$0.084 		       & 0.262 & 31 \\
$L_{bol}-T_{spec}$    & \refsim & Scatter  & \phantom{$-$}$-$0.391$\pm$0.016  & $-$0.618$\pm$0.060		        & $-$0.366$\pm$0.141 		  & $-$0.340$\pm$0.081 & $-$0.147$\pm$0.089 		       & 0.211 & 31 \\ 
$L_{bol}-T_{spec}$    & \agn~8.0 & Scatter  & \phantom{$-$}$-$0.398$\pm$0.030  & $-$0.253$\pm$0.115		        & $-$0.663$\pm$0.249 		  & $-$0.456$\pm$0.154 & $-$0.054$\pm$0.158 		       & 0.771 & 31 \\ 
$L_{bol}-T_{spec}$    & \agn~8.5 & Scatter  & \phantom{$-$}$-$0.274$\pm$0.025  & $-$0.458$\pm$0.093		        & $-$0.433$\pm$0.188 		  & $-$0.725$\pm$0.119 & $-$0.054$\pm$0.117 		       & 0.522 & 31 \\
\hline
$L_{bol}-T_{spec,cor}$ & \nocool & Median & \phantom{$-$-}44.478$\pm$0.014     & \phantom{$-$}1.353$\pm$0.056 & \phantom{$-$}2.277$\pm$0.096  & \phantom{$-$}2.370$\pm$0.070 & \phantom{$-$}0.246$\pm$0.061 		       & 0.132 & 29 \\  
$L_{bol}-T_{spec,cor}$    & \refsim & Median & \phantom{$-$-}43.470$\pm$0.016     & \phantom{$-$}1.839$\pm$0.057 & \phantom{$-$}2.233$\pm$0.139  & \phantom{$-$}2.614$\pm$0.078 & \phantom{$-$}0.201$\pm$0.086 		       & 0.189 & 31 \\  
$L_{bol}-T_{spec,cor}$    & \agn~8.0 & Median & \phantom{$-$-}43.420$\pm$0.018     & \phantom{$-$}1.760$\pm$0.069 & \phantom{$-$}4.016$\pm$0.150  & \phantom{$-$}3.489$\pm$0.092 & $-$0.610$\pm$0.096 		       & 0.279 & 31 \\ 
$L_{bol}-T_{spec,cor}$    & \agn~8.5 & Median & \phantom{$-$-}43.035$\pm$0.023     & \phantom{$-$}2.294$\pm$0.088 & \phantom{$-$}2.945$\pm$0.178  & \phantom{$-$}3.148$\pm$0.112 & $-$0.285$\pm$0.113 		       & 0.463 & 31 \\ 
\hline
$L_{bol}-T_{spec,cor}$ & \nocool & Scatter  & \phantom{$-$}$-$0.591$\pm$0.032  & $-$0.547$\pm$0.130		        & $-$0.293$\pm$0.221 		  & \phantom{$-$}0.024$\pm$0.161 & \phantom{$-$}0.046$\pm$0.141 		       & 0.699 & 29 \\
$L_{bol}-T_{spec,cor}$    & \refsim & Scatter  & \phantom{$-$}$-$0.427$\pm$0.019  & $-$0.633$\pm$0.069		        & $-$0.309$\pm$0.167 		  & $-$0.010$\pm$0.094 & $-$0.005$\pm$0.104 		       & 0.273 & 31 \\
$L_{bol}-T_{spec,cor}$    & \agn~8.0 & Scatter  & \phantom{$-$}$-$0.488$\pm$0.018  & $-$0.420$\pm$0.070		        & $-$0.570$\pm$0.152 		  & $-$0.059$\pm$0.094 & $-$0.128$\pm$0.098 		       & 0.287 & 31 \\ 
$L_{bol}-T_{spec,cor}$    & \agn~8.5 & Scatter  & \phantom{$-$}$-$0.497$\pm$0.061  & \phantom{$-$}0.052$\pm$0.233		        & $-$1.089$\pm$0.472 		  & $-$0.352$\pm$0.297 & \phantom{$-$}0.304$\pm$0.300 		       & 3.239 & 31 \\ 
\hline
\end{tabular}
\label{table:LTbrokenpowerlaw}
\end{table*}

In order to do an analogous analysis for the bolometric X-ray luminosity--temperature scaling relation (for both core-excised and non-core-excised X-ray spectroscopic temperatures), we apply a method similar to the one described in Section~\ref{sec:fitting}. In brief, we obtain a value for the median observable and the scatter at fixed temperature about the median relation as a function of temperature and redshift as well as simple functional forms for their temperature and redshift dependencies by fitting log-normal distributions in a few temperature bins at several redshifts. Specifically, the median scaling relation at the same nine different redshifts as previously was obtained by fitting a spline to the median bolometric X-ray luminosity--temperature relation computed in ten equally logarithmically spaced temperature bins in the range $0.5\le k_{B}T\le5.0$ keV. The best-fitting spline is used to factor out the median relation when fitting a log-normal distribution to the distribution of $L_X$ in four temperature bins (chosen to be $0.5\le k_{B}T\le1.3$ keV, $1.3\le k_{B}T\le2.17$ keV, $2.17\le k_{B}T\le3.0$ keV and $3.0\le k_{B}T\le8.0$ keV). The rest of the procedure is identical to the one described in Section~\ref{sec:fitting}.

\begin{figure}
\begin{center}
\includegraphics[width=1.0\hsize]{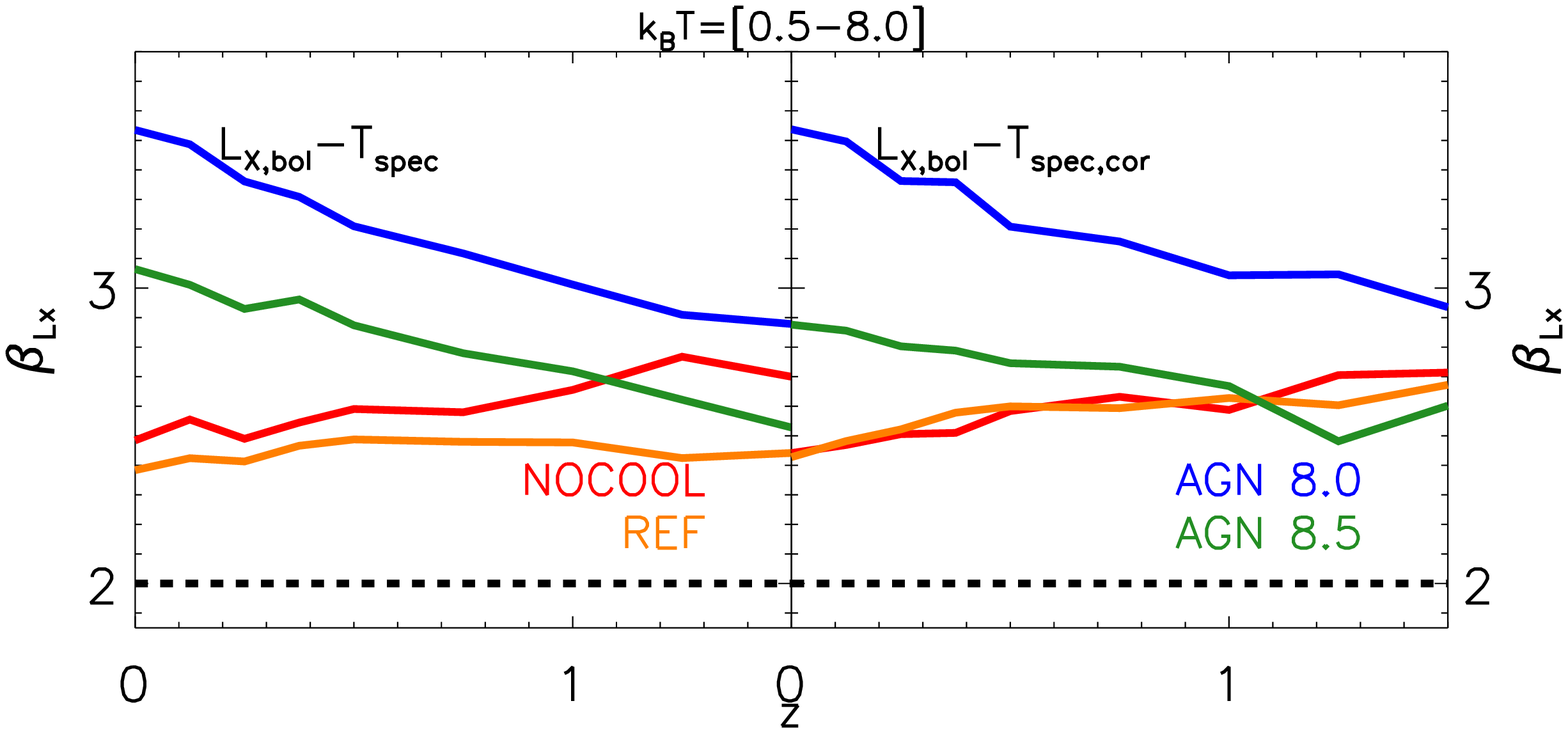}
\caption{Evolution of the temperature slope from $z=0$ to $z=1.5$ for the temperature--bolometric X-ray luminosity scaling relations (for both core-excised and non-core-excised X-ray spectroscopic temperatures). In each panel, we plot the redshift evolution of the best-fitting power-law indices obtained by fitting the broken power-law given by equation~\eqref{eq:LTpower} at each individual redshift. The solid curves (red, orange, blue and green) correspond to the different simulations and the horizontal dashed lines to the self-similar expectation, respectively. The bolometric X-ray luminosity--X-ray spectroscopic temperature (be it core excised or not) relation is steeper than self-similar for all the models considered and this independently of redshift.}
\label{fig:LTslopeevo}
\end{center}
\end{figure}

\begin{figure}
\begin{center}
\includegraphics[width=1.0\hsize]{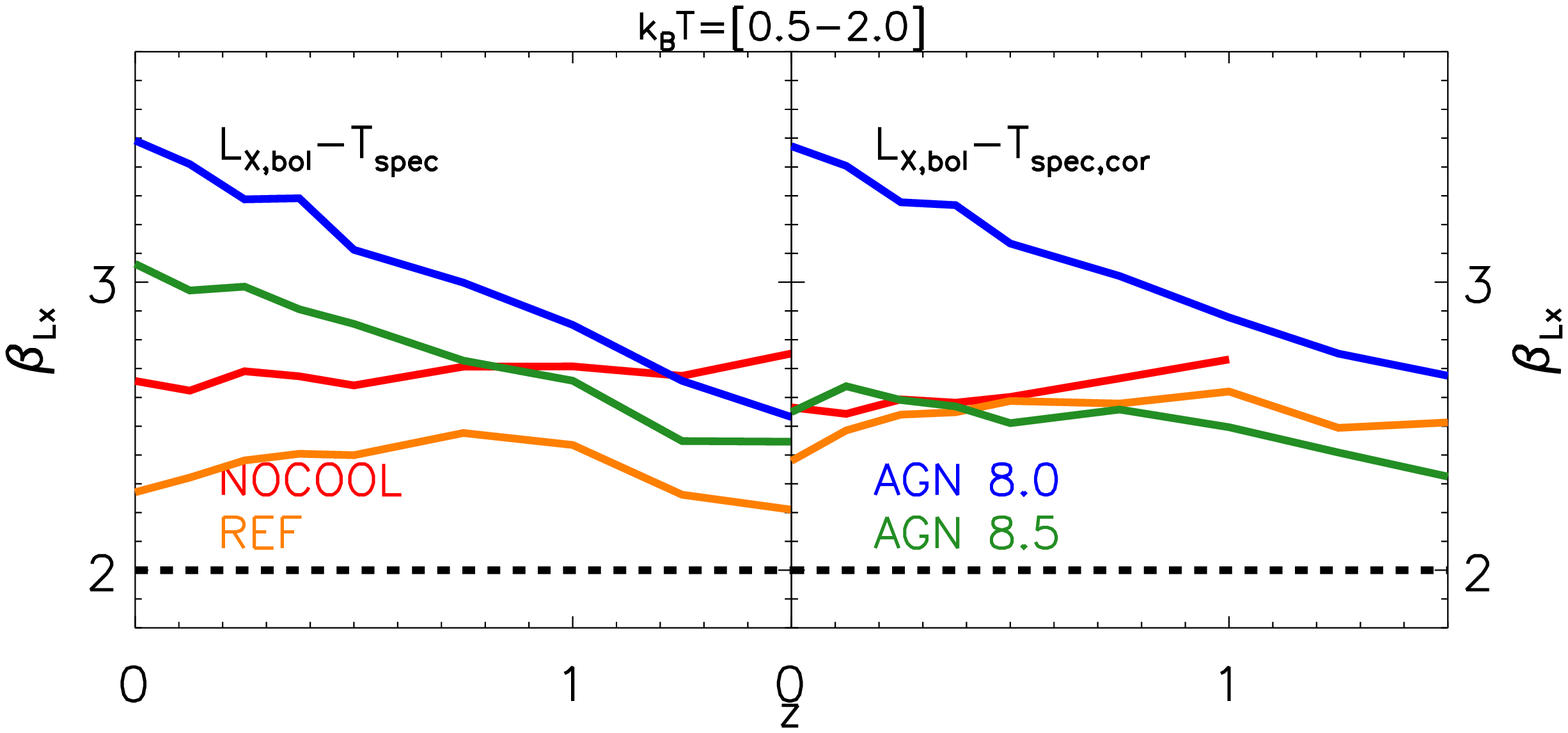}
\includegraphics[width=1.0\hsize]{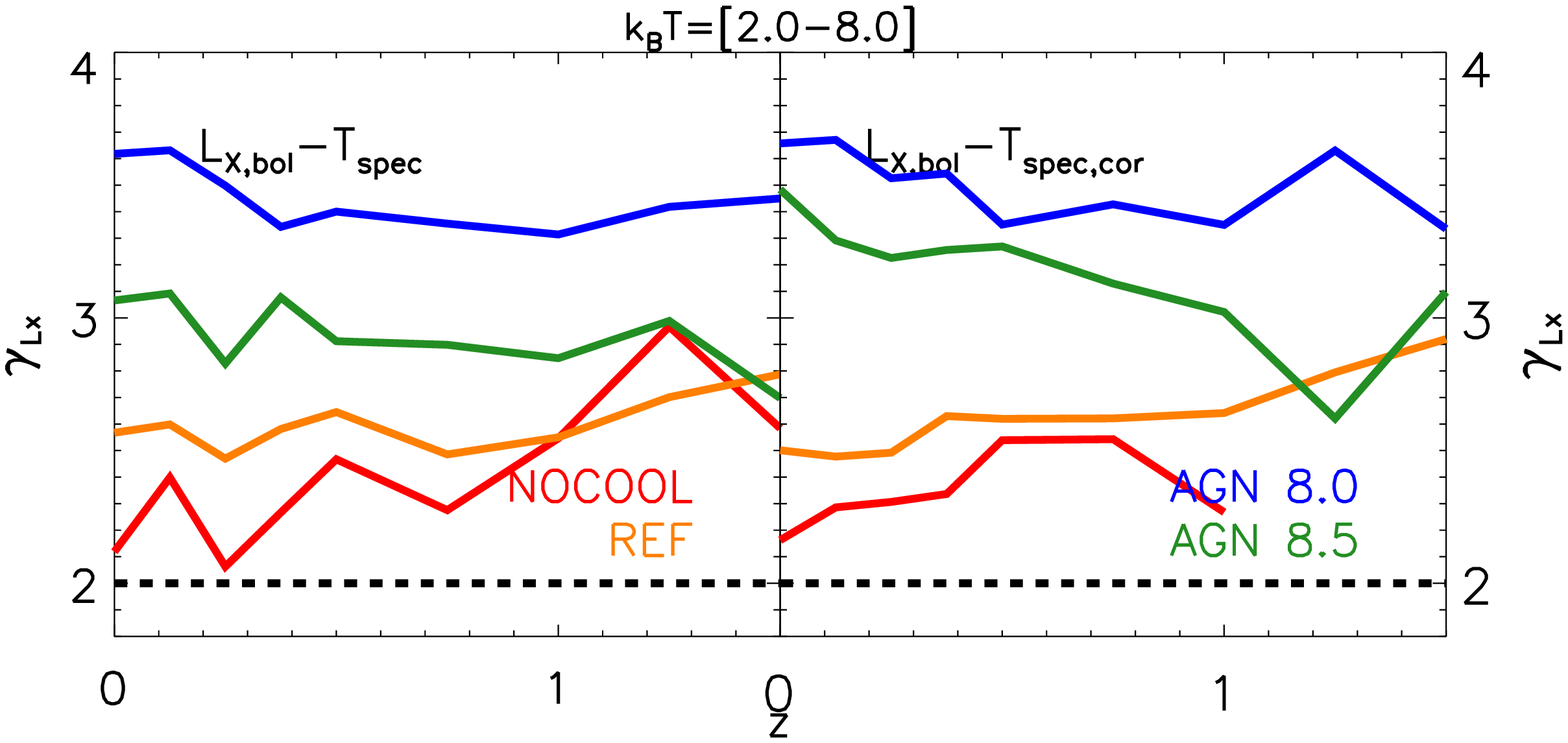}
\caption{Evolution of the temperature slope from $z=0$ to $z=1.5$ for the temperature--bolometric X-ray luminosity scaling relations (for both core-excised and non-core-excised X-ray spectroscopic temperatures). In each subpanel, we plot the redshift evolution of the low-temperature (\emph{top} panel) and high-temperature (\emph{bottom} panel) best-fitting power-law indices obtained by fitting the broken power-law given by equation~\eqref{eq:LTbroken} at each redshift independently. The solid curves (red, orange, blue and green) correspond to the different simulations and the horizontal dashed lines to the self-similar expectation, respectively. The bolometric X-ray luminosity--X-ray spectroscopic temperature (be it core excised or not) relation is steeper than self-similar for all the models considered and this independently of temperature and redshift.}
\label{fig:LTslopeevobroken}
\end{center}
\end{figure}

\begin{figure}
\begin{center}
\includegraphics[width=1.0\hsize]{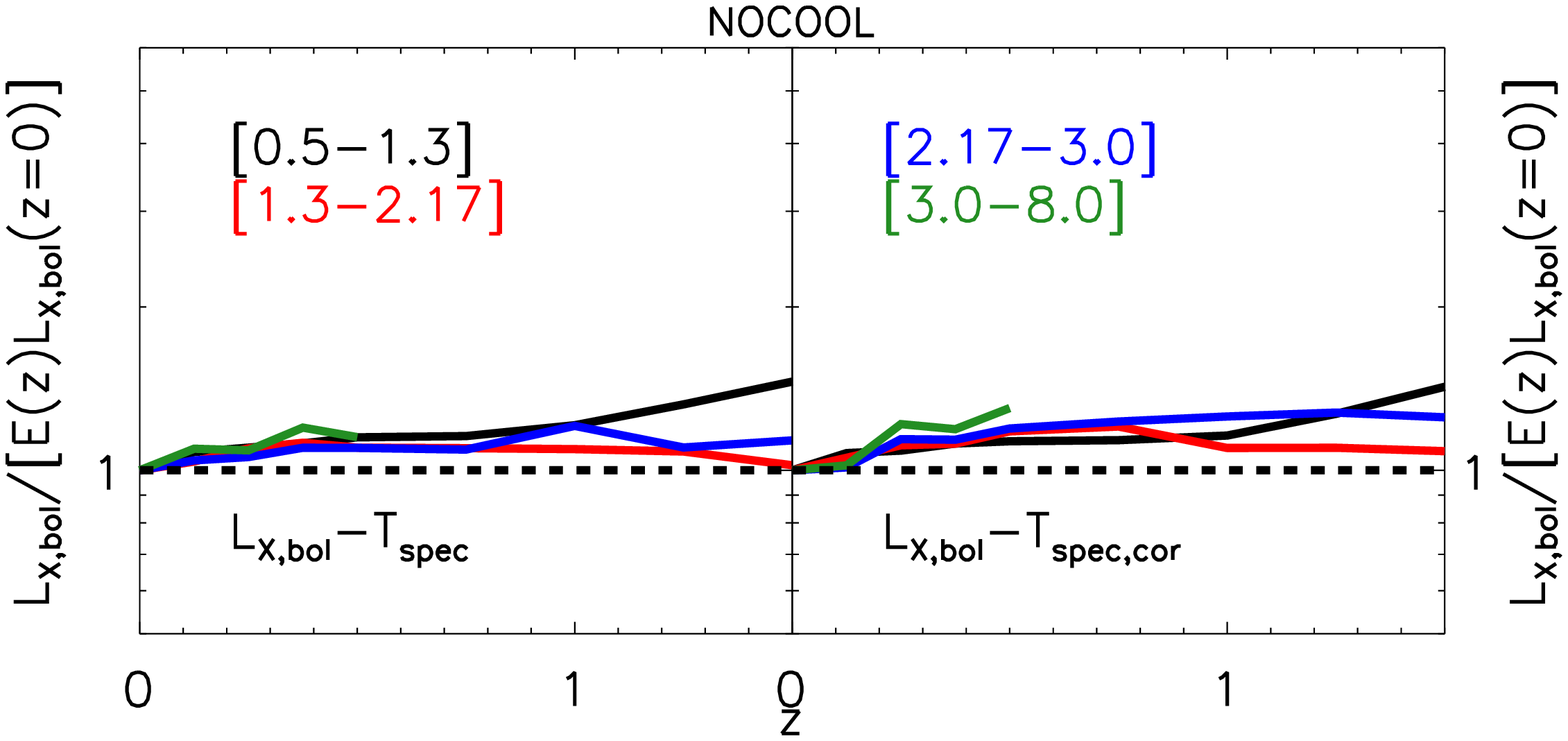}
\includegraphics[width=1.0\hsize]{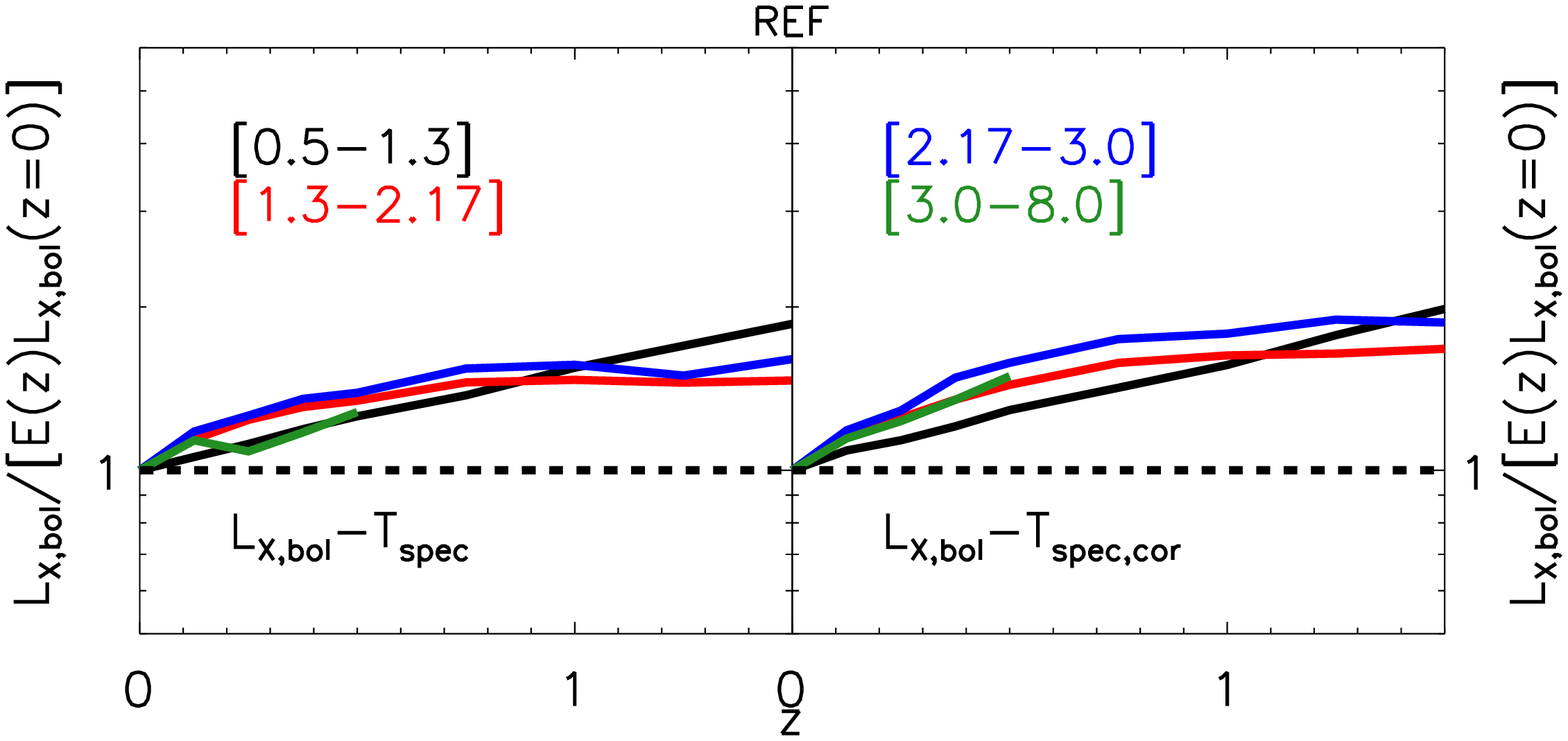}
\includegraphics[width=1.0\hsize]{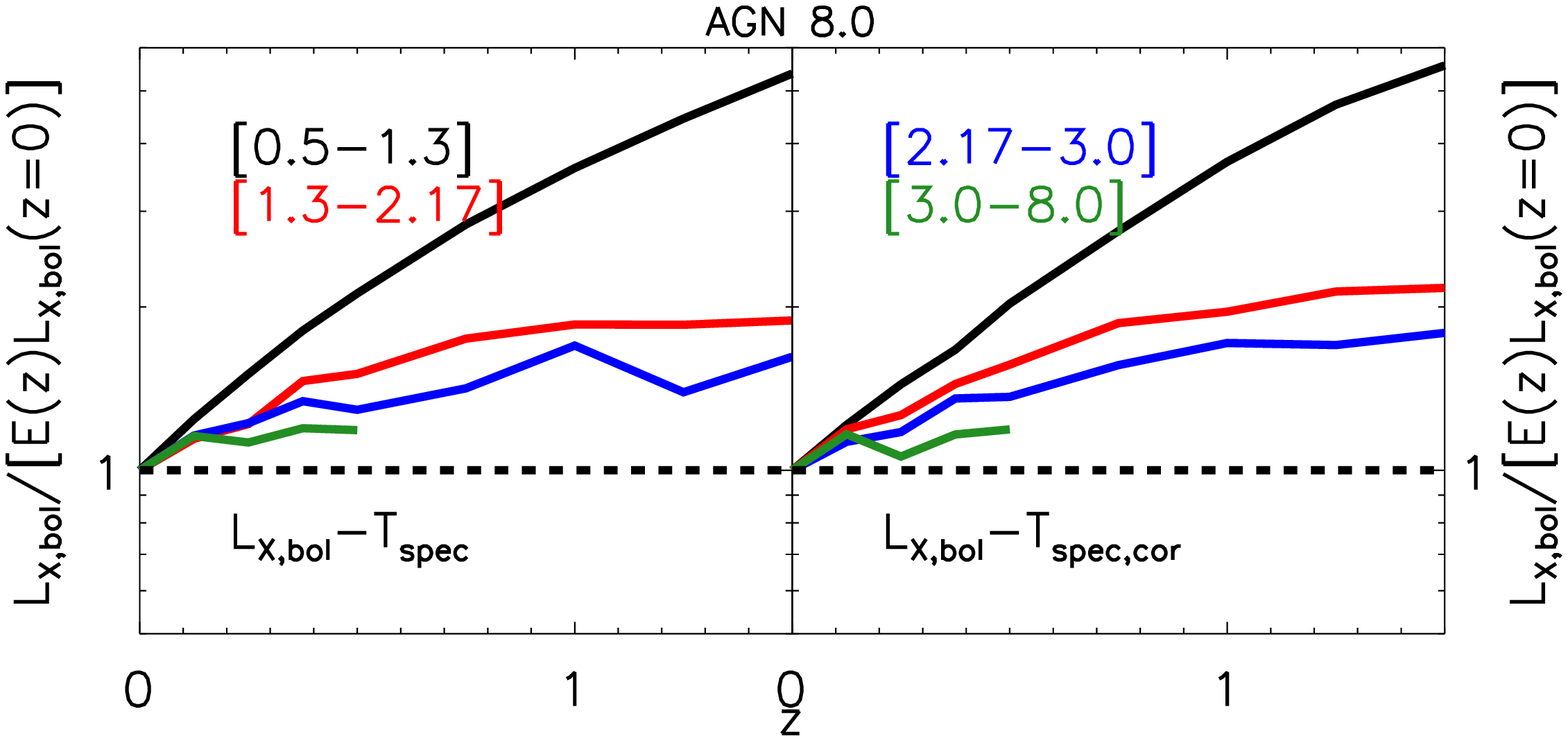}
\includegraphics[width=1.0\hsize]{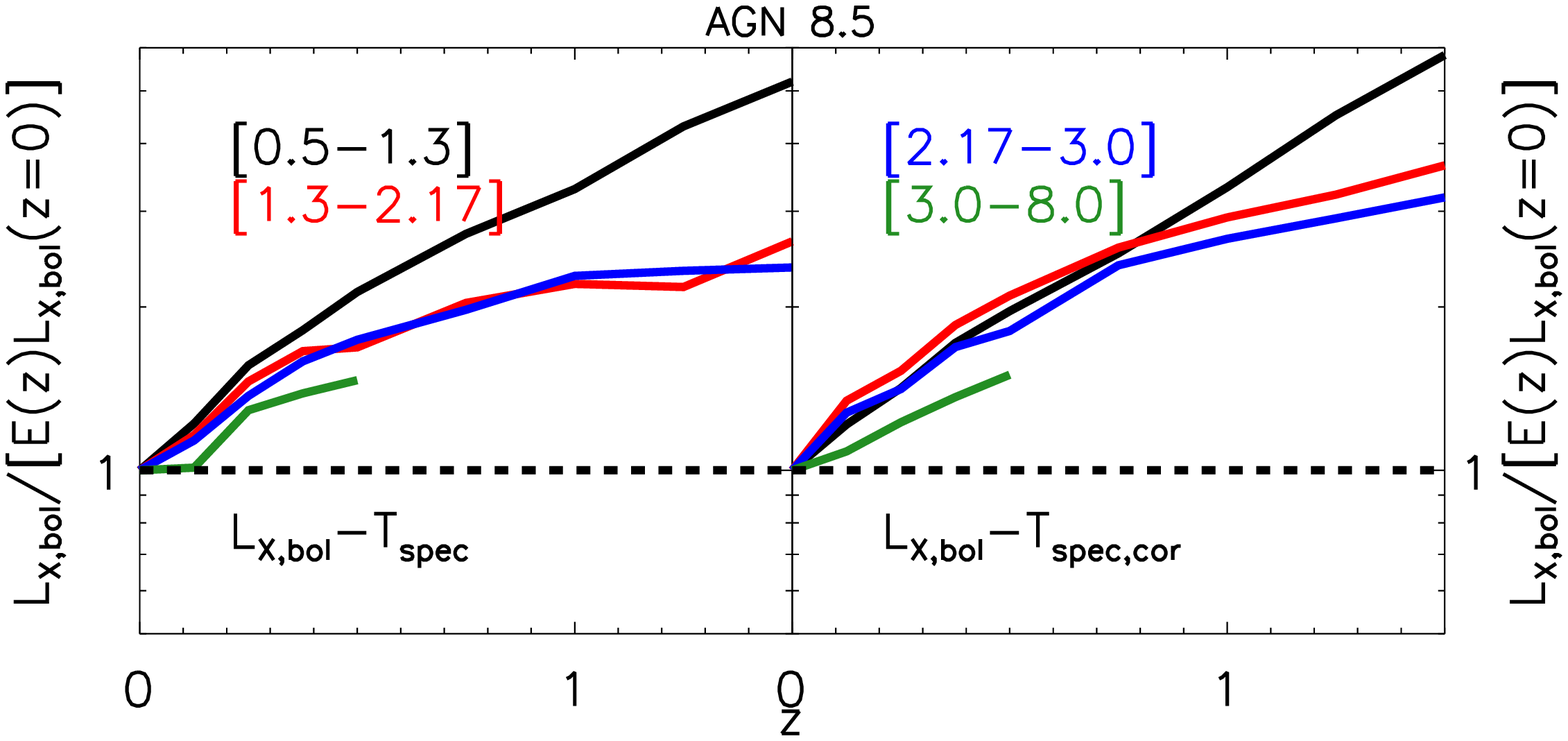}
\caption{Evolution of the normalisation from $z=0$ to $z=1.5$  for the temperature--bolometric X-ray luminosity scaling relations (for both core-excised and non-core-excised X-ray spectroscopic temperatures) for each of the simulations. The normalisations of each scaling relation in the four temperature bins (denoted by solid lines of different colours) have been normalised by the self-similar expectation for the redshift evolution at fixed temperature (shown as an horizontal dashed line). The amplitude of the bolometric X-ray luminosity--X-ray spectroscopic temperature relation evolves positively for all the physical models with an amplitude which is strongly temperature dependent, slightly redshift dependent, strongly sensitive to the non-gravitational physics of galaxy formation (it becomes more positive with increasing feedback intensity) and slightly sensitive to cosmology (slightly faster evolution in the \wmap7 cosmology, especially for the highest temperature bin).}
\label{fig:LTnormevo}
\end{center}
\end{figure}

\begin{figure*}
\begin{center}
\includegraphics[width=0.497\hsize]{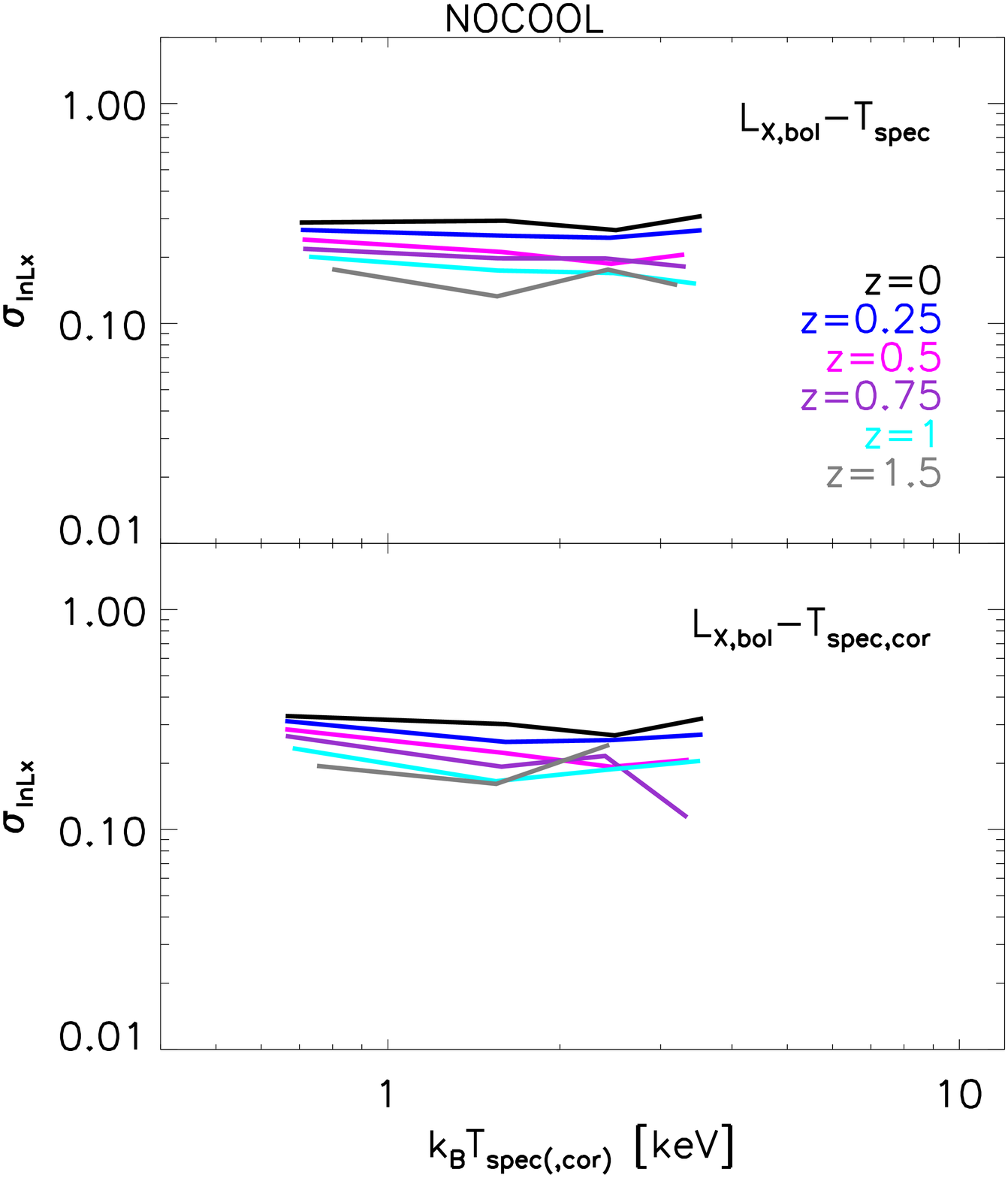}
\includegraphics[width=0.497\hsize]{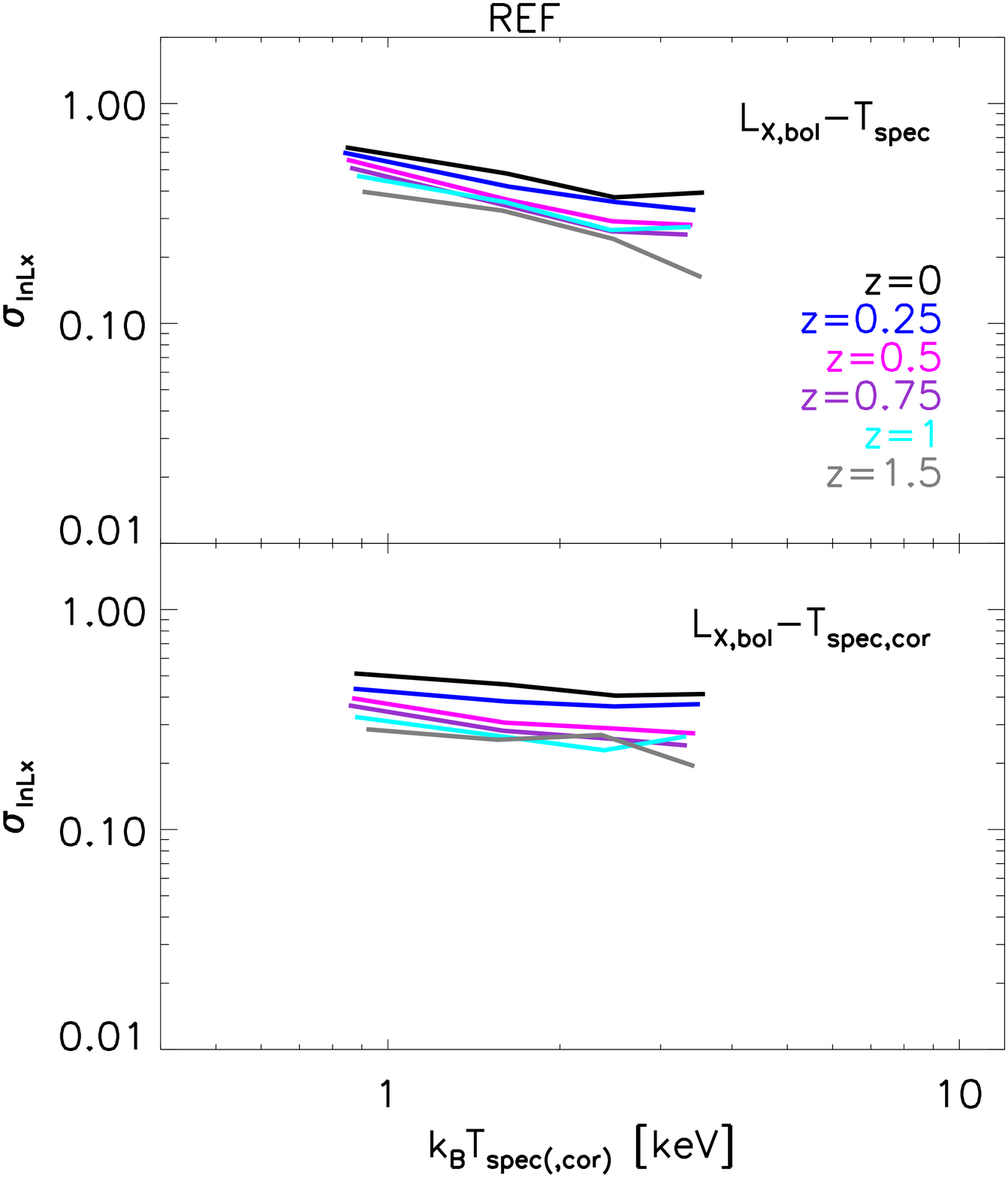}
\includegraphics[width=0.497\hsize]{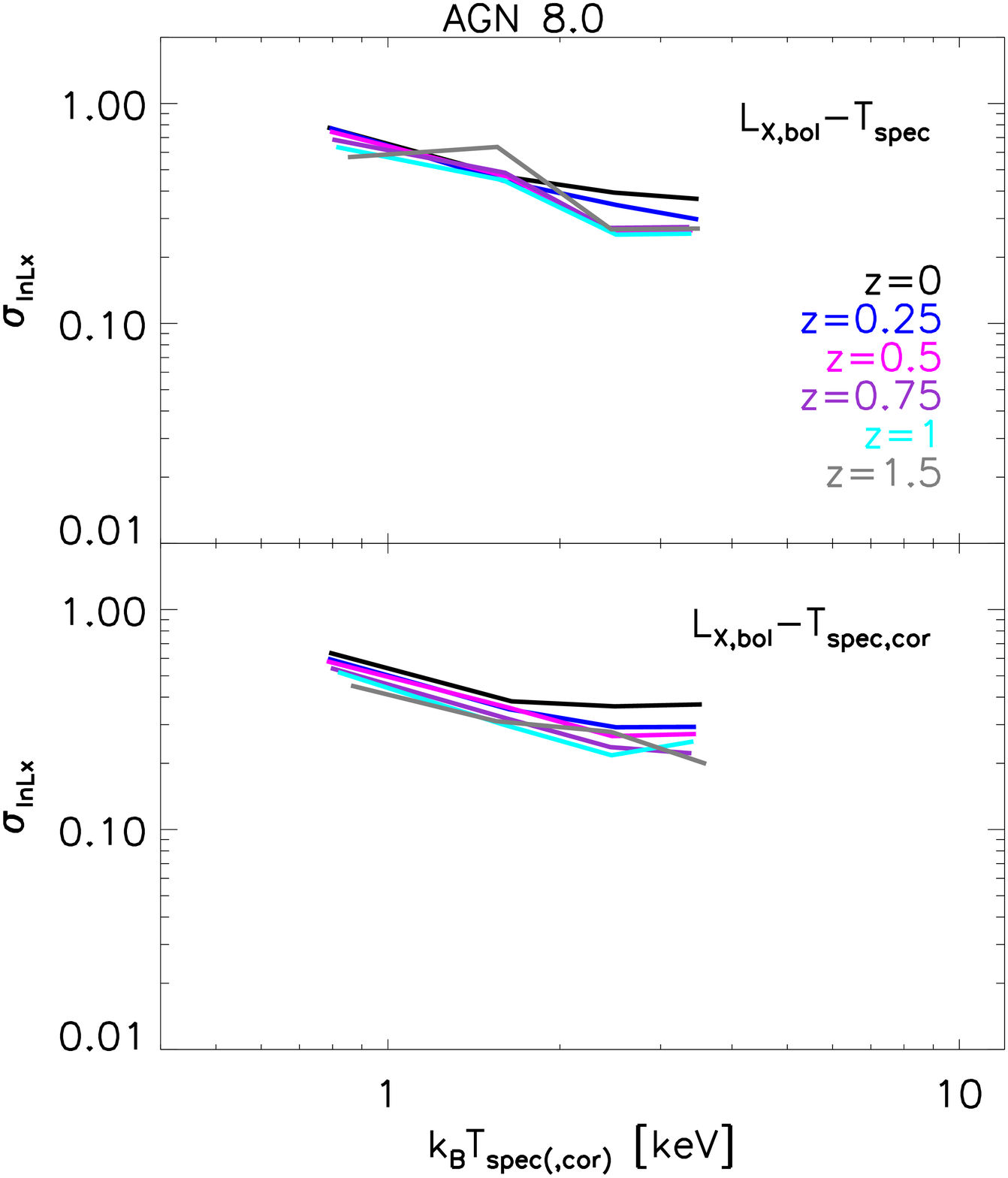}
\includegraphics[width=0.497\hsize]{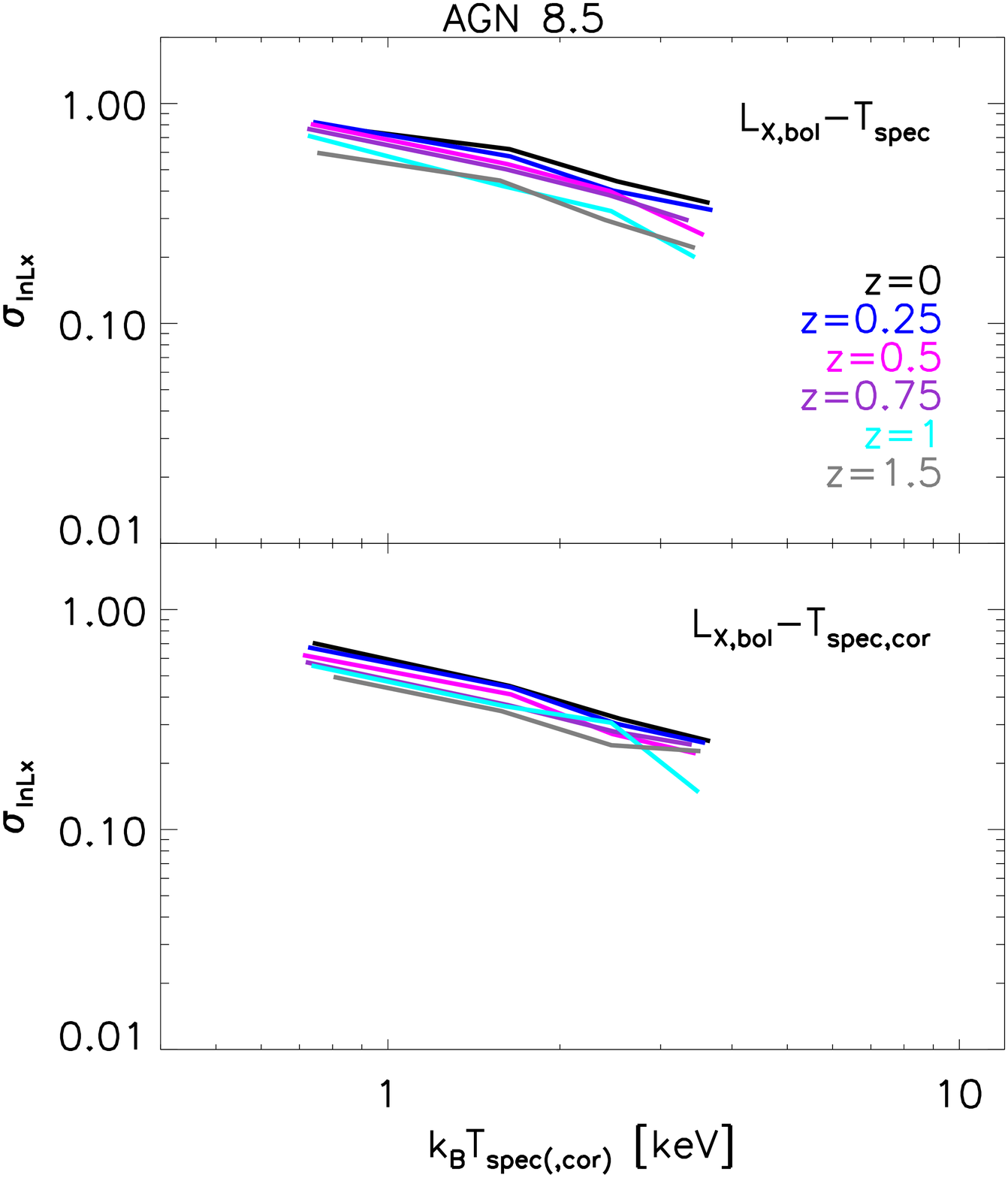}
\caption{Evolution of the log-normal scatter from $z=0$ to $z=1.5$ for the temperature--bolometric X-ray luminosity scaling relations (for both core-excised and non-core-excised X-ray spectroscopic temperatures). For each simulation and each scaling relation, we plot the log-normal scatter as a function of $T_{spec(,cor)}$ and denote the redshift using lines of different colours. The log-normal scatter of the bolometric X-ray luminosity--X-ray spectroscopic temperature relation varies only mildly with temperature (typically changing by a factor of $\approx2-3$ over $\sim1$ decade in temperature), is somewhat sensitive to non-gravitational physics (the amplitude tends to increase with increasing feedback intensity), but displays a moderately strong redshift dependence (it tends to decrease with increasing redshift).}
\label{fig:LTscatter}
\end{center}
\end{figure*}

We can then fit an evolving power-law of the form:
\begin{equation}
L_X=10^AE(z)^\alpha\left(\frac{k_BT}{2~\textrm{keV}}\right)^\beta,
\label{eq:LTpower}
\end{equation}
an evolving broken power-law of the form:
\begin{equation}
L_X=10^{A'}E(z)^{\alpha'}\left(\frac{k_BT}{2~\textrm{keV}}\right)^{\epsilon'}
\label{eq:LTbroken}
\end{equation}
where 
\begin{equation}
\epsilon'=
\begin{cases}
\beta' & \text{if } k_BT\le2~\textrm{keV} \\
\gamma' & \text{if } k_BT>2~keV, 
\end{cases}
\end{equation}
and finally an evolving broken power-law with a redshift dependent low-mass power-law index of the form:
\begin{equation}
L_X=10^{A''}E(z)^{\alpha''}\left(\frac{k_BT}{2~\textrm{keV}}\right)^{\epsilon''}
\label{eq:LTbrokenextra}
\end{equation}
where 
\begin{equation}
\epsilon''=
\begin{cases}
\beta''+\delta''E(z) & \text{if } k_BT\le2~\textrm{keV} \\
\gamma'' & \text{if } k_BT>2~keV
\end{cases}
\end{equation}
to the median relation and the log-normal scatter about it as a function of both temperature and redshift.  The results of the evolving power-law and broken power-law with an evolving low-mass power-law index fitting for all the physical models are presented in Tables~\ref{table:LTpowerlaw} and~\ref{table:LTbrokenpowerlaw}, respectively.

In Figs.~\ref{fig:LTslopeevo} and~\ref{fig:LTslopeevobroken}, we show the evolution of the mass slopes from $z=0$ to $z=1.5$ for the bolometric X-ray luminosity--X-ray spectroscopic temperature (be it core excised or not) for each of the four physical models. Fig.~\ref{fig:LTslopeevo} shows the redshift evolution of the temperature slope obtained by fitting the power-law given by equation~\eqref{eq:LTpower} at each individual redshift whereas Fig.~\ref{fig:LTslopeevobroken} shows the redshift evolutions of the low-temperature (\emph{top} panel) and high-temperature (\emph{bottom} panel) temperature slopes resulting from the fitting of the broken power-law given by equation~\eqref{eq:LTbroken} at each redshift independently. The solid curves (red, orange, blue and green) corresponding to the different simulations and the horizontal dashed lines to the self-similar expectation, respectively. The bolometric X-ray luminosity--X-ray temperature (be it core excised or not) relation is steeper than the self-similar expectation of 2 (see equation~\ref{eq:ssLxbol-T}) for bolometric X-ray luminosity for all the models considered here and this independently of temperature and redshift. The bolometric X-ray luminosity--X-ray temperature relation therefore qualitatively behaves in exactly the same way as the bolometric X-ray luminosity--total mass (see Section~\ref{sec:slopeevo} and Figs.~\ref{fig:slopeevo} and~\ref{fig:slopeevobroken}). Hence, the deviations from self-similarity are most likely due to the same reasons, namely that the gas does not trace the dark matter affecting both density and temperature, which are also potentially further affected by non-thermal pressure support. 

In Fig.~\ref{fig:LTnormevo}, we show the evolution of the normalisation from $z=0$ to $z=1.5$ for the bolometric X-ray luminosity--X-ray spectroscopic temperature (be it core excised or not) relation for the four physical models considered here. The normalisations of each scaling relation in the four temperature bins (denoted by solid lines of different colours) have been normalised by the self-similar expectation for the redshift evolution at fixed mass (shown as an horizontal dashed line). As was the case for the bolometric X-ray luminosity--total mass relation and the models which include AGN feedback (see Section~\ref{sec:normevo} and Fig.~\ref{fig:normevo}), the amplitude of the bolometric X-ray luminosity--X-ray temperature relation evolves positively for all the physical models with an amplitude which is strongly temperature dependent, slightly redshift dependent (it slightly flattens out with increasing redshift) and strongly sensitive to the non-gravitational physics of galaxy formation (it becomes more positive with increasing feedback intensity). The reversal of the direction of the deviation from self-similarity for the models without AGN feedback (from negative for the bolometric X-ray luminosity--total mass relation to positive for the bolometric X-ray luminosity--X-ray spectroscopic temperature relation is straightforwardly explained by the negative evolution of the mass--temperature relation. \citet{Short2010} also found positive evolution for zoom simulations which include a semi-analytic prescription for AGN feedback.

In Fig.~\ref{fig:LTscatter}, we show the evolution of the log-normal scatter (at fixed temperature) from $z=0$ to $z=1.5$ for the bolometric X-ray luminosity--X-ray spectroscopic temperature (be it core excised (\emph{bottom} subpanels) or not (\emph{top} subpanels)) relation for the four physical models considered here. For each simulation and each scaling relation, we plot the log-normal scatter as a function of $T_{spec(,cor)}$ and denote the redshift using lines of different colours. As was the case for the bolometric X-ray luminosity--total mass relation (see Section~\ref{sec:scatter} and Fig.~\ref{fig:scatter}), the log-normal scatter of the bolometric X-ray luminosity--X-ray spectroscopic temperature relation varies only mildly with temperature (typically changing by a factor of $\approx2-3$ over $\sim1$ decade in temperature), is somewhat sensitive to non-gravitational physics (the amplitude tends to increase with increasing feedback intensity), but displays a moderately strong redshift dependence (it tends to decrease with increasing redshift).

All the differences between the behaviours of the bolometric X-ray luminosity--mass and bolometric X-ray luminosity--X-ray spectroscopic temperature can be easily explained by the behaviour of the mass--temperature relation.

Finally, it is worth noting that the calibrated \calsim~model has a much steeper bolometric X-ray luminosity--X-ray spectroscopic temperature relation than the cosmo-OWLS \agn~models apart at the high-temperature end and that the bolometric X-ray luminosity--X-ray spectroscopic temperature relation evolves slightly faster in the \wmap7 cosmology, especially for the highest temperature bin.

\bsp	
\label{lastpage}
\end{document}